\documentclass[
 aip,
 amsmath,amssymb,
 reprint,%
]{revtex4-1}

\usepackage{graphicx}% Include figure files
\usepackage{dcolumn}% Align table columns on decimal point
\usepackage{bm}% bold math
\usepackage[utf8]{inputenc}
\usepackage[T1]{fontenc}
\usepackage{mathptmx}
\usepackage{etoolbox}
\usepackage{multirow}
\usepackage{color}

\makeatletter
\newcommand{\rmnum}[1]{\romannumeral #1}
\newcommand{\Rmnum}[1]{\expandafter\@slowromancap\romannumeral #1@}

\def\@email#1#2{%
 \endgroup
 \patchcmd{\titleblock@produce}
  {\frontmatter@RRAPformat}
  {\frontmatter@RRAPformat{\produce@RRAP{*#1\href{mailto:#2}{#2}}}\frontmatter@RRAPformat}
  {}{}
}%
\makeatother
\begin{document}

\preprint{AIP/123-QED}

\title{\textcolor{black}{Efficient generation of divergent and collimated hot electrons via a novel multi-beam two-plasmon decay and stimulated Raman scattering mechanism}
%Angular variations in hot electron generation through two-plasmon decay instability driven by dual symmetric laser beams
}
% Force line breaks with \\
\author{K.Y. Meng}
\affiliation{Department of Plasma Physics and Fusion Engineering and CAS Key Laboratory of Geospace Environment, University of Science and Technology of China, Hefei, Anhui 230026, People’s Republic of China.}
 
\author{Z.H.Cai}
\affiliation{Department of Plasma Physics and Fusion Engineering and CAS Key Laboratory of Geospace Environment, University of Science and Technology of China, Hefei, Anhui 230026, People’s Republic of China.}

\author{J.Li}%
\thanks{Corresponding author: junlisu@ustc.edu.cn}
% \email{junlisu@ustc.edu.cn}
\affiliation{Department of Plasma Physics and Fusion Engineering and CAS Key Laboratory of Geospace Environment, University of Science and Technology of China, Hefei, Anhui 230026, People’s Republic of China.}%
\affiliation{ 
Collaborative Innovation Center of IFSA (CICIFSA), Shanghai Jiao Tong University, Shanghai 200240, People’s Republic of China%\\This line break forced with \textbackslash\textbackslash
}%

\author{C.Yao}
\affiliation{Department of Plasma Physics and Fusion Engineering and CAS Key Laboratory of Geospace Environment, University of Science and Technology of China, Hefei, Anhui 230026, People’s Republic of China.}
\affiliation{Institute of Applied Physics and Computational Mathematics, Beijing 100088, People’s Republic of
China%\\This line break forced% with \\
}%}

\author{L.Hao}
\affiliation{Institute of Applied Physics and Computational Mathematics, Beijing 100088, People’s Republic of
China%\\This line break forced% with \\
}%}

\author{F.X.Zhou}
\affiliation{%
Department of Modern Mechanics, University of Science and Technology of China, Hefei, Anhui
230026, People’s Republic of China.}%

\author{R. Yan}
\thanks{Corresponding author: ruiyan@ustc.edu.cn}
%\email{ruiyan@ustc.edu.cn}
\affiliation{%
Department of Modern Mechanics, University of Science and Technology of China, Hefei, Anhui
230026, People’s Republic of China.}%
\affiliation{ 
Collaborative Innovation Center of IFSA (CICIFSA), Shanghai Jiao Tong University, Shanghai 200240, People’s Republic of China%\\This line break forced with \textbackslash\textbackslash
}%

\author{J. Zheng}
\affiliation{Department of Plasma Physics and Fusion Engineering and CAS Key Laboratory of Geospace Environment, University of Science and Technology of China, Hefei, Anhui 230026, People’s Republic of China.}%
\affiliation{ 
Collaborative Innovation Center of IFSA (CICIFSA), Shanghai Jiao Tong University, Shanghai 200240, People’s Republic of China%\\This line break forced with \textbackslash\textbackslash
}%

\date{\today}% It is always \today, today,
             %  but any date may be explicitly specified

\begin{abstract}

In inertial confinement fusion (ICF) implosions, the preheating risks associated with hot electrons generated by laser plasma instabilities (LPI) are contingent upon the angular characteristics of these hot electrons for a given total energy.
Using particle-in-cell simulations, we \textcolor{black}{reveal a novel multi-beam collaborative mechanism of two-plasmon decay (TPD) and stimulated Raman scattering (SRS), and} investigate the angular variations of hot electrons generated from \textcolor{black}{this shared TPD-SRS (STS) instability} %, one of the major LPIs that produce hot electrons, 
driven collectively by dual laser beams %symmetric to the density gradient 
with varying incident angles $\theta_{in}$ ($24^\circ$ to $55^\circ$ at the incident plane)
for typical ICF conditions.
\textcolor{black}{In the simulations with $\theta_{in}\gtrsim44^\circ$, \textcolor{black}{STS} emerges as the dominant mechanism responsible for hot electron generation, leading to} a wide angular distribution of hot electrons \textcolor{black}{that exhibit both pronounced divergent and collimated components. The common Langmuir wave associated with \textcolor{black}{STS} plays a crucial role in accelerating both components.}
%TSI emerges as the dominant mechanism of generating hot electrons, and causes }the hot electrons spread widely in angle with \textcolor{black}{pronounced divergent and collimated components. The common Langmuir wave associated with TSI plays a key role in accelerating both the two components.
%The divergent hot electrons, exhibiting side peaks at up to $\sim72^\circ$ to the density gradient, are produced by the divergentTSI common plasma waves.} 
%Conversely, the collimated hot electrons with narrow angular spreadsare produced for all the $\theta_{in}$. 
%Notably, \textcolor{black}{they are still produced by two divergent TSI common plasma waves, which form collimated moving electric fields through interference.} 
By properly modeling the \textcolor{black}{STS} common wave gains, we establish scaling relations between these gains and the energies of collimated and divergent hot electrons. These relations reveal that the divergent hot electrons are more sensitive to variations in gain compared to the collimated electrons. Additionally, the calculated gains qualitatively predict the asymmetry in hot electron angular distributions when the density gradients deviate from the bisector of the laser beams.
Our findings offers insights for hot electron generation with multiple beams, potentially complementing previous experiments that underscore the critical role of overlapped intensity from symmetric beams within the same cone and the dominance of dual-beam coupling.

\end{abstract}

\maketitle

\section{\label{sec:level1}Introduction}

In inertial confinement fusion (ICF), a spherical capsule is imploded by multiple high-power laser beams to compress the central fuels to extreme density and pressure, thereby initiating self-sustained fusion reactions\cite{Atzeni2004}. During the implosion, hot electrons (with energy above $\sim50$ keV) produced by laser plasma instabilities(LPI)\cite{Kruer2003} can preheat the uncompressed fuel and degrade the implosion\cite{christopherson2021,Solodov2022}. In most ICF schemes\cite{Nuckolls1972,He2016,Zhang2020,Lan2022}, such hot electrons need to be reduced to improve the implosion, except for shock ignition\cite{Betti2007} (SI), in which hot electrons generated during the spike pulse can enhance the shock pressure\cite{Baton2020,Tentori2021} and enhance the compression. 
\textcolor{black}{While ignition has been achieved using the indirect-drive scheme on the NIF laser facility\cite{Abu-shawareb2024} with hot electron threats minimized by unique experimental configurations, the pursuit of higher energy gains\cite{Batani2023,Sui2024} and alternative ignition schemes\cite{Gopalaswamy2024,Williams2024,He2016,Zhang2020,Lan2022,Betti2007} still demands effective control and thorough characterization of hot electron effects.}
Thus, reliably characterizing the properties of hot electrons is crucial to \textcolor{black}{evaluating} their potential detrimental or beneficial impacts.
%Currently, although ignition has been achieved for indirect-drive scheme on NIF laser facility\cite{Abu-shawareb2024}, the pursuit for high-gain design and other ignition schemes\cite{He2016,Zhang2020,Lan2022} still require proper control and characterization of hot electron effects.}
%Therefore, it is essential to characterize the properties of hot electron to assess their detrimental or beneficial effects. 

Previous ICF experiments have explored the hot electron energy and temperature scaling relations under various laser-plasma conditions\cite{Michel2013,Froula2012}. 
One primary source of these hot electrons \textcolor{black}{has been identified as } two-plasmon decay (TPD), a major LPI in both direct\cite{Seka2009a,Turnbull2020b} and indirect\cite{Regan2010} ICF schemes. 
In TPD, one laser wave decays into two electron plasma waves (EPW), efficiently transferring energy to the EPWs\cite{Kruer2003}, which can accelerate background electrons. % moving near the EPW phase velocity of the EPWs.
In addition to experimental studies, scaling laws have been examined through first-principle particle-in-cell (PIC) simulations\cite{Cao2022,Cao2023}, considering both TPD and stimulated Raman scattering (SRS)—another key LPI that generates hot electrons\cite{Rosenberg2018,Dewald2016}. In SRS, a laser wave decays into an EPW and an electromagnetic wave (EMW)\cite{Kruer2003}. \textcolor{black}{In prior direct-drive ICF experiments, TPD has been recognized as the primary source of hot electrons\cite{Seka2009a,Michel2013,turnbull2020,Turnbull2020b}, except under NIF experimental conditions involving high electron thermal temperature (4-5 keV) and long density scale length (500-700$\mu m$), where SRS dominates hot electron generation\cite{Rosenberg2018,Wen2015}. }

\textcolor{black}{
Various approaches have been developed to mitigate hot electron generation, including advancements in target design\cite{Follett2016} and broadening of laser bandwidths\cite{Yao2024,Lei2024,Wang2024,Liu2024,Ma2021,Follett2021,Gao2020,Zhao2017}. \textcolor{black}{These studies primarily focus on the suppression of the total hot electron energy.} Nevertheless, uncertainties \textcolor{black}{remain concerning} the angular distribution of hot electrons.}
Given hot electron energy and temperature, the impact of hot electrons depends on their angular characteristics. In conventional center hot-spot ignition for both direct\cite{christopherson2021} and indirect drive\cite{Dewald2016} ICF, directional hot electrons that propagate towards the capsule center pose a greater preheating threat than those moving laterally.
%cause much stronger preheating threats compared to those going sideways. 
Similarly, the SI scheme benefits from directional hot electrons to augment the shock pressure. 
Conversely, in the emerging double-cone ignition (DCI) scheme\cite{Zhang2020}, hot electrons generated in the compression cone and moving laterally may preheat the ignition cone, forming local pre-plasmas that reduce the energy coupling efficiency of the picosecond ignition laser pulse to the compressed fuel\cite{Macphee2010,Li2013}.

Experiments have demonstrated that the angular distribution of hot electrons varies with laser illumination geometry and plasma conditions.
%In past experiments, the hot electron angular characteristics vary with laser illuminating geometry and laser plasma conditions.
Early studies\cite{Ebrahim1980} involving single-beam $CO_2$ laser pulses interacting with long-scale-length plasmas have revealed that hot electrons are concentrated at $45^\circ$ relative to the laser incident direction in the polarization plane. 
%Early experiments\cite{Ebrahim1980} of single-beam $CO_2$ laser pulse interacting with long-scale-length plasmas have demonstrated hot electrons peaked at $45^{\circ}$ to the laser incident direction in the polarization plane.
In contrast, more recent experiments\cite{Yaakobi2013} on the OMEGA laser facility have observed a wide divergence of hot electrons during implosions of spherical targets illuminated by 60 laser beams, \textcolor{black}{which agrees well with the PIC simulations\cite{Yan2012,Yan2009}}.
%In contrast, more recent experiments\cite{Yaakobi2013} performed on OMEGA laser facility have demonstrated wide divergence of hot electrons during the implosions of spherical targets illuminated by all the 60 laser beams. 
In subsequent SI-relevant experiments\cite{Zhang2020a} on the OMEGA-EP laser facility, where intense single-beam laser pulses (with intensities of $5-10\times10^{15} W/cm^2$) interacted with planar targets, the measured hot electrons associated with both TPD and SRS\cite{Li2020} exhibited directional characteristics. \textcolor{black}{A more comprehensive experimental study\cite{Liu2017} using single laser beam at varying intensities has shown that the angular distributions of hot electrons varies depending on the specific laser and target configurations.
}
%In the later SI-relevant experiments\cite{Zhang2020} on OMEGA-EP laser facility, when single-beam intense laser pulse (with intensity $5-10\times10^{15}W/cm^2$) interacts with planar targets, the measured electrons associated with both TPD and SRS\cite{Li2020} exhibit directional features. 
The broad angular spread of hot electrons is generally attributed to the wide EPW spectrum in the nonlinear phase of TPD\cite{Myatt2014,Yan2014}, \textcolor{black}{while the SRS hot electrons are considered more directional\cite{Yan2014}.}
%Among these results, the wide angular spread of hot electrons are usually considered relevant to the broad EPW spectra in the nonlinear stage of TPD\cite{Myatt2014}. 
Nevertheless, the mechanisms behind the generation of hot electrons with various angular features remain insufficiently understood.
%\textcolor{black}{However, the generation mechanisms of the hot electrons with different angular features 
%have not been well understood.} 
A key factor influencing these angular characteristics is the laser beam geometry, as the daughter EPWs in TPD and SRS are primarily driven by wave vector matching across multiple laser beams in current ICF experiments\cite{Myatt2014,Michel2015}.
%\textcolor{black}{Apparently, one of the main factors acting on the angular features is the laser beam geometry, since the daughter EPWs of TPD and SRS are mostly driven by matching the wave vector with multiple laser beams in current ICF experiments\cite{Myatt2014,Michel2015}.}

In this article, 
we investigate the angular variations of \textcolor{black}{hot electrons} in relation to laser illumination geometry using fully kinetic PIC simulations and theoretical analysis.
Our results demonstrate that laser geometry is the primary determinant of the angular features of hot electrons, \textcolor{black}{which are mainly driven by a novel shared TPD-SRS (STS) mechanism for moderate or high incident angles.
These hot electrons exhibit both pronounced divergent and collimated components, differing from previous understanding, where SRS mainly produces collimated hot electrons and TPD generates widely spread hot electrons. }
In our study, we specifically focus on dual laser beams symmetric to the density gradient, \textcolor{black}{as the theoretical model we developed under symmetric configurations is also capable of accounting for asymmetry.}  
There are two key reasons for emphasizing the symmetric dual beam cases.
First, 
these configurations can be extended to more beams within the same incident cone, where the overlapped intensity plays a dominant role in hot electron generation\cite{Michel2013,Regan2010}. 
Second, typical OMEGA-EP experiments have shown the dominance of dual-beam coupling rather than more beams in TPD growth\cite{Michel2012a}. 
We study incident angle $\theta_{in}$ ranging from $24^{\circ}$ to $55^{\circ}$, as $\theta_{in}\sim60^{\circ}$ brings the laser reflection density near the quarter-critical density $n_c/4$, lowering the TPD threshold$--$a scenario already addressed in previous work\cite{Lian2022,Zhou2023}. 
Higher $\theta_{in}$ values shift the reflection density away from $n_c/4$ to lower density region, making TPD instability unlikely.

This article is organized as follows. Section II describes the simulation setup. In Section III, we provide simulation results with theoretical explanations and discuss the underlying physics.
%we give explanations of these results and discuss the underlying physics. 
\textcolor{black}{Section IV offers further discussion on the applicability of our results to realistic experimental conditions, 
%\textcolor{red}{including the asymmetric configurations}. 
Section V presents the conclusions of this study.}

\section{\label{sec:level2}Simulation setup}
%\subsection{Setup}

\begin{figure}
\includegraphics[width=0.48\textwidth]{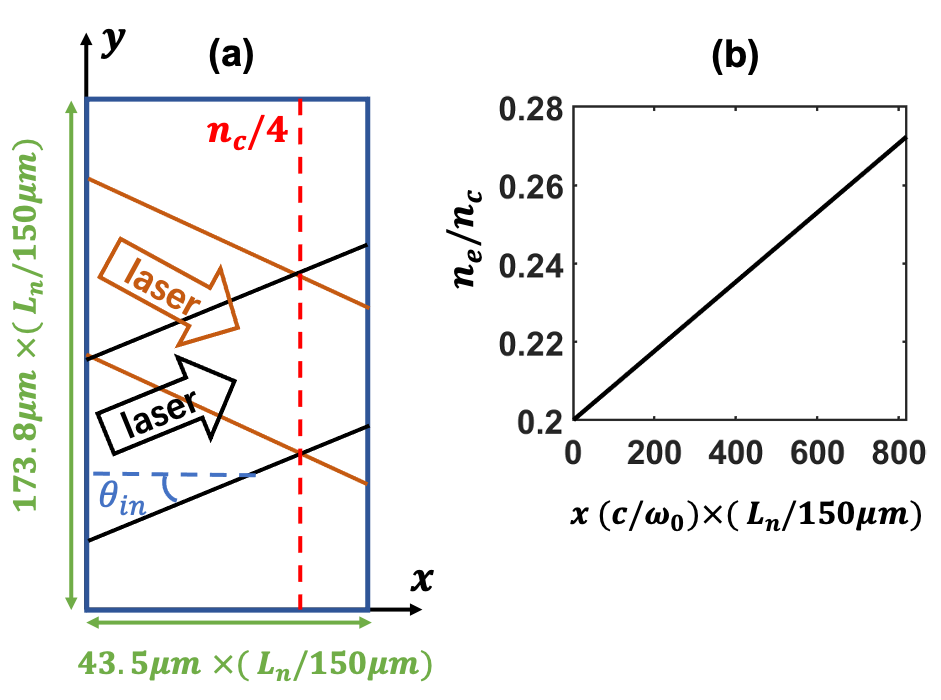}% Here is how to import EPS art
\caption{\label{fig:setup} (a) Schematic setup used in 2D PIC simulations. 
%The marked domain size corresponds to the plasma density scale length $L_n=150\mu m$. 
Note that the figure is not drawn to scale. (b)The density profile along x-direction. }
\end{figure}

\begin{table}
\caption{\label{tab:simu_para} The simulation parameter sets}
\begin{ruledtabular}
\begin{tabular}{ccccccc}
Index&$I_{ovr}$($\times 10^{14}W/cm^2$)&$\theta_{in}(^\circ)$ &$L_n(\mu m)$&$T_e(keV)$&$T_i(keV)$&$\eta$\\
\hline
\multirow{3}{*}{\rmnum{1}} & \multirow{3}{*}{5.0} & 24,32,40,  & \multirow{3}{*}{150} & \multirow{3}{*}{3} &\multirow{3}{*}{1.5} &\multirow{3}{*}{1.04}\\
                           &                    & 44,48,52,  &                      &                   & \\
                           &                    & 55  &  
                             &                   & \\
%                           &                    & 44  &  
%                           &                   & \\
%                           &                    & 48  &                      &                   & \\
%                           &                    & 52  &                      &                   & \\
%                           &                    & 55  &                      &                   & \\

%\hline
%\multirow{1}{*}{\rmnum{2}} & \multirow{1}{*}{18} & \textcolor{black}{24,40,52}  & \multirow{1}{*}{150} & \multirow{1}{*}{3} &\multirow{1}{*}{1.5} &\multirow{1}{*}{3.6}\\

\hline

\multirow{1}{*}{\rmnum{2}} & \multirow{1}{*}{3.6} & 24,40,52  & \multirow{1}{*}{150} & \multirow{1}{*}{2} &\multirow{1}{*}{1} &\multirow{1}{*}{1.12}\\
 %                          &                    & 40  &                      &                   & \\
 %                          &                    & 52  &                      &                   & \\
\hline
%\rmnum{13} & 6 & 40 & 75 & 3 &1.5\\
%\rmnum{14} & 6 & 52 & 75 & 3 &1.5\\
\multirow{1}{*}{\rmnum{3}} & \multirow{1}{*}{10.0} & \textcolor{black}{24,40,52}  & \multirow{1}{*}{75} & \multirow{1}{*}{3} &\multirow{1}{*}{1.5} &\multirow{1}{*}{1.04}\\
 %                          &                    & \textcolor{black}{40}  &                      &                   & \\
 %                          &                    & \textcolor{black}{52}  &                      &                   & \\
\hline
%\rmnum{15} & 1.5 & 40 & 300 & 3 &1.5\\
%\rmnum{16} & 1.5 & 52 & 300 & 3 &1.5\\
\multirow{1}{*}{\rmnum{4}} & \multirow{1}{*}{2.5} & \textcolor{black}{24,40,52}  & \multirow{1}{*}{300} & \multirow{1}{*}{3} &\multirow{1}{*}{1.5} &\multirow{1}{*}{1.04}\\
%                           &                    & \textcolor{black}{40}  &                      &                   & \\
%                           &                    & \textcolor{black}{52}  &                      &                   & \\
\hline
%\rmnum{15} & 1.5 & 40 & 300 & 3 &1.5\\
%\rmnum{16} & 1.5 & 52 & 300 & 3 &1.5\\
\multirow{1}{*}{\rmnum{5}} & \multirow{1}{*}{5.0} & \textcolor{black}{52}  & \multirow{1}{*}{150} & \multirow{1}{*}{2} &\multirow{1}{*}{1} &\multirow{1}{*}{1.57}\\
%                           &                    & \textcolor{black}{40}  &                      &                   & \\
%                           &                    & \textcolor{black}{52}  &                      &                   & \\
                           \hline
%\rmnum{15} & 1.5 & 40 & 300 & 3 &1.5\\
%\rmnum{16} & 1.5 & 52 & 300 & 3 &1.5\\
\multirow{1}{*}{\rmnum{6}} & \multirow{1}{*}{6.3} & \textcolor{black}{52}  & \multirow{1}{*}{150} & \multirow{1}{*}{2} &\multirow{1}{*}{1} &\multirow{1}{*}{1.97}\\
%                           &                    & \textcolor{black}{40}  &                      &                   & \\
%                           &                    & \textcolor{black}{52}  &                      &                   & \\
\end{tabular}
\end{ruledtabular}
\end{table}

We perform a series of two-dimensional PIC simulations \textcolor{black}{using OSIRIS\cite{Fonseca2002}} code to investigate the interactions of CH plasmas with linear density profiles and dual incident laser beams with incident angles of $\pm\theta_{in}$ relative to the direction of the  density gradient, as shown in figure (\ref{fig:setup}). 
\textcolor{black}{In these simulations, the domain sizes and the laser beam widths are scaled proportionally to the plasma density scale length $L_n$ in order to preserve the same plasma density range and maintain the configuration of the laser beam overlap region. 
The laser beams are consistently focused on the center of the $n_c/4$ surface, with transverse Gaussian profiles characterized by 
%FWHM $\sim1882 c/\omega_0$ ($\sim100\mu m$) 
an e-folding width of approximately $1130 c/\omega_0$ ($\sim60\mu m$) multiplied by $L_n/150\mu m $ 
for the wave field, where $c$ and $\omega_0$ represent the speed of light in a vacuum and the laser frequency, respectively, for a wavelength $\lambda=1/3$ $\mu m$.}
This beam width sufficiently covers the overlap regions necessary for the growth of TPD. 
Notably, for the largest $\theta_{in}=55^{\circ}$, the beams begin to overlap at 0.215$n_c$, which is below the Landau cutoff density for TPD growth. 

The overlapped laser intensity $I_{ovr}$ (twice of the single beam intensity $I_{sig}$), incident angle $\theta_{in}$, initial electron density $n_e(x)$, density scale length $L_n=n_e/(dn_e/dx)$ at $n_e=n_c/4$, as well as electron and ion temperatures $T_e$, $T_i$ are chosen within specific \textcolor{black}{ranges of (2.5, 10) $\times10^{14}W/cm^2$,  $(24^{\circ},55^{\circ})$, $(0.2,0.272)n_c$, $(75,300)\mu m$, $(2,3) keV$ and  $(1,1.5) keV$}, respectively, as listed in table (\ref{tab:simu_para}).
\textcolor{black}{We define $I_{sig}=I_{ovr}/2$ as the beam intensity averaged along transverse direction (y-direction) within the e-folding width of the wave field with Gaussian transverse profiles. }
%The ion temperatures $T_i$ are always initialized at $T_e/2$. 
%\textcolor{black}{The laser beams are focused at the $n_c/4$ surface with Gaussian profiles along the transverse direction. The beam width is chosen so that the two beams overlap sufficiently within the density range $(0.2,0.26)n_c$.}
These conditions are typical for direct-drive ICF experiments with the TPD threshold\cite{Simon1983} factor $\eta=I_{ovr}L_{n}\lambda/T_{e}/81.86$ from 1.04 (right above threshold $\eta=1$) to 1.97, where $I_{ovr}$, $L_{n}$, $\lambda$ and $T_{e}$ should be in unit of $10^{14}W/cm^2$, $\mu m$, $\mu m$ and $keV$, respectively. 
We deliberately avoid higher $\eta$, or longer $L_n$ and higher $T_e$ to focus on a regime dominated by multi-beam effects rather than single-beam processes. 
\textcolor{black}{In our simulations, the reflectivity caused by SRS and stimulated Brillouin scattering (SBS) remains below 3\% and 7\%, respectively, except for the SBS reflectivity of 11\% for the simulation (iii) of table \ref{tab:simu_para} with $\theta_{in}=52^\circ$}. 

At time $t=0$, the p-polarized laser pulses are launched at the left boundary ($x=0$) of the simulation domain.  
For the simulations with $L_n=150\mu m$, the domain size is $L_x\times L_y=819.2c/\omega_0\times3276.8c/\omega_0$ (or $43.5\mu m\times173.8\mu m$) consisting of $4096\times16384$ square cells with size $dx=dy=0.2c/\omega_0$ ($0.011\mu m$). Each cell is initially distributed 200 numerical particles (100 electrons, 50 fully ionized Carbon ions and 50 Hydrogen ions). It is worth noting that the cell size remains the same for all the simulations with varying domain sizes. 
The simulations progress with a time step of $dt=0.1414\omega_0^{-1}=0.025fs$ until $\sim6ps$ \textcolor{black}{when TPD has mostly reached a quasi-steady state}. 
We employ open field boundaries and thermal particle boundaries along both x- and y-directions. During the simulations, we record all the hot electrons crossing the boundaries with energies exceeding 50 keV. The electromagnetic fields are diagnosed 
%in two modes. The first mode is ``snapshot'' that demonstrates the spatial field distributions and all existing wave modes at certain times. The second mode is 
using the time Fourier transform (TFFT) module\cite{Wen2019,Cao2022} with time windows of $\sim 1ps$. This module records the time Fourier-transformed fields throughout the entire space but only within specific frequency ranges: (0.45,0.55)$\omega_0$ for Langmuir waves, (0.95,1.05)$\omega_0$ for light waves. 
With frequency filtering, this method provides cleaner signals than the conventional spatial Fourier transform of the ``snapshots", \textcolor{black}{which is also employed in our analysis,} and can distinguish the directions of different wave modes. 
%Of course, we still employ conventional ``snapshots" diagnostics of all the particles and fields.

%We make Particle-in-cell(PIC) simulations to submit the different mechanism to get the hot-e with different angle driven by the TPD and SRS common waves. As Fig.~\ref{fig:png1}, we set a $45.8\mu m\times 183.1\mu m$ space. Then we set two pulses(wave length $\lambda=351nm$) inject CH plasma from the left boundary symmetrically. And we set a linear density profile. Table~\ref{tab:table1} shows the parameters we use in the simulations. Here the heavy ion plasma is a kind of hypothetical particle, the mass is 1000 times to Hydrogen, but the charge is equal to Hydrogen. To research the hot-e would affect the implosion, we just diagnose those electrons over 50keV. 

\section{Results and discussions}

\iffalse

\textcolor{black}{Figure arrangements:\\
For simulation index i: \\
(1)x-t figure of $\delta n_e$: 24 degree, 40 degree, 52 degree;\\
Point: different TPD modes location in both linear and nonlinear stage. \\
(2)linear stage spectra of 3 angles (tfft); \\
nonlinear spectra of 3 angles (tfft). \\
theoretical coupling curves \\
Point: nonlinear spectra are connected (similar) to linear spectra; wave coupling methods differ for different angles (forward, intermediate, backward); \\
(3) hot electron angular distributions, $f_{hot}$ of all incident angles (separate $f_{hot, forward}$ and $f_{hot, side}$?) \\
points: hot electron angular spectra are connected to the nonlinear TPD modes; large incident angles causes divergent hot electrons but do not affect forward electrons much (the mechanism of coherent side EPW generating forward hot electrons, time correlation of them); hot electron temperatures?\\
}

\textcolor{black}{theory:\\
calculation of convective gain theory can explain the peak modes (any physics picture for the near but not perpendicular SRS?);\\
The theory of peak modes also works for different TPD $\eta$, $I$, $T_e$, $L_n$ (simulations ii to v). \\
The correlation of hot electrons with peak modes does not change? \\
}

\fi

\subsection{Two major characteristics of TPD growth for varying $\theta_{in}$}

\begin{figure}
\includegraphics[width=0.42\textwidth]{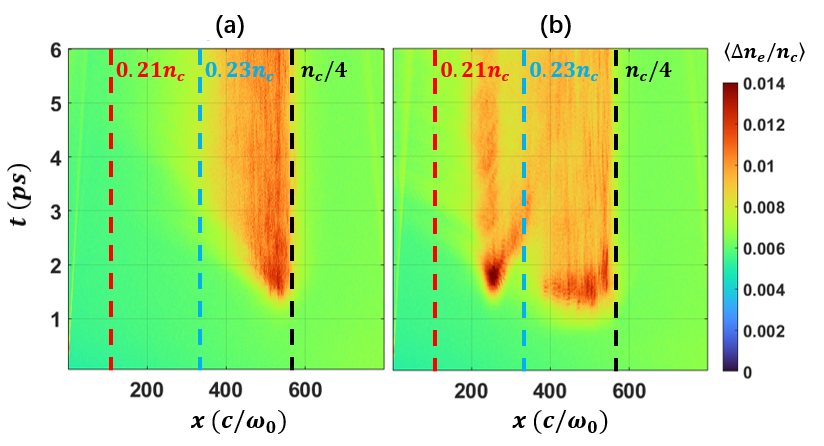}% Here is how to import EPS art
\caption{\label{fig:dne_x_t} The space (x-direction) and time evolution of the electron density perturbation $<\Delta n_e>$ for the incident angle $24^\circ$ and $52^\circ$ in simulation series (i) [Table \ref{tab:simu_para}]. The $x-$locations of $0.21n_c$, $0.23n_c$ and $n_c/4$ are marked by vertical dashed lines in red, blue and black colors, respectively.}
\end{figure}

To illustrate the effects of $\theta_{in}$ on TPD growth, we first present the results of the simulations with identical physical parameters but varying $\theta_{in}$, as listed in table \ref{tab:simu_para}(i). 
These conditions correspond to conventional TPD threshold\cite{Simon1983} factor $\eta=1.04$, slightly above the threshold of $\eta=1$ to represent a typical TPD-dominant case. 
As $\theta_{in}$ increases from $24^\circ$ to $55^\circ$, 
the TPD growth primarily exhibits two distinct characteristics for relatively \textcolor{black}{smaller ($\theta_{in}<40^\circ$) }and \textcolor{black}{larger ($\theta_{in}>44^\circ$)} incident angles, respectively. 
%both linear and nonlinear spectra, together with the hot electron energy and angular distributions as well. 
The two characteristics can be effectively demonstrated by typical cases of $\theta_{in}=24^\circ$ and $52^\circ$. For these cases, the space (x-direction) and time evolution of electron density perturbations, mainly produced mainly by the TPD daughter EPWs, are plotted in figure (\ref{fig:dne_x_t}a) and (\ref{fig:dne_x_t}(b), respectively. 
The color scale in the two figures represents $<\Delta n_e>(x,t)$, which is the perturbation level at certain location $x$ and time $t$: 
%And it is defined as the standard deviation of $n_e$ around the initial density along $y$-direction:

\begin{align}
<\Delta n_e>(x,t) = \sqrt{\sum_{i_y=1}^{N_y} [n_e(x,dy\cdot i_y,t)-n_e(x,dy\cdot i_y,0)]^2/N_y}
\end{align}
where $i_y$ is the cell index along $y$-direction, and $N_y$ is total the cell number along $y$-direction of the simulation domain. 

The growth of TPD modes at different times can be demonstrated by $<\Delta n_e>(x,t)$ in figure (\ref{fig:dne_x_t}). 
%The TPD growth exhibit contrasting behaviors for the two cases.
For $\theta_{in}=24^\circ$, TPD modes first grow near $n_c/4$, then diffuse to $\sim0.23n_c$ and reach the quasi-steady state after $\sim 3ps$. By contrast, TPD modes for $\theta_{in}=52^\circ$ initially grow in a much broader space range from $0.22n_c$ to $n_c/4$, and this range remains until TPD reaches a quasi-steady state. 

\begin{figure}
\includegraphics[width=0.45\textwidth]{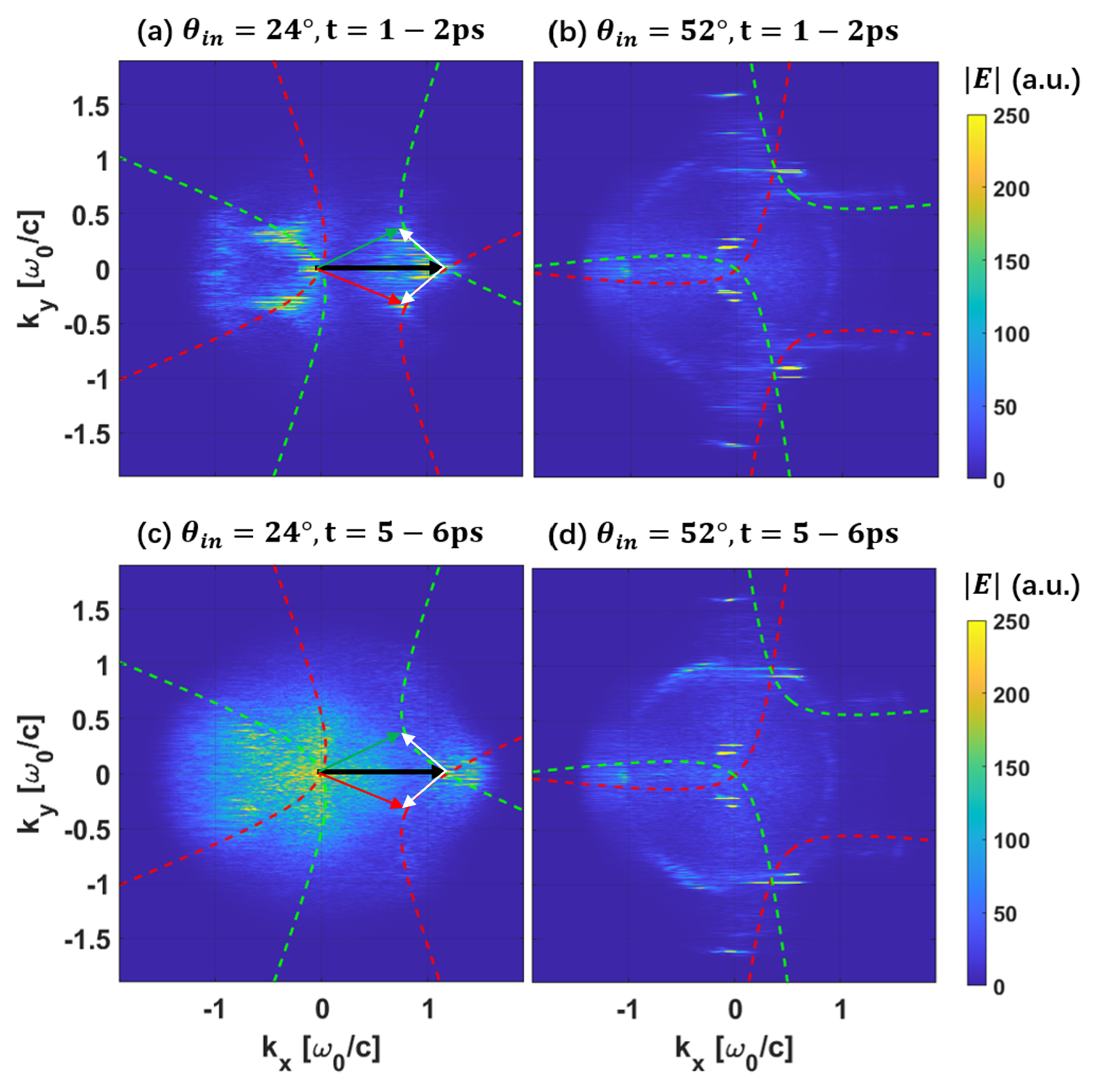}% Here is how to import EPS art
\caption{\label{fig:absE} The space spectra of $|E(k_x,k_y)|$ defined by equation (\ref{eqn:absE}) for the simulation (i) [table \ref{tab:simu_para}] during the time of 1-2ps (a)(b) and 5-6ps(c)(d), and with $\theta_{in}=24^\circ$ (a)(c) and $52^\circ$ (b)(d). The TPD hyperbola rotated by $\pm\theta_{in}$ are plotted as dashed curves, representing the locations corresponding to maximum TPD growth rates for the two incident laser beams. In (a)(c), the colored arrows mark the wave vectors of the laser beams (green and red), common EPW (black) and the other TPD modes (white). The corresponding wave vectors for (b)(d) are given in figure \ref{fig:TPD_SRS}(b).
%\textcolor{black}{(Considering to add a Landau cutoff circle and a $v_{ph}=c$ circle in all the figures. Add the laser wave vector arrows?)}
}
\end{figure}

The different behaviors of $<\Delta n_e>(x,t)$ correspond to distinct EPWs of TPD daughter waves with frequency $\sim\omega_0/2$. 
The space spectra of the electric fields with frequency from $0.45\omega_0$ to $0.55\omega_0$ are plotted in figure \ref{fig:absE} for two time intervals: $1$ to $2 ps$ and $5$ to $6 ps$, representing the linear and nonlinear stages, respectively. 
%for both $\theta_{in}=24^\circ$ and $52^\circ$ cases. 
To equally reflect the electric fields along $x-$ and $y-$ directions, we define  $|E|$ as the magnitude of electric fields in $k_x$ and $k_y$ phase space with frequency from $0.45\omega_0$ to $0.55\omega_0$:

\begin{align}
\label{eqn:absE}
|E|=|E(k_x,k_y)|=\sqrt{E_x(k_x,k_y)^2+E_y(k_x,k_y)^2}
\end{align}
where $E_x$ and $E_y$ are the electric fields along $x-$ and $y-$ directions. 
For $\theta_{in}=24^\circ$, TPD grows significantly near the crossing point of the two TPD hyperbolas of the two incident beams in the linear stage [figure (\ref{fig:absE}a)], agreeing with the typical growth mechanism of sharing a forward-going common wave\cite{Zhang2014}. In the later nonlinear stage, when the quasi-steady state is formed, the spectrum broadens due to nonlinear saturations\cite{Myatt2012}, though the common wave signals remain dominant. 

The excitation of such common wave relies on the crossing of the hyperbolas in the forward direction (positive $x-$direction) [figure (\ref{fig:absE}a)(\ref{fig:absE}c)]
%\textcolor{black}{(Draw arrows in figure 2a c? The arrow colors need to be changed.)}
, which is not satisfied for $\theta_{in}=52^\circ$ [figure (\ref{fig:absE}b)(\ref{fig:absE}d)]. In this case, the hyperbolas intersect at the sides with EPWs observed near the intersections. 
One might intuitively guess they are common waves co-driven by two laser beams via TPD. However, this cannot be the case, as the common wave cannot simultaneously act as the daughter wave with relatively longer $\vec{k}$ (common wave frequency $\omega_{co}>\omega_0/2$) for one laser beam and shorter $\vec{k}$ ($\omega_{co}<\omega_0$/2) for the other beam when both beams have the same frequency $\omega_0$. 

We also observe electric field modes at $k_x\sim0$ with discrete peaks at different $k_y$'s. The modes with $k_x\sim0$ and $k_y\sim1.6k_0$ near the hyperbola have much greater $|k|$ than all the modes for $\theta_{in}=24^\circ$. According to the TPD theory, these modes should locate at much lower density than the modes for $\theta_{in}=24^\circ$, agreeing with the results in figure (\ref{fig:dne_x_t}). 
Moreover, the overall electric field spectra for $\theta_{in}=52^\circ$ are dominated by separated peaks rather than broad signal islands for $\theta_{in}=24^\circ$, even in the nonlinear stage. This causes multi-peak angular distributions of hot electrons, which are presented \textcolor{black}{in section \ref{sec:hot_e}}.

\subsection{The excitation mechanism of side TPD modes for relatively larger $\theta_{in}$ }
\label{sec:TPD_mechanism}

For the two characteristics described in the previous section, the one corresponds to $\theta_{in}=24^\circ$ is driven through a forward-going common wave and has been well understood\cite{Zhang2014}. 
Thus, we focus on the underlying physics of the side modes excitation for $\theta_{in}=52^\circ$. 
We find that the side EPWs are still common waves, but they are shared by TPD driven by one beam and SRS by the other, instead of two TPDs, \textcolor{black}{so we name it shared TPD-SRS (STS) instability in this article}. 
This \textcolor{black}{STS} mechanism can explain most major modes in figure (\ref{fig:absE}b)(\ref{fig:absE}d), especially those side modes with $k_x\sim 0$.

To better illustrate the \textcolor{black}{STS} mechanism, we first recall the basic TPD and SRS theories of wave coupling and maximum growth conditions. As typical three-wave coupling processes, both TPD and SRS require matching conditions of frequency $\omega_{0,1,2}$ and wave vector $\vec{k}_{0,1,2}$ to be satisfied for the instability to grow:
\begin{align}
\label{eqn:match_k}
\vec{k_0}&=\vec{k_1}+\vec{k_2} \\
\label{eqn:match_w}
\omega_0&=\omega_1+\omega_2
\end{align}
where the subscripts 0, 1 and 2 refer to the laser wave and two daughter waves, respectively. The $\omega_{0,1,2}$ and $\vec{k}_{0,1,2}$ must follow the dispersion relations of electromagnetic waves (EMWs) [equation (\ref{eqn:DR_EMW})] or EPWs [equation (\ref{eqn:DR_EPW})], depending on the wave forms:
\begin{align}
\label{eqn:DR_EMW}
\omega^2&=\omega_{pe}^2+k^2c^2 \\
\label{eqn:DR_EPW}
\omega^2&=\omega_{pe}^2+3k^2v_{te}^2,
\end{align}
where $\omega_{pe}$ is the plasma frequency depending on the density $n_e$, $v_{te}=\sqrt{T_e/m_e}$ denotes the electron thermal speed, where $m_e$ is the electron mass. 

\begin{figure}
\includegraphics[width=0.4\textwidth]{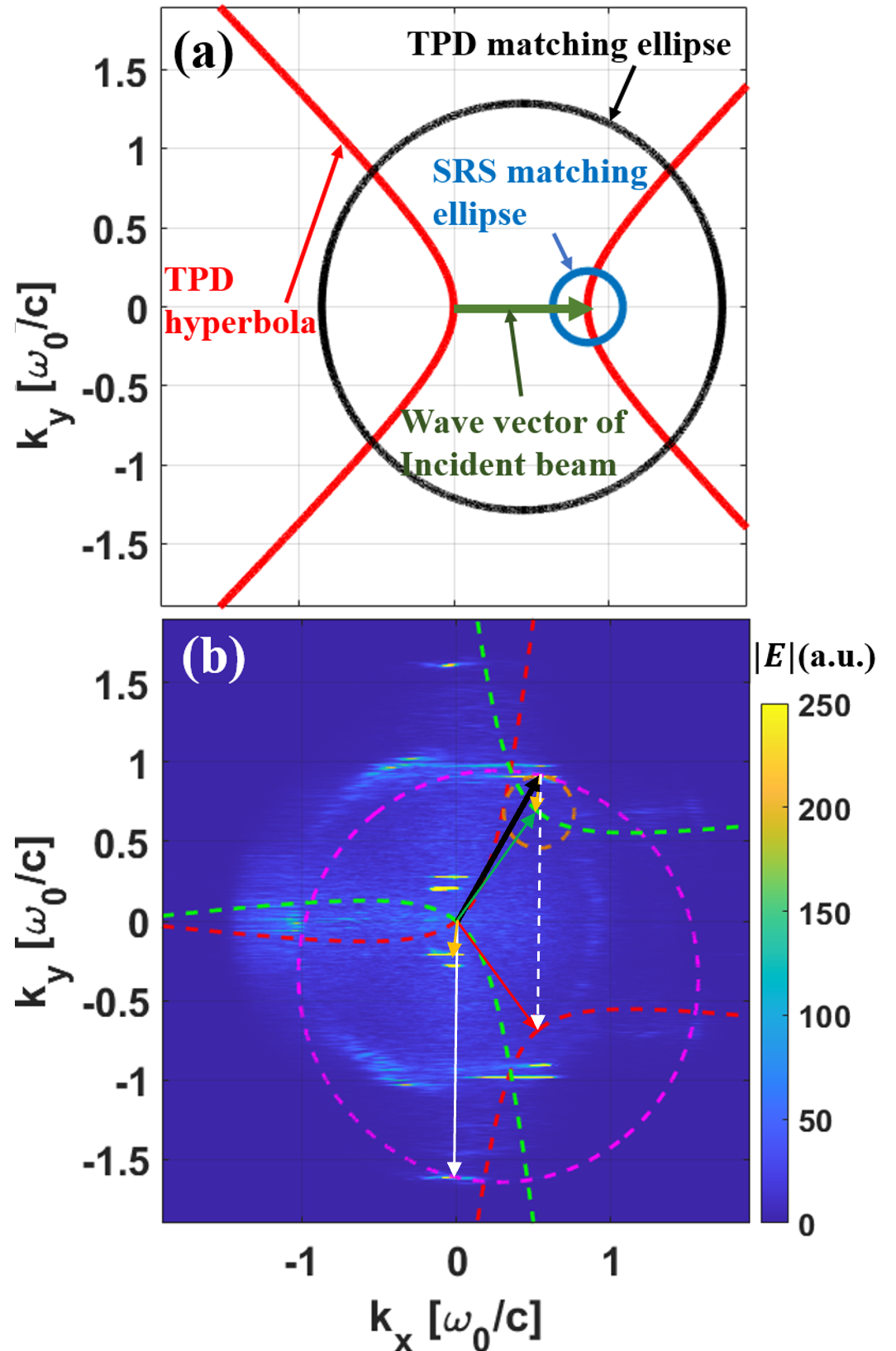}% Here is how to import EPS art
\caption{\label{fig:TPD_SRS} (a) Schematic view of the matching ellipses of TPD (black) and SRS (blue) in $k_x$ and $k_y$ phase space for a laser beam with wave number (green) $k_0\sim0.87\omega_0/c$ for a certain $n_e$
%, corresponding to plasma density $n_e=n_c/4$
. 
The red hyperbola $k_y^2=k_x(k_x-k_0)$ corresponds to the wave numbers with the greatest TPD growth rates for different $n_e$. For relatively higher $n_e$, the ellipses are smaller. (b) The space spectra of $|E(k_x,k_y)|$ defined by equation (\ref{eqn:absE}) for the simulation (i) [table \ref{tab:simu_para}] with $\theta_{in}=52^\circ$ during the simulation time of 5-6ps.
%\textcolor{black}{(we should put vector arrows on it to show the matching. More instructions should be given to say the hyperbola and ellipses are from (a).)}
}
\end{figure}

Given $\omega_{pe}$ and $v_{te}$ (or $n_e$ and $T_e$), the equations of dispersion relations and matching conditions determine two matching ellipses of TPD and SRS [figure (\ref{fig:TPD_SRS}a)]. 
The vectors starting from the origin ending on the ellipses represent all the possible EPW daughter waves satisfying the matching equation (\ref{eqn:match_k}) to (\ref{eqn:DR_EPW}). The vectors from the end of $k_0$ (the horizontal green arrow) to the blue ellipse mark the corresponding wave vectors of scattered lights via SRS. 

With the above theory of matching ellipses, we can conveniently identify all the common waves that can be co-driven by multiple incident laser beams. In figure (\ref{fig:TPD_SRS}b), We redraw the $|E|$ spectrum for $\theta_{in}=52^\circ$ case during 5-6 ps [figure (\ref{fig:absE}d)] with the TPD ellipse (purple dashed line) for the laser beam pointing down-right and the SRS ellipse (brown dashed line) for the laser beam pointing up-right for $n_e=0.217n_c$ and $T_e=3 keV$ [table \ref{tab:simu_para}(i)]. The two ellipses intersect in $k$ space, suggesting the EPW pointing from the origin to the intersection can be co-driven by the two laser beams via TPD and SRS.%, while neither of the two beams are strong enough to excite TPD or SRS alone.
%Please note that the choices of $n_e$ and $T_e$ here are examples corresponding to strongly driven common waves.

Nonetheless, the formation of such common waves does not guarantee strong growth, which requires high gain of the driven modes. Actually, the matching of the common waves broadly exists for a considerable density range between $0.21n_c$ and $0.23n_c$, wherever the two matching ellipses intersect. However, only one dominant common wave among these matching modes can grow significantly, as shown in the simulation results [figure (\ref{fig:TPD_SRS}b)].

To determine the relative significance of all the matching modes, we adopt the Rosenbluth gain\cite{Rosenbluth1973} $G_R$, which physically means the amplification factor of a delta-function seed in nonuniform plasmas:

\begin{align}
\label{eqn:Rosenbluth}
G_{R}=exp[\frac{\pi\gamma_0^2}{\kappa'V_1V_2}]
\end{align}
where $\gamma_0$ is the temporal growth rate of the instability in a homogeneous plasma, $V_1$ and $V_2$ are the group velocities of the two daughter waves, and $\kappa'=\frac{d}{dx}(k_{0x}-k_{1x}-k_{2x})$ denotes the $k$ mismatch factor along the $x-$direction (along the density gradient). The above equation (\ref{eqn:Rosenbluth}) applies to both TPD and SRS with all the variables varying for different $n_e$ and $T_e$. For simplicity, we do not write this equation separately for TPD and SRS, and all the following equations apply for both TPD and SRS unless otherwise stated.

%To provide clarity, we point out that the equation (\ref{eqn:Rosenbluth}) of Rosenbluth gain was developed with WKB approximations\cite{Rosenbluth1973}, which basically requires that all the involved EPWs and EMWs are not close to their critical densities. This condition, however, does not necessarily hold for the dominant wave modes in our theoretical [the peak in figure (\ref{fig:TPD_SRS_Gain}a)] and numerical [figure (\ref{fig:TPD_SRS}b)] results, where the scattered light of SRS is sufficiently close to its own critical density to violate the WKB approximation\cite{Michel2023}. This makes the SRS absolutely unstable and the calculated convective gain from the equation (\ref{eqn:Rosenbluth}) no longer reflects the original meaning of the amplification factor of a delta-function seed. However, we still follow the Rosenbluth gain formula [equation (\ref{eqn:Rosenbluth})] here. First, except for those dominant modes, this gain formula works well for most matching modes at other densities and can properly reflect their gains. Second, the ``erroneous" results of this formula for the dominant absolute SRS modes are actually overwritten by the effects of finite width of the laser beams, as we will explain in the following few \textcolor{black}{paragraphs} [equation (\ref{eqn:GRGW}) and figure (\ref{fig:TPD_SRS_Gain}b)]. 

Using Rosenbluth gain theory, the total gain of the \textcolor{black}{STS} common waves driven simultaneously by TPD and SRS can be evaluated by considering the contributions from both TPD and SRS:

\begin{align}
\label{eqn:Gtot1}
G_{tot} = G_{S}\times G_{T}
\end{align}
where $G_{S}$ and $G_{T}$ are the gains of SRS and TPD, respectively.
In 1-dimension cases, they are basically Rosenbluth gains described in equation (\ref{eqn:Rosenbluth}). 
Please note that achieving $G_R$ relies on the complete amplifications process of initial seeds traveling along the density gradient for a finite distance.
However, in our simulations using Gaussian beams with finite beam widths, the amplification processes of TPD and SRS seeds which have finite transverse velocities are limited by the transverse size of the laser overlap region [figure (\ref{fig:setup}a)]. Namely, the unstable seeds may leave the laser overlap region and stop growing before reaching $G_R$.
Therefore, 
%$G_R$ represents the actual amplification factor only if the unstable seeds remain in the overlap region during the entire growth period. Otherwise, 
the gain should be truncated to the amplification within the laser overlap region, and the gains $G_S$ and $G_T$
%of complete formula of $G_{tot}$ 
should be similarly written as

\begin{align}
\label{eqn:GRGW}
G_{S} \text{ or } G_T = min(G_R, G_w)
\end{align}
where 
\begin{align}
\label{eqn:GW}
G_w = exp[\frac{\gamma_0L_{tr}}{max(|V_{1y}|,|V_{2y}|)}]
\end{align}
is the gain assuming the unstable seed continues growing by growth rate $\gamma_0$ without saturation during propagation until leaving the laser overlap region from the sides. 
Here, $L_{tr}$ is the transverse size of the laser overlap region, and $V_{1y}$ and $V_{2y}$ respectively represent the group velocities of the two daughter waves along the transverse ($y-$) direction. 
Please note that that the instability growth rate $\gamma_0$ for homogeneous plasmas is also the temporal growth rate of the seed propagating in inhomogeneous plasmas\cite{Rosenbluth1973}. Thus, the minimum value of $G_R$ and $G_w$ should represent the overall gain of the common waves.

\begin{figure}[h]
\includegraphics[width=0.4\textwidth]{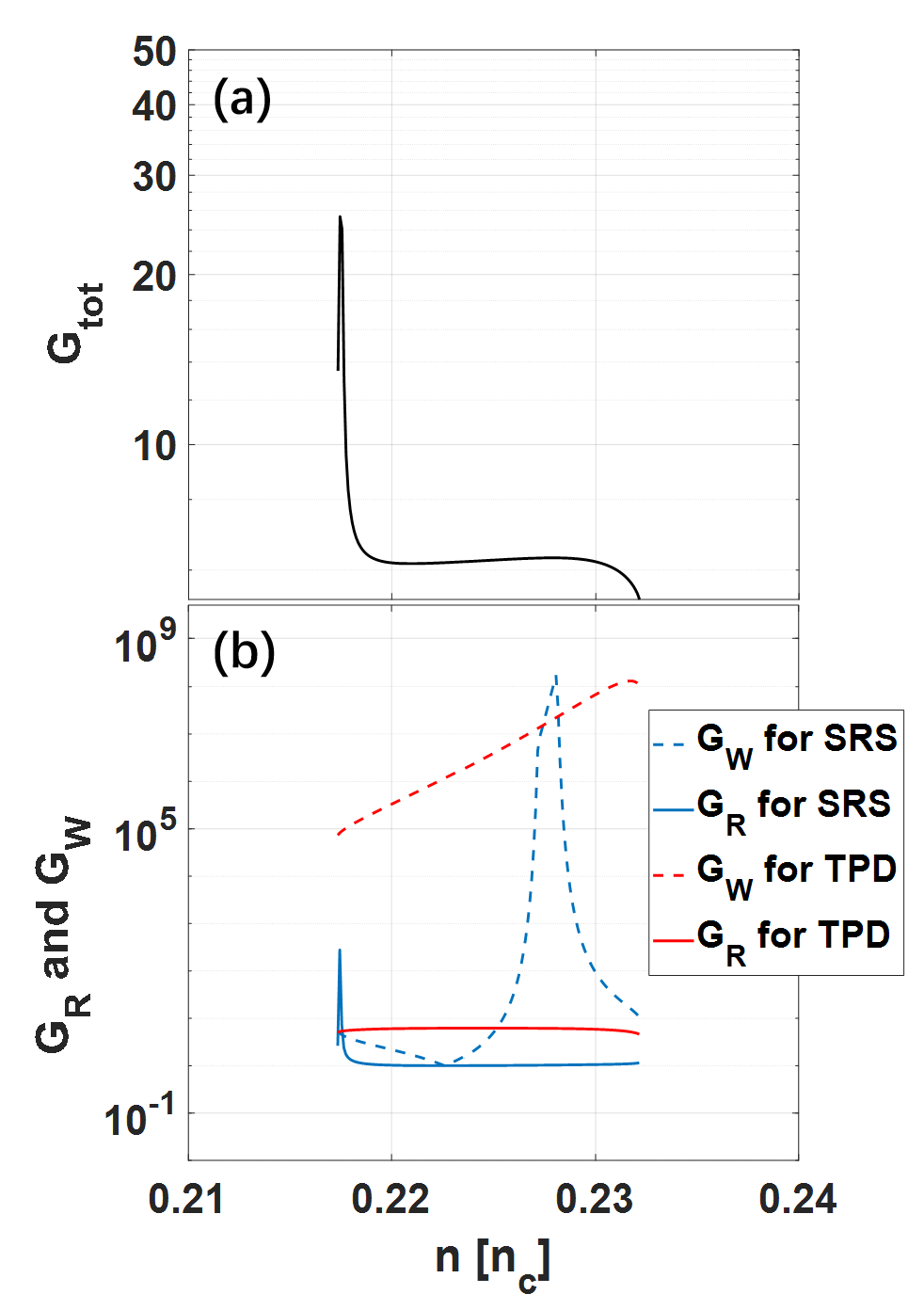}% Here is how to import EPS art
\caption{\label{fig:TPD_SRS_Gain} The calculated (a) $G_{tot}$, (b) $G_R$ and $G_w$ of TPD and SRS for simulation parameter set (i) in table \ref{tab:simu_para} with $\theta_{in}=52^\circ$. The Rosenbluth gain $G_R$ of SRS (blue solid line) is greater than $G_w$ (blue dashed line) near the narrow peak, and does not contribute to $G_{tot}$.}
\end{figure}

\begin{figure}[h]
\includegraphics[width=0.5\textwidth]{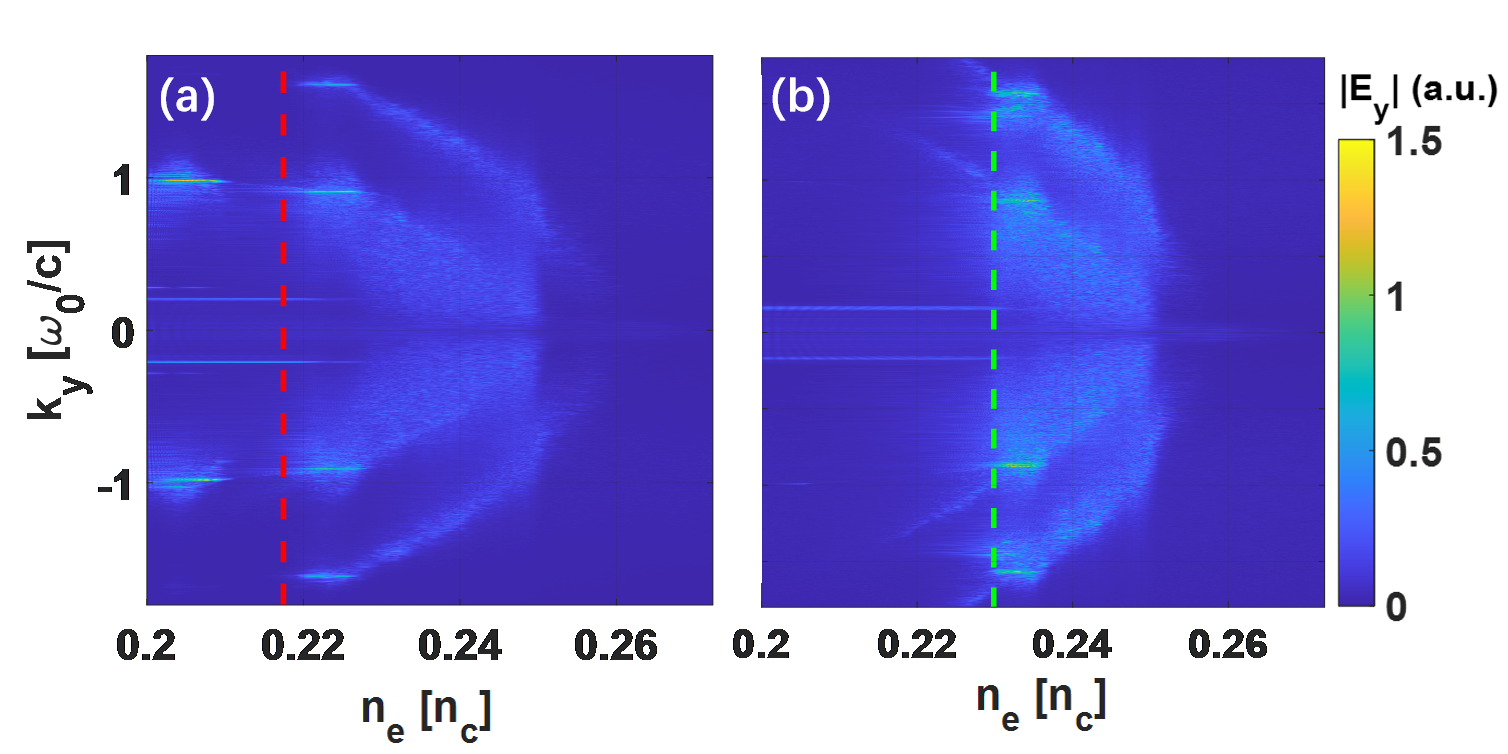}% Here is how to import EPS art
\caption{\label{fig:npt_gain_4} The $k_y-n_e$ spectra for (a) simulation (\rmnum{1}) and (b) simulation (\rmnum{5}) [table \ref{tab:simu_para}] with $\theta_{in}=52^\circ$ during 5-6 ps. The vertical dashed lines mark the location of the maximum $G_{tot}$ [figure (\ref{fig:TPD_SRS_Gain}a)].
%The agreement between the common wave signals of $E_y$ during the time of 5-6ps in the spectrum for simulation and the location of common wave mode with the highest gain for our calculation(dashed line) when $\theta_{in}=52^\circ$ for simulation \rmnum{1}(a) and simulation \rmnum{5}(b).
}
\end{figure}

With the above analysis, we calculate $G_{tot}$ as a function of $n_e$ with $T_e=3keV$, $L_n=150\mu m$ [table \ref{tab:simu_para}(i)] with $\theta_{in}=52^\circ$. The results are plotted in \textcolor{black}{figure \ref{fig:TPD_SRS_Gain}(a)}. We find that the $G_{tot}$ reaches the highest peak within a narrow density region around \textcolor{black}{$0.217n_c$}
with $k_x\approx0.54\text{ to } 0.57\omega_0/c$ and $k_y\approx 0.91\omega_0/c$. Both the location and the wave number agree very well with the strongest EPW modes observed in \textcolor{black}{figure (\ref{fig:TPD_SRS}b) and \textcolor{black}{(\ref{fig:npt_gain_4}a)}}. 
\textcolor{black}{One may notice another neighboring peak close to the common wave with similar $k_x$ but slightly larger $k_y\approx0.98\omega_0/c$. This signal is caused by the absolute SRS mode driven by single beams, and is discussed at the end of this section.}

%\begin{figure}[h]
%\includegraphics[width=0.5\textwidth]{figures/npt_gain_4.png}% Here is how to import EPS art
%\caption{\label{fig:npt_gain_4} (a)The $G_{tot}$ according to our calculation when $\theta_{in}=48^\circ$, $\theta_{in}=52^\circ$ in simulation (\rmnum{1}) and $\theta_{in}=52^\circ$ in simulation (\rmnum{5}). The agreement between the common wave signals in the spectrum for simulation and the mode with the highest gain for our calculation when $\theta_{in}=48^\circ$(b), $\theta_{in}=52^\circ$(c) for simulation (\rmnum{1}) and $\theta_{in}=52^\circ$(d) for simulation (\rmnum{5}).\textcolor{black}{(time of bcd)}}
%\end{figure}

The formation of the common-wave peak relies on the local maximum point of SRS gain $G_S$
% due to $\kappa'=\frac{d}{dx}(k_{0x}-k_{1x}-k_{2x})\approx0$
, as the TPD gain $G_T$ does not show any local maximum point [\textcolor{black}{figure (\ref{fig:TPD_SRS_Gain}b)]}. However, 
although the SRS Rosenbluth gain $G_R$ reaches an extremely high peak around the same location due to $\kappa'=\frac{d}{dx}(k_{0x}-k_{1x}-k_{2x})\approx0$, it is not the cause of $G_S$ peak, as it is overwritten by the local $G_w$, as shown by the blue solid and dashed lines in figure (\ref{fig:TPD_SRS_Gain}b).

Additionally, we clarify that the local maximum $G_w$ is not the one with SRS scattered light pointing strictly parallel ($0^\circ$) to the y-direction
%, for the gain of this mode is restricted strongly by $L_{tr}$
. Instead, the strongest SRS scattered light travels along \textcolor{black}{$\theta_s\sim\pm95^\circ$} to the positive x-direction, according to our calculations. This qualitatively agrees with the simulation results in figure \ref{fig:TPD_SRS}(b), where the SRS scattered light covers the band region starting from ($k_x\sim0$, $k_y\sim0.21\omega_0/c$) and extending to negative $k_x$, maintaining the same $k_y$ during the propagation. This phenomenon is more clearly depicted in the $k_y-n_e$ spectra of $E_y$ in \textcolor{black}{figure (\ref{fig:npt_gain_4})}, where the horizontal lines with $k_y\sim0.21\omega_0/c$ represent the propagation of the scattered light towards the negative x-direction.
The extension of these lines towards the left \textcolor{black}{boundary} of the simulation domain is not expected for the scattered light with $\theta=90^\circ$, as such light 
would propagate exclusively along the y-direction, 
with no component in the x-direction. 

\textcolor{black}{In addition to the example discussed above, our theoretical predictions of the peak common wave locations align well with the PIC simulation results under various laser plasma conditions. 
Figure (\ref{fig:npt_gain_4}) presents the $k_y-n_e$ phase spaces of $E_y$ overlaid with the peak locations of $G_{tot}$ for the simulation (i) and (v) [table \ref{tab:simu_para}] with $\theta_{in}=52^\circ$. In the two $k_y-n_e$ figures, the modes near (a) $0.22n_c$ and (b) $0.23n_c$ with $k_y\sim0.9\omega_0/c$ are the \textcolor{black}{STS} common EPWs, while those at the same density but with much larger $k_y$ correspond to the other paired TPD modes. These figures show that the $G_{tot}$ peak locations consistently match the theoretical \textcolor{black}{STS} common wave locations.}

In figure \textcolor{black}{(\ref{fig:npt_gain_4}a)}, we observe additional modes in the lower density region, exhibiting slightly larger $k_y$ than the common waves. In the $k_x-k_y$ phase space, these modes are situated 
\textcolor{black}{ at ($k_x\sim0.54\omega_0/c$, $k_y\sim0.98\omega_0/c$)} [figure \ref{fig:absE}(d) and \ref{fig:TPD_SRS}(b)], near the band signals of the common waves.
%we also find other two bands \textcolor{black}{with scattered light at ($kx=?$, $k_y=?$) and the EPWs at ($kx=?$, $k_y=?$). 
%Compared to the common waves, these bands have similar $k_x$ but obviously greater $|k_y|$}. 
%They all align with x-direction due to the constant $k_y$ during the propagation.
%Spatially, they are located in the lower density region compared to the common waves \textcolor{black}{[figure \ref{fig:npt_gain_4}(?)]}. 
%\textcolor{black}{(At somewhere else, we need to mention these bands all align with x-direction due to the constant $k_y$ during the propagation.)}
We identify these modes as EPWs of absolute SRS modes\cite{afeyan1985}, with corresponding scattered light at ($k_x\sim0$, $k_y\sim0.28\omega_0/c$), driven by single laser beams near the incident plane at $n_e\geq0.2n_c$. 
%Following equation (\ref{eqn:Rosenbluth}), (\ref{eqn:GRGW}) and (\ref{eqn:GW}), we calculate the $G_R$ and $G_w$ for the SRS driven by a single beam at $n_e\sim0.2n_c$, and the effective $G_s$ is plotted in \textcolor{black}{figure ??}. The peak of $G_s$ stays in the region where 
\textcolor{black}{This absolute SRS occurs when $\kappa'$ close to zero. For the simulation (i) with $\theta_{in}=52^\circ$ at $n_e=0.2n_c$, our calculations of absolute SRS indicate ($k_x\approx0.53\omega_0/c$, $k_y\approx 0.97\omega_0/c$) for the EPW modes and ($k_x\approx0.02\omega_0/c$, $k_y\approx 0.27\omega_0/c$) for the scattered light, which are in good agreement with the above simulation results.
Absolute SRS modes can develop 
when the laser-plasma conditions surpass the instability 
threshold\cite{afeyan1985}.} 
%which is described by equation (48) of the reference [\citen{Afeyan1985}]. 
\textcolor{black}{In our simulations, they occur mildly in some cases for which the threshold is slightly exceeded while not observed in others, suggesting they play a negligible role in both hot 
electron generation and laser reflection.  }
%exceeding the threshold while others stay below the threshold. In those with these modes 

%\begin{align}
%a_0
%\end{align}

\subsection{The angular characteristics of hot electrons for different $\theta_{in}$}
\label{sec:hot_e}

\begin{figure}[h]
\includegraphics[width=0.4\textwidth]{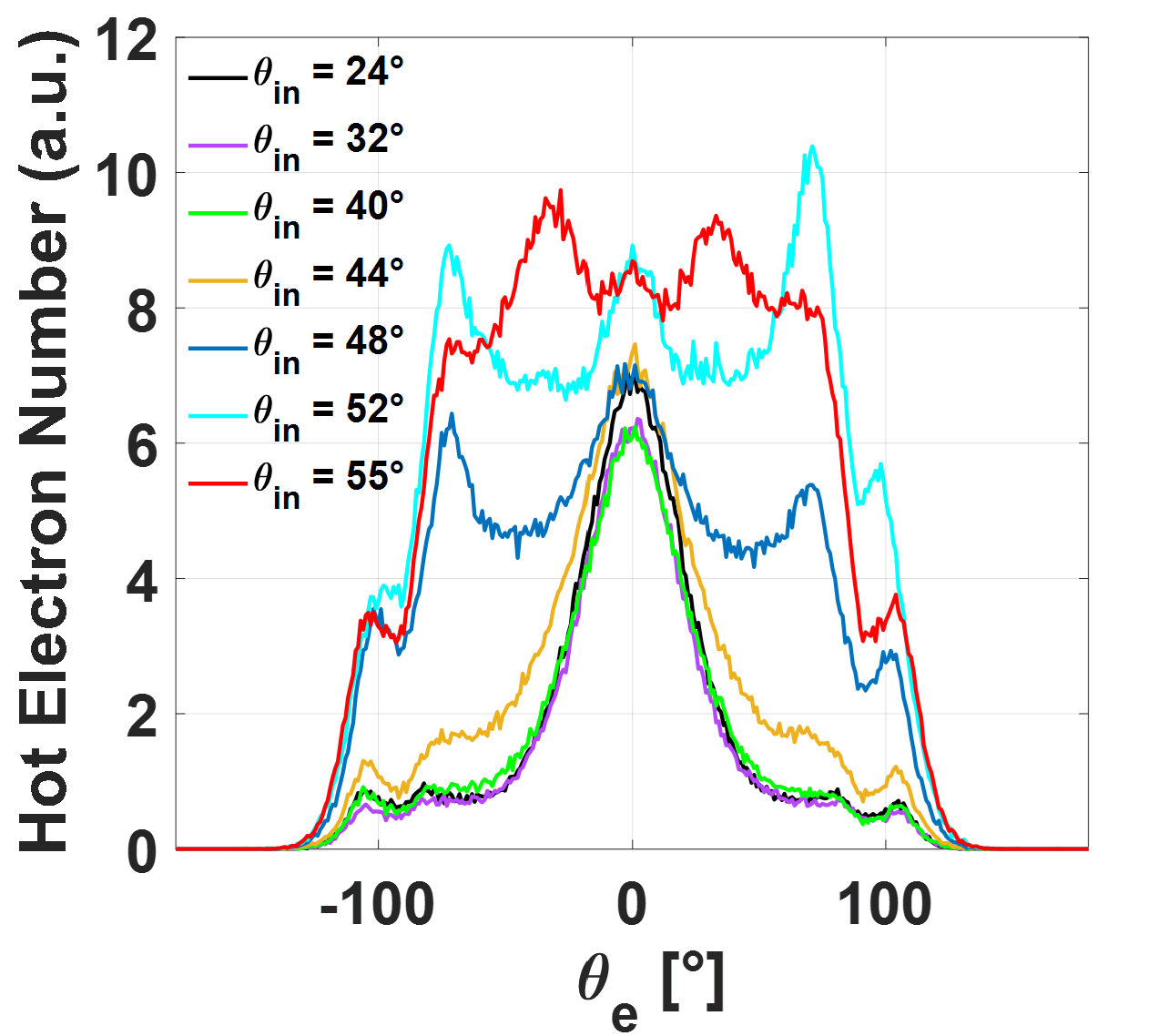}% Here is how to import EPS art
\caption{\label{fig:angle_d} The angular distribution of hot electrons for the simulation (i) in table \ref{tab:simu_para} with different laser incident angle $\theta_{in}$'s.
%\textcolor{black}{(This caption needs revisions. We may plot the amplitude of the side TPD modes for different angles. Y axis label: Hot electron number (a.u.))}
}
\end{figure}

The driven EPWs can effectively accelerate background thermal electrons with velocity close to the EPW phase velocity $v_{ph}$. Therefore, the hot electron characteristics are closely tied to the EPW spectra. For smaller $\theta_{in}=24^\circ$, the EPW spectra exhibit a two-island structure, as shown in figure (\ref{fig:absE}c). The main island, centered around the origin, contains EPWs with a broad range of wave vectors, which complicates the analysis of the resulting hot electrons. Fortunately, most of these modes have  $|k|<0.5\omega_0/c$, corresponding to $v_{ph}\ge c$, which is significantly greater than the electron thermal velocities, thus inefficient for generating hot electrons. Instead, the smaller island with \textcolor{black}{$k_x>1\omega_0/c$} and $k_y\sim 0$, leading to \textcolor{black}{$v_{ph}\sim0.5c$}, 
can efficiently capture and accelerate background electrons with temperatures around a few $keV$. 
Correspondingly, as shown in figure (\ref{fig:angle_d}), the hot electrons accelerated in the case of $\theta_{in}=24^\circ$ exhibit a collimated feature in the angular distribution. 
Also, the same feature is observed for $\theta_{in}$ =32$^\circ$ when the side TPD and SRS common waves are not significantly driven.

For larger $\theta_{in}=52^\circ$, the EPW spectra display a multi-peak structure (figure \ref{fig:absE}d and \ref{fig:TPD_SRS}b). Thus, the hot electron characteristics should be determined by these peaks. According to our analysis in the previous section, these peaks mainly correspond to the modes of TPD and SRS co-driven by duel beams %\textcolor{black}{(also mention single beam SRS?)}
, and primarily point sideways. Specifically, the common EPW propagates at $\sim60^\circ$ to the x-direction, while the other TPD daughter wave has $k_x\sim 0$ and \textcolor{black}{$k_y\sim 1.6\omega_0/c$}, perpendicular to the x-direction. 
These two EPWs should generate hot electrons with high divergent angles. 
This feature is reflected in the hot electron angular distributions for $\theta_{in}\geq 48^\circ$ \textcolor{black}{[figure (\ref{fig:angle_d})]} which exhibit pronounced side peaks at \textcolor{black}{$\sim 72^\circ$}. 

Furthermore, the hot electron angular distributions also show a peak at $0^\circ$ along the x-direction, corresponding to collimated hot electrons. However, the EPW spectrum does not indicate any EPW modes traveling along the x-direction [figure (\ref{fig:TPD_SRS}b)]. 
To address this discrepancy, we find that the collimated hot electrons can be generated by two common EPW modes traveling oblique to x-direction at symmetric angles. Owing to symmetric laser beams, the common EPWs pointing up-right and down-right share the same $k_{cx}$ (the wave number along x-direction) with positive and negative $k_{cy}$ (the wave number along y-direction). 
The common EPWs along the two directions can form a forward-going phase of electric field, which can capture and accelerate background electrons similar to an EPW traveling along the x-direction. This can be demonstrated by a brief derivation. We assume $E_{cx1}(x,t)$ and $E_{cx2}(x,t)$ represent the electric fields along the x-direction of the two common EPWs, given by:

\begin{align}
    E_{cx1}(x,t) = E_{cx0}sin(k_{cx}x+k_{cy}y-\omega_ct+\phi_{c1})  \\
    E_{cx2}(x,t) = E_{cx0}sin(k_{cx}x-k_{cy}y-\omega_ct+\phi_{c2})  
%    E_{cx}(x,t)= E_{cx1}(x,t) + E_{cx2}(x,t)
\end{align}
where $E_{cx0}$ and $\omega_c$ are the field amplitude and frequency of the common waves, respectively, with random phases $\phi_{c1}$ and $\phi_{c2}$. The total electric field $E_{cx}$ caused by the common EPWs becomes

\begin{align}
\label{eqn:Ecx}
    E_{cx} &= E_{cx1}(x,t) + E_{cx2}(x,t) \nonumber \\
           &=2E_{cx0}
           sin(k_{cx}x-\omega_ct+\phi_{c1}/2+\phi_{c2}/2) \nonumber \\
           &\times cos(k_{cy}y +\phi_{c1}/2-\phi_{c2}/2)
\end{align}
which consists of a forward-propagating electric field $sin(k_{cx}x-\omega_ct)$ with phase velocity $v_{cph}=\omega_c/k_{cx}$, modulated along the y-direction by $cos(k_{cy}y)$. 
%Such modulated electric field can be observed in the simulations, as shown in \textcolor{black}{(do fft-filter-ifft? figure ??)}. 
For this field $E_{cx}$, the background electrons can be captured and accelerated as if they are being driven by an EPW with $v_{ph}=v_{cph}$ along the x-direction.

\begin{figure}[h]
\includegraphics[width=0.4\textwidth]{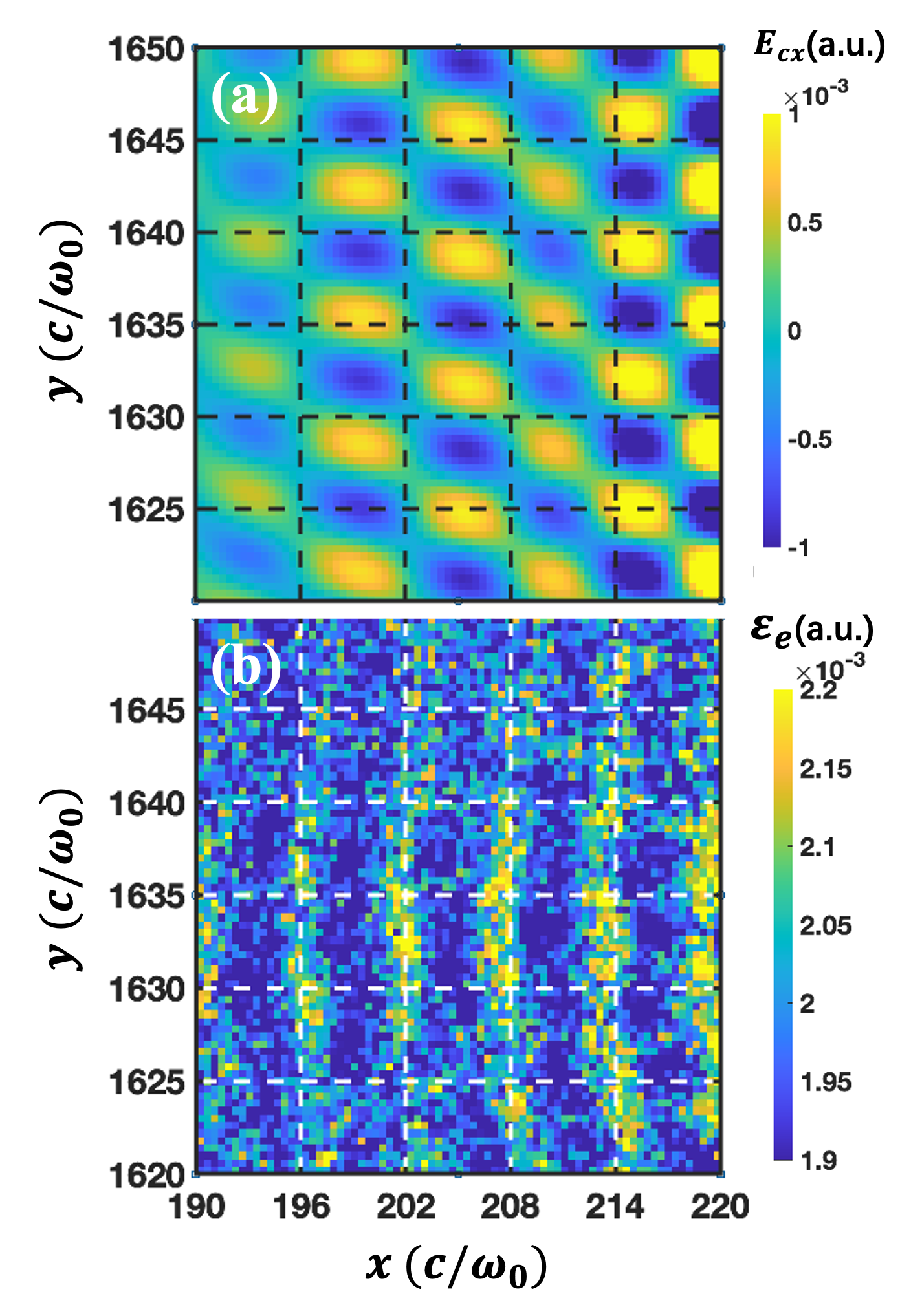}% Here is how to import EPS art
\caption{\label{fig:Ecx} (a) The $E_{cx}$ fields [equation (\ref{eqn:Ecx})] of two common waves and \textcolor{black}{(b)} electron energy density $\epsilon_e$ [equation (\ref{eqn:epsilon})]
in the overlap region near $n_e=0.217n_c$ at $t=1.9ps$ for the PIC simulation (i) [Table \ref{tab:simu_para}] with $\theta_{in}=52^\circ$. The dashed grids in the figures illustrate the alignment between the (a) $E_{cx}$ structures and (b) the accelerated electrons. The data in (a) are obtained by filtering the space spectra of the common waves out of the total electric field using Fourier and inverse-Fourier transforms. }
\end{figure}

The electric field structure described by the equation (\ref{eqn:Ecx}) is observed in the overlap region of the common waves propagating up-right and down-right. This region is situated around $n_e\sim0.217n_c$ or $x=193c/\omega_0$ and $y=1640c/\omega_0\sim L_y/2$. 
In this area,
although the fields of common waves are always mixed with the fields from the incident laser beams, their spatial spectra are clearly separated. Therefore, we can isolate the spatial distribution of only the 
common wave fields by applying spectral filtering 
in the Fourier-transformed space and then performing an inverse Fourier transform. 
%we can obtain the space distribution of only the common waves fields by performing spectra filtering in the space Fourier transform results and then conducting inverse Fourier transform. 
Figure (\ref{fig:Ecx}a) presents the filtered common wave fields at $t=1.9ps$, where both longitudinal and transverse modulations are visible around ($k_x = 0.54 \omega_0/c$, $k_y = 0.91 \omega_0/c$), in agreement with the waveform predicted by equation (\ref{eqn:Ecx}).
%which clearly exhibit both longitudinal and transverse modulations at around ($k_x=0.54\omega_0/c$, $k_y=0.91\omega_0/c$) as expected, agreeing with the waveform of equation (\ref{eqn:Ecx}).
This common wave field should propagate along x-direction and is capable of capturing and accelerating background electrons to collimated hot electrons. 

The captured electrons being accelerated by this field can be visualized through the energy density profile of the electrons, defined as:

\begin{align}
\label{eqn:epsilon}
\epsilon_e(i,j)= \sum_{k=1}^{N_{ij}}w_e(\gamma-1)
\end{align}
where $\epsilon_e$ represents the electron energy density in the cell indexed by $(i,j)$. The number of electrons in the cell is denoted by $N_{ij}$, with $w_e$ and $\gamma$ referring to the charge weight and the relativistic gamma factor of the $k$-th electron, respectively.
In the absence of the accelerated background electrons, $\epsilon_e$ is proportional to $n_e T_e$, which is approximately $1.9 \times 10^{-3}$ at $n_e \sim 0.217n_c$.
%When no background electrons are accelerated, $\epsilon_e \propto n_eT_e$ is approximately $1.9\times10^{-3}$ at $n_e\sim0.217n_c$. 
When common waves begin accelerating electrons, the local $\epsilon_e$ will increase in regions where these electrons are located. 
Figure (\ref{fig:Ecx}b) depicts the spatial distribution of $\epsilon_e$, where the yellow bands with $\epsilon_e>1.9\times10^{-3}$ corresponds to the areas where accelerated electrons are located.

By comparing the common wave field structure \textcolor{black}{[figure~(\ref{fig:Ecx}a)] to the locations of the accelerated electrons [figure (\ref{fig:Ecx}b)], using the dashed grids as a reference}, it becomes evident that the hot electrons are concentrated in bands positioned between the peaks of the common wave fields. 
%By comparing the common wave fields to the electrons being accelerated in figure (\ref{fig:Ecx}a) and (\ref{fig:Ecx}b), we find that the hot electrons gather like bands that are located between the peaks of the common wave fields. 
The parallel yellow bands along the y-direction 
suggest that these electrons are being accelerated by the wave fields and are moving along the x-direction. 
These findings indicate that the symmetric, divergent common waves can produce collimated hot electrons.

%The parallel yellow bands along y-direction suggest that these electrons are accelerated by the wave fields and move along x-direction.
%These results show that the symmetric divergent common waves can generate collimated hot electrons.

%More evidence about the connections between the forward-going hot electrons and the common waves can be given by the temporal relevance. \textcolor{black}{Figure \ref{fig:hote_collimated_and_e1} demonstrate the time evolution of instantaneous hot electron energy and the energy of common waves for simulation(i) with $\theta_{in}=52^\circ$. }They clearly show the forward-going hot electrons and the common wave energy are relevant in time, supporting our analysis.

%\textcolor{black}{(describe the relevance of side hot electrons and side EPW modes).}

\subsection{The energy scaling of hot electrons with different \textcolor{black}{divergent} angles}

\begin{figure}[h]
\includegraphics[width=0.45\textwidth]{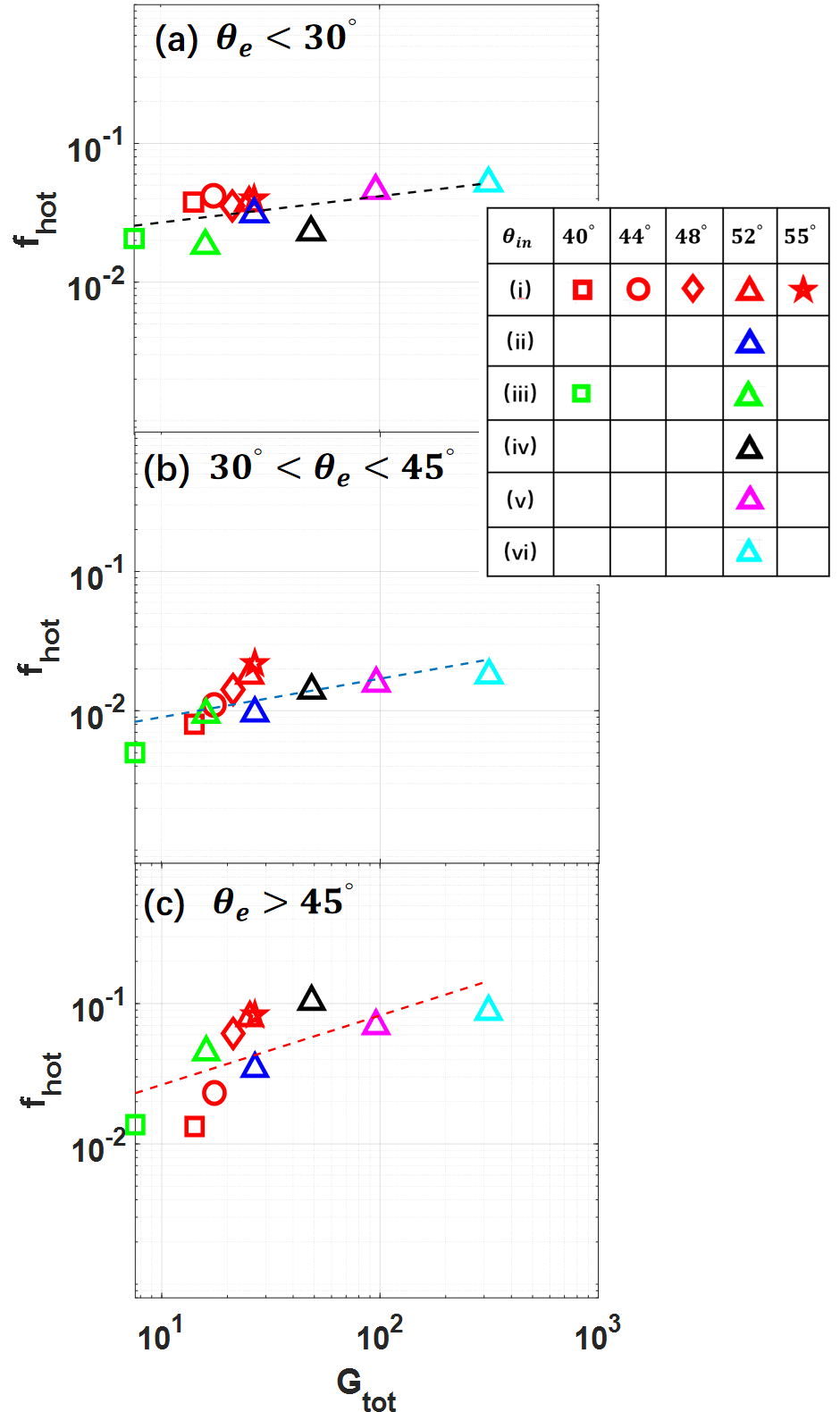}% Here is how to import EPS art
\caption{\label{fig:f_hot_gain} Scaling relations between energy of hot electrons ($f_{hot}$ normalized by incident beams) with (a) $\theta_e<30^\circ$, (b) $30^\circ<\theta_e<45^\circ$ and (c) $\theta_e>45^\circ$) and $G_{tot}$ of common wave driven by TPD and SRS.}
\end{figure}

As we discussed in the previous section, hot electrons with different divergent angles are generated through varied mechanisms. Thus, they may follow different energy scalings. 
According to the divergent angle $\theta_e$, we classify the hot electrons into three categories: collimated ($\theta_e<30^\circ$), intermediate ($30^\circ<\theta_e<45^\circ$) and divergent ($\theta_e>45^\circ$). 
Figure (\ref{fig:f_hot_gain}) plots the energy fractions of these groups against 
$G_{tot}$ for the simulations listed in 
table \ref{tab:simu_para}, specifically for 
$\theta_{in}\geq 40^\circ$.
Smaller $\theta_{in}$ values correspond to the dominant forward dual-TPD common wave mechanism [section \ref{sec:TPD_mechanism}], rather than the side TPD-SRS common waves, which $G_{tot}$ is based upon.
%The energy fractions of the three groups are plotted towards $G_{tot}$ in figure (\ref{fig:f_hot_gain}) for the simulations in table \ref{tab:simu_para} with $\theta_{in}\geq 40^\circ$, as smaller $\theta_{in}$ corresponds to the dominant mechanism of forward dual-TPD common waves [section \ref{sec:TPD_mechanism}], instead of the side TPD-SRS common waves, based on which $G_{tot}$ is calculated.
Among these simulations, the results of simulation (ii) and (iv) with $\theta_{in}\geq 40^\circ$ are not given, as the $f_{hot}$ has not reached quasi-steady state till the end of the simulation time due to the weak instability growth.

Our findings show that both the collimated and divergent hot electrons tend to have higher energy than the intermediate group, consistent with the multi-peak angular distributions in figure (\ref{fig:angle_d}). 
%mostly have higher energy than the intermediate components, agreeing with the multi-peak angular distributions in figure (\ref{fig:angle_d}). 
In this log-log plot, the $f_{hot}$ values of all three groups increase quasi-linearly with $G_{tot}$, but with notably different slopes. This behavior suggests the following approximate scaling:
%of all the three groups increases quasi-linearly as $G_{tot}$ going up, but with notably different slopes. This indicates the following approximate scaling
\begin{align}
\label{eqn:scaling}
f_{hot}\approx CG_{tot}^\alpha
\end{align}
where \textcolor{black}{$\alpha=$ 0.19, 0.28 and 0.49 }, and the coefficient $C=$ $1.7\times10^{-2}$, $4.8\times10^{-3}$ and $8.5\times10^{-3}$ for the collimated, intermediate and divergent groups of hot electrons, respectively.

The different $\alpha$ values indicate that $f_{hot}$ exhibits varying sensitivity to $G_{tot}$, which could be related to the differences in acceleration efficiency 
among the groups. 
%suggest different sensitivity of $f_{hot}$ to $G_{tot}$, which can be relevant to the acceleration efficiency of the hot electrons in the three groups. 
Divergent electrons are accelerated by the common wave and the paired side TPD mode. In the example of the simulation (i) with $\theta_{in}=52^\circ$, the two modes have $|k|\sim 1.1\omega_0/c$ and $1.6\omega_0/c$, corresponding to phase velocities of $0.45c\sim5.9v_{te}$ and $0.31c\sim4.0v_{te}$, respectively. 
In contrast, according to the equation (\ref{eqn:Ecx}), the collimated hot electrons are accelerated by the moving electric field with phase velocity of $v_{cph}=\omega_c/k_{cx}\sim0.88c\sim11.5v_{te}$, much faster than that for divergent hot electrons. 
These results show that divergent electrons 
are more easily accelerated due to the lower 
phase velocity of EPW modes compared to 
collimated electrons. 
This difference may account for the varying sensitivity of $f_{hot}$ in response to changes in the amplitude of EPW modes, which is directly related to $G_{tot}$. A detailed study on the reasons behind these different $\alpha$ 
values will be the focus of future research.

\section{Considering Realistic Experimental Conditions}
\subsection{\textcolor{black}{Effects of asymmetric incidence}}

\begin{figure}[h]
\includegraphics[width=0.35\textwidth]{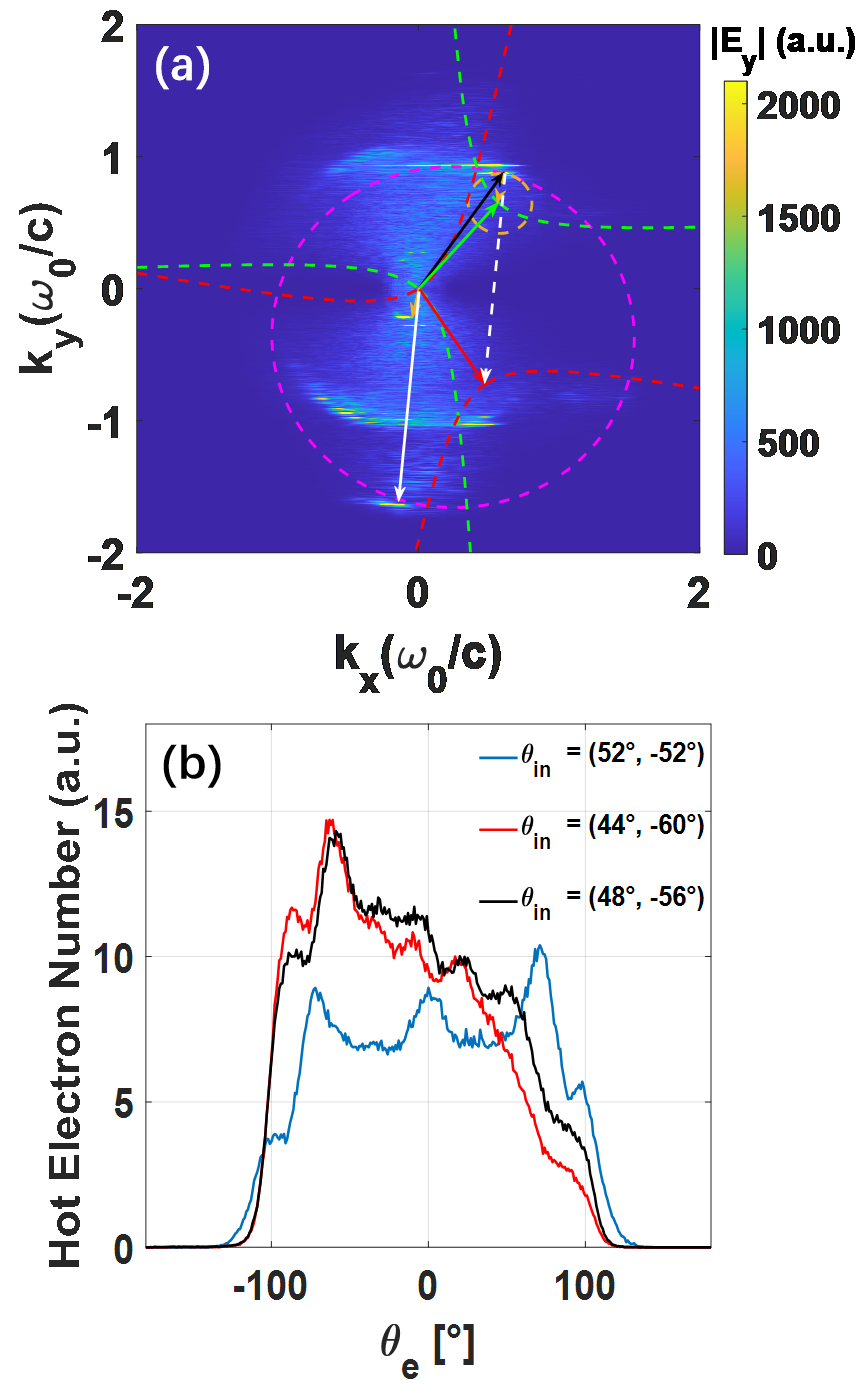}% Here is how to import EPS art
\caption{\label{fig:asymmetry} \textcolor{black}{(a) The space spectra of $E_y$ for the simulation (b) [table \ref{tab:G_tot_asymmetry}] during the
simulation time of 5-6ps. (b) The angular distributions of hot electrons for all the simulations in table \ref{tab:G_tot_asymmetry} with different $\theta_{in}$'s.}}
\end{figure}

\textcolor{black}{In realistic experiments on large-scale laser facilities, despite the symmetric configuration of the beams, the orientation of density contours evolves in space and time during the target surface ablation. As a result, it is common for laser beams to deviate from symmetry relative to the normal of the density contours near $n_c/4$. Therefore, hot electron generation under asymmetric incidence is crucial for accurately assessing $f_{hot}$ in experimental evaluations.
In such scenarios, the matching ellipses [Figure (\ref{fig:TPD_SRS}a)] for both TPD and SRS mechanisms continue to define the common wave modes responsible for the \textcolor{black}{STS} mechanism. The cumulative wave amplification at various densities can still be scaled by the total gain, $G_{tot}$, as derived from equations (\ref{eqn:Rosenbluth}) to (\ref{eqn:GW}).}

\textcolor{black}{We calculate two new cases using the same physical parameters as case (i) in table (\ref{tab:simu_para}) but with the laser incident angles directed up-right and down-right ($\theta_{in,up}$, $\theta_{in,down}$) changed from symmetric ($52^\circ$, $-52^\circ$) to asymmetric configurations of ($48^\circ$, $-56^\circ$) and ($44^\circ$, $-60^\circ$). These shifts move the beams away from the density gradient direction by $4^\circ$ and $8^\circ$, as detailed in table (\ref{tab:G_tot_asymmetry}). }

\begin{table}
\caption{\label{tab:G_tot_asymmetry}\textcolor{black}{The incident angles and calculated $G_{tot,up}$ and $G_{tot,down}$ for asymmetric incident laser beams.}}
\centering
%\resizebox{\columnwidth}{!}{%
\begin{tabular}{c c c c c}
\toprule
 Index&$\theta_{in,up}(^\circ)$ &$\theta_{in,down}(^\circ)$ &$G_{tot,up}$ &$G_{tot,down}$   \\
%\midrule
\hline
(a)&52 & -52 & 25 & 25 \\
\hline
(b)&48& -56  & 28 & 23 \\
\hline
(c)&44& -60  & 32 & 20 \\
\hline
%\bottomrule
\end{tabular}
%}
\end{table}

\textcolor{black}{The computed total gains, $G_{tot}$, for the common waves propagating up-right and down-right, denoted as $G_{tot,up}$ and $G_{tot,down}$, are presented in table (\ref{tab:G_tot_asymmetry}) for all the three combinations of ($\theta_{in,up}$, $\theta_{in,down}$). 
From these results, we observe that the asymmetric incident beams lead to variations of up to $28\%$ in $G_{tot,up}$ and $G_{tot,down}$ when compared to the symmetric case.
These variations results in differing amplitudes for the two common wave modes and their associated TPD modes, leading to asymmetric angular distributions of hot electrons. As illustrated in Figure (\ref{fig:asymmetry}a), the amplitude of the common wave propagating up-right (black arrow) is noticeably stronger than that of the wave propagating down-right. For the stronger up-right common wave, the corresponding SRS-scattered light (orange arrow) and the paired TPD mode (white solid arrow), both of which point near the negative $k_y$ direction, exhibit significantly greater strength than those associated with the other down-right common wave.}

\textcolor{black}{Due to the asymmetry in $G_{tot,up}$ and $G_{tot,down}$, the angular distributions of hot electrons exhibit an asymmetric shape, as illustrated in figure (\ref{fig:asymmetry}b)]. 
Interestingly, although $G_{tot,up}>G_{tot,down}$ [table (\ref{tab:G_tot_asymmetry})] suggests a stronger common wave propagating in the up-right direction (positive angles), a greater number of hot electrons are produced along negative angles [figure (\ref{fig:asymmetry}b)]. 
This apparent contradiction arises because the hot electrons are generated by both the common wave and the paired TPD modes. 
The two waves travel oppositely in y-direction, meaning that hot electrons with positive angles are influenced by the stronger common wave propagating up-right and the weaker TPD mode paired with the down-right common wave, and \textit{vice versa}.
%so the hot electrons with positive angles are contributed by the stronger common wave pointing up-right and the weaker TPD mode paired with the weaker common wave pointing down-right, and \textit{vice versa}.
As a result, it is not immediately clear whether more hot electrons will be generated at positive or negative angles. However, the results indicate a greater concentration of hot electrons at negative angles, implying that hot electron generation is more sensitive to the TPD modes with higher $|k|$ and lower phase velocity.}

\textcolor{black}{Moreover, the total hot electron energy increases by approximately 20\% for asymmetric $\theta_{in}$'s, as the surplus of hot electrons on one side outweighs the reduction on the other.}

%A natural question is to what extent can our PIC simulation results in ps time scale infer the realistic implosion process in several ns.
%It is the fact that there is a huge gap between the two time scales, which can intuitively cause doubts over the applicability of our results to the realistic experiment. 
%To this question, the answer it that, as long as a quasi-steady state can be achieved in the short-time-scale process, the quantity at the quasi-steady state can be reasonably coupled to the long-time-scale process via the slow-varying hydro quantities, as the long-time-scale process does not care about the short-term variations before reaching quasi-steady states.
%Following this principle, previous study has achieved proper predictions of the time evolution of hot electron energy in ns scale experiments\cite{Follett2017} using the results from a group of hybrid simulations (running for 20ps) describing the hot electron generation for different laser plasma conditions through the entire laser pulse and focal spot.  

%Back to our simulations, since the results have clearly reached quasi-steady states [figure (\ref{fig:dne_x_t})], our results should properly reflect the hot electron characteristics for the corresponding physics parameters and beam geometry of lasers and plasmas.

\subsection{\textcolor{black}{Experimental observability of the \textcolor{black}{STS} mechanism}}

\textcolor{black}{Since the scattered light in the \textcolor{black}{STS} mechanism is essentially enhanced SRS scattered light, it should be detectable in experiments. For the simulation (i) [table \ref{tab:simu_para}] with symmetric incident angle $\theta_{in}=52^\circ$, if we assume $n_e$ only varies along $x-$ direction in space, the scattered light with a frequency of $\omega\sim0.512\omega_0$ and wave vector component $k_y\sim0.212\omega_0/c$ will ultimately have a scattering angle of only $\sim24^\circ$ relative to the negative x-direction after exiting into the vacuum. 
This scattered light has the same frequency as the collective Thomson scattering of the incident light on the common EPW mode. Once diagnosed, it should mix with the double-band structure of the TPD $\omega_0/2$ spectra, and strengthen the blue-shift component. This qualitatively agrees with the observations of significantly stronger blue-shift components compared to red-shift components off the target normal direction in previous OMEGA experiments\cite{Seka2014}.
}

\subsection{\textcolor{black}{The actual incident angles in the vacuum}}

\textcolor{black}{It is important to clarify that all $\theta_{in}$ values in this article refer to the incident angles at the density $n_e \sim 0.2n_c$ on the incident plane of our PIC simulations. These angles typically differ from the actual vacuum incident angles, $\theta_{in,vac}$, which are the true parameters defining the beam geometry. We now discuss the relationship between $\theta_{in,vac}$ and $\theta_{in}$ by analyzing two extreme cases.}

\textcolor{black}{In the first case, where the plasma density is spherically symmetric with the incident beams directed toward the center, we have $\theta_{in,vac} \sim \theta_{in}$. In contrast, for a 1D planar geometry where $n_e$ varies only along the x-axis, the relationship between $\theta_{in,vac}$ and $\theta_{in}$ can be determined by the fact that $k_y$ is constant and by using the light wave dispersion relation at $n_e = 0.2n_c$:}

\begin{align}
\label{eqn:theta_vac}
\theta_{in,vac}\approx arcsin[0.89sin\theta_{in}]
\end{align}

\textcolor{black}{This yields pairs of ($\theta_{in,vac}$, $\theta_{in}$) such as ($50^\circ$, $60^\circ$), ($43^\circ$, $50^\circ$), and ($35^\circ$, $40^\circ$). Generally, for a given $\theta_{in,vac}$, the effective incident angle for TPD should lie between $\theta_{in,vac}$ and the calculated $\theta_{in}$ from equation (\ref{eqn:theta_vac}).}

\section{conclusion}

In this study, we investigate the angular variations of hot electrons generated by multiple laser beams with varying incident angles. We identify a novel multi-beam collaborative process involving two-plasmon decay (TPD) and stimulated Raman scattering (SRS), termed \textcolor{black}{STS}, which produces hot electrons with distinct divergent and collimated components.

We demonstrate that the key factor influencing the acceleration of these components is the common Langmuir wave associated with \textcolor{black}{STS}. By properly modeling the common wave gain, we obtain scaling relations between the gains and the energy fractions of hot electrons at different angles. Our results show that the energy of divergent hot electrons is more sensitive to variations in the common wave gain.
Furthermore, when the density gradient deviates from the bisector of the laser beams— a common occurrence in large-scale laser facilities—our study reveals that asymmetric beam incidence results in unequal common wave gains, leading to asymmetric angular distributions of hot electrons. The stronger peak is observed in the direction opposite to the laser beam with the larger incident angle. \textcolor{black}{These findings highlight potential preheating concerns due to divergent and collimated hot electrons in laser-plasma configurations with high gains in future DCI experiments.}

%Finally, we point out that the scattered light of the TSI mechanism qualitatively agrees with previous experimental observations.
%Our study shows that this scattered light shares the frequency of collective Thomson scattering from the common EPW mode. 
%Although the scattered light has large onsite scatter angle of $\sim95^\circ$, after reaching vacuum, the scattered light propagation direction shifts close to the density gradient, approximately $24^\circ$ in our simulations, with 1D plasma density variations. 
%This scattered light should mix with the double-band structure of the TPD $\omega_0/2$ spectra and enhance the blue-shift component, which is consistent with previous OMEGA experiments\cite{Seka2014} that observed significantly stronger blue-shift than red-shift components when viewed off the target normal.

Finally, we observe that the scattered light associated with the \textcolor{black}{STS} mechanism qualitatively agrees with previous experimental findings. 
This scattered light shares the frequency of collective Thomson scattered light from the common wave. 
Although the onsite scattering angle is large (approximately $95^\circ$), the direction of the scattered light shifts towards the density gradient when propagating towards vacuum, with \textcolor{black}{final} angle of about $24^\circ$ in our simulations for 1D plasma density variations. This scattered light merges with the double-band structure of the TPD $\omega_0/2$ spectrum, enhancing the blue-shift component, consistent with prior OMEGA experiments\cite{Seka2014} that observed significantly stronger blue-shift than red-shift signals when viewed off the target normal.

\begin{acknowledgments}
This research was supported by the National Natural Science Foundation of China (NSFC) (Grant Nos. 12275269, 12175229, 12375243, 12275032, 12388101), by the Strategic Priority Research Program of Chinese Academy of Sciences (Grant Nos. XDA25010200 and XDA25050400). The numerical simulations in this paper were conducted on Hefei advanced computing center.
\end{acknowledgments}

\section*{References}
%\nocite{*}
%\bibliographystyle{plain}
\bibliography{aipsamp}% Produces the bibliography via BibTeX.

%merlin.mbs aipnum4-1.bst 2010-07-25 4.21a (PWD, AO, DPC) hacked
%Control: key (0)
%Control: author (8) initials jnrlst
%Control: editor formatted (1) identically to author
%Control: production of article title (0) allowed
%Control: page (1) range
%Control: year (1) truncated
%Control: production of eprint (0) enabled
\providecommand{\noopsort}[1]{}\providecommand{\singleletter}[1]{#1}%
\begin{thebibliography}{59}%
\makeatletter
\providecommand \@ifxundefined [1]{%
 \@ifx{#1\undefined}
}%
\providecommand \@ifnum [1]{%
 \ifnum #1\expandafter \@firstoftwo
 \else \expandafter \@secondoftwo
 \fi
}%
\providecommand \@ifx [1]{%
 \ifx #1\expandafter \@firstoftwo
 \else \expandafter \@secondoftwo
 \fi
}%
\providecommand \natexlab [1]{#1}%
\providecommand \enquote  [1]{``#1''}%
\providecommand \bibnamefont  [1]{#1}%
\providecommand \bibfnamefont [1]{#1}%
\providecommand \citenamefont [1]{#1}%
\providecommand \href@noop [0]{\@secondoftwo}%
\providecommand \href [0]{\begingroup \@sanitize@url \@href}%
\providecommand \@href[1]{\@@startlink{#1}\@@href}%
\providecommand \@@href[1]{\endgroup#1\@@endlink}%
\providecommand \@sanitize@url [0]{\catcode `\\12\catcode `\$12\catcode
  `\&12\catcode `\#12\catcode `\^12\catcode `\_12\catcode `\%12\relax}%
\providecommand \@@startlink[1]{}%
\providecommand \@@endlink[0]{}%
\providecommand \url  [0]{\begingroup\@sanitize@url \@url }%
\providecommand \@url [1]{\endgroup\@href {#1}{\urlprefix }}%
\providecommand \urlprefix  [0]{URL }%
\providecommand \Eprint [0]{\href }%
\providecommand \doibase [0]{http://dx.doi.org/}%
\providecommand \selectlanguage [0]{\@gobble}%
\providecommand \bibinfo  [0]{\@secondoftwo}%
\providecommand \bibfield  [0]{\@secondoftwo}%
\providecommand \translation [1]{[#1]}%
\providecommand \BibitemOpen [0]{}%
\providecommand \bibitemStop [0]{}%
\providecommand \bibitemNoStop [0]{.\EOS\space}%
\providecommand \EOS [0]{\spacefactor3000\relax}%
\providecommand \BibitemShut  [1]{\csname bibitem#1\endcsname}%
\let\auto@bib@innerbib\@empty
%</preamble>
\bibitem [{\citenamefont {Atzeni}\ and\ \citenamefont {Meyer-ter
  Vehn}(2004)}]{Atzeni2004}%
  \BibitemOpen
  \bibfield  {author} {\bibinfo {author} {\bibfnamefont {S.}~\bibnamefont
  {Atzeni}}\ and\ \bibinfo {author} {\bibfnamefont {J.}~\bibnamefont {Meyer-ter
  Vehn}},\ }\href@noop {} {\emph {\bibinfo {title} {{The Physics of Inertial
  Fusion: BeamPlasma Interaction, Hydrodynamics, Hot Dense Matter}}}}\
  (\bibinfo  {publisher} {Oxford University Press},\ \bibinfo {address}
  {Oxford},\ \bibinfo {year} {2004})\BibitemShut {NoStop}%
\bibitem [{\citenamefont {W.L.Kruer}(2003)}]{Kruer2003}%
  \BibitemOpen
  \bibfield  {author} {\bibinfo {author} {\bibnamefont {W.L.Kruer}},\
  }\href@noop {} {\emph {\bibinfo {title} {{The Physics of Laser Plasma
  Interactions}}}}\ (\bibinfo  {publisher} {Westview Press},\ \bibinfo
  {address} {Boulder CO},\ \bibinfo {year} {2003})\BibitemShut {NoStop}%
\bibitem [{\citenamefont {Christopherson}\ \emph {et~al.}(2021)\citenamefont
  {Christopherson}, \citenamefont {Betti}, \citenamefont {Forrest},
  \citenamefont {Howard}, \citenamefont {Theobald}, \citenamefont {Delettrez},
  \citenamefont {Rosenberg}, \citenamefont {Solodov}, \citenamefont {Stoeckl},
  \citenamefont {Patel}, \citenamefont {Gopalaswamy}, \citenamefont {Cao},
  \citenamefont {Peebles}, \citenamefont {Edgell}, \citenamefont {Seka},
  \citenamefont {Epstein}, \citenamefont {Wei}, \citenamefont {Gatu~Johnson},
  \citenamefont {Simpson}, \citenamefont {Regan},\ and\ \citenamefont
  {Campbell}}]{christopherson2021}%
  \BibitemOpen
  \bibfield  {author} {\bibinfo {author} {\bibfnamefont {A.~R.}\ \bibnamefont
  {Christopherson}}, \bibinfo {author} {\bibfnamefont {R.}~\bibnamefont
  {Betti}}, \bibinfo {author} {\bibfnamefont {C.~J.}\ \bibnamefont {Forrest}},
  \bibinfo {author} {\bibfnamefont {J.}~\bibnamefont {Howard}}, \bibinfo
  {author} {\bibfnamefont {W.}~\bibnamefont {Theobald}}, \bibinfo {author}
  {\bibfnamefont {J.~A.}\ \bibnamefont {Delettrez}}, \bibinfo {author}
  {\bibfnamefont {M.~J.}\ \bibnamefont {Rosenberg}}, \bibinfo {author}
  {\bibfnamefont {A.~A.}\ \bibnamefont {Solodov}}, \bibinfo {author}
  {\bibfnamefont {C.}~\bibnamefont {Stoeckl}}, \bibinfo {author} {\bibfnamefont
  {D.}~\bibnamefont {Patel}}, \bibinfo {author} {\bibfnamefont
  {V.}~\bibnamefont {Gopalaswamy}}, \bibinfo {author} {\bibfnamefont
  {D.}~\bibnamefont {Cao}}, \bibinfo {author} {\bibfnamefont {J.~L.}\
  \bibnamefont {Peebles}}, \bibinfo {author} {\bibfnamefont {D.~H.}\
  \bibnamefont {Edgell}}, \bibinfo {author} {\bibfnamefont {W.}~\bibnamefont
  {Seka}}, \bibinfo {author} {\bibfnamefont {R.}~\bibnamefont {Epstein}},
  \bibinfo {author} {\bibfnamefont {M.~S.}\ \bibnamefont {Wei}}, \bibinfo
  {author} {\bibfnamefont {M.}~\bibnamefont {Gatu~Johnson}}, \bibinfo {author}
  {\bibfnamefont {R.}~\bibnamefont {Simpson}}, \bibinfo {author} {\bibfnamefont
  {S.~P.}\ \bibnamefont {Regan}}, \ and\ \bibinfo {author} {\bibfnamefont
  {E.~M.}\ \bibnamefont {Campbell}},\ }\bibfield  {title} {\enquote {\bibinfo
  {title} {Direct {{Measurements}} of {{DT Fuel Preheat}} from {{Hot
  Electrons}} in {{Direct-Drive Inertial Confinement Fusion}}},}\ }\href@noop
  {} {\bibfield  {journal} {\bibinfo  {journal} {Physical Review Letters}\
  }\textbf {\bibinfo {volume} {127}},\ \bibinfo {pages} {055001} (\bibinfo
  {year} {2021})}\BibitemShut {NoStop}%
\bibitem [{\citenamefont {Solodov}\ \emph {et~al.}(2022)\citenamefont
  {Solodov}, \citenamefont {Rosenberg}, \citenamefont {Stoeckl}, \citenamefont
  {Christopherson}, \citenamefont {Betti}, \citenamefont {Radha}, \citenamefont
  {Stoeckl}, \citenamefont {Hohenberger}, \citenamefont {Bachmann},
  \citenamefont {Epstein}, \citenamefont {Follett}, \citenamefont {Seka},
  \citenamefont {Myatt}, \citenamefont {Michel}, \citenamefont {Regan},
  \citenamefont {Palastro}, \citenamefont {Froula}, \citenamefont {Campbell},\
  and\ \citenamefont {Goncharov}}]{Solodov2022}%
  \BibitemOpen
  \bibfield  {author} {\bibinfo {author} {\bibfnamefont {A.~A.}\ \bibnamefont
  {Solodov}}, \bibinfo {author} {\bibfnamefont {M.~J.}\ \bibnamefont
  {Rosenberg}}, \bibinfo {author} {\bibfnamefont {M.}~\bibnamefont {Stoeckl}},
  \bibinfo {author} {\bibfnamefont {A.~R.}\ \bibnamefont {Christopherson}},
  \bibinfo {author} {\bibfnamefont {R.}~\bibnamefont {Betti}}, \bibinfo
  {author} {\bibfnamefont {P.~B.}\ \bibnamefont {Radha}}, \bibinfo {author}
  {\bibfnamefont {C.}~\bibnamefont {Stoeckl}}, \bibinfo {author} {\bibfnamefont
  {M.}~\bibnamefont {Hohenberger}}, \bibinfo {author} {\bibfnamefont
  {B.}~\bibnamefont {Bachmann}}, \bibinfo {author} {\bibfnamefont
  {R.}~\bibnamefont {Epstein}}, \bibinfo {author} {\bibfnamefont {R.~K.}\
  \bibnamefont {Follett}}, \bibinfo {author} {\bibfnamefont {W.}~\bibnamefont
  {Seka}}, \bibinfo {author} {\bibfnamefont {J.~F.}\ \bibnamefont {Myatt}},
  \bibinfo {author} {\bibfnamefont {P.}~\bibnamefont {Michel}}, \bibinfo
  {author} {\bibfnamefont {S.~P.}\ \bibnamefont {Regan}}, \bibinfo {author}
  {\bibfnamefont {J.~P.}\ \bibnamefont {Palastro}}, \bibinfo {author}
  {\bibfnamefont {D.~H.}\ \bibnamefont {Froula}}, \bibinfo {author}
  {\bibfnamefont {E.~M.}\ \bibnamefont {Campbell}}, \ and\ \bibinfo {author}
  {\bibfnamefont {V.~N.}\ \bibnamefont {Goncharov}},\ }\bibfield  {title}
  {\enquote {\bibinfo {title} {{Hot-electron preheat and mitigation in
  polar-direct-drive experiments at the National Ignition Facility}},}\ }\href
  {\doibase 10.1103/physreve.106.055204} {\bibfield  {journal} {\bibinfo
  {journal} {Physical Review E}\ }\textbf {\bibinfo {volume} {106}},\ \bibinfo
  {pages} {055204} (\bibinfo {year} {2022})}\BibitemShut {NoStop}%
\bibitem [{\citenamefont {Nuckolls}\ \emph {et~al.}(1972)\citenamefont
  {Nuckolls}, \citenamefont {Wood}, \citenamefont {Thiessen},\ and\
  \citenamefont {Zimmerman}}]{Nuckolls1972}%
  \BibitemOpen
  \bibfield  {author} {\bibinfo {author} {\bibfnamefont {J.}~\bibnamefont
  {Nuckolls}}, \bibinfo {author} {\bibfnamefont {L.}~\bibnamefont {Wood}},
  \bibinfo {author} {\bibfnamefont {A.}~\bibnamefont {Thiessen}}, \ and\
  \bibinfo {author} {\bibfnamefont {G.}~\bibnamefont {Zimmerman}},\ }\bibfield
  {title} {\enquote {\bibinfo {title} {{Laser compression of matter to
  super-high densities: Thermonuclear (CTR) applications}},}\ }\href {\doibase
  10.1038/239139a0} {\bibfield  {journal} {\bibinfo  {journal} {Nature}\
  }\textbf {\bibinfo {volume} {239}},\ \bibinfo {pages} {139--142} (\bibinfo
  {year} {1972})}\BibitemShut {NoStop}%
\bibitem [{\citenamefont {He}\ \emph {et~al.}(2016)\citenamefont {He},
  \citenamefont {Li}, \citenamefont {Fan}, \citenamefont {Wang}, \citenamefont
  {Liu}, \citenamefont {Lan}, \citenamefont {Wu},\ and\ \citenamefont
  {Ye}}]{He2016}%
  \BibitemOpen
  \bibfield  {author} {\bibinfo {author} {\bibfnamefont {X.~T.}\ \bibnamefont
  {He}}, \bibinfo {author} {\bibfnamefont {J.~W.}\ \bibnamefont {Li}}, \bibinfo
  {author} {\bibfnamefont {Z.~F.}\ \bibnamefont {Fan}}, \bibinfo {author}
  {\bibfnamefont {L.~F.}\ \bibnamefont {Wang}}, \bibinfo {author}
  {\bibfnamefont {J.}~\bibnamefont {Liu}}, \bibinfo {author} {\bibfnamefont
  {K.}~\bibnamefont {Lan}}, \bibinfo {author} {\bibfnamefont {J.~F.}\
  \bibnamefont {Wu}}, \ and\ \bibinfo {author} {\bibfnamefont {W.~H.}\
  \bibnamefont {Ye}},\ }\bibfield  {title} {\enquote {\bibinfo {title} {{A
  hybrid-drive nonisobaric-ignition scheme for inertial confinement fusion}},}\
  }\href {\doibase 10.1063/1.4960973} {\bibfield  {journal} {\bibinfo
  {journal} {Physics of Plasmas}\ }\textbf {\bibinfo {volume} {23}},\ \bibinfo
  {pages} {082706} (\bibinfo {year} {2016})}\BibitemShut {NoStop}%
\bibitem [{\citenamefont {Zhang}\ \emph
  {et~al.}(2020{\natexlab{a}})\citenamefont {Zhang}, \citenamefont {Wang},
  \citenamefont {Yang}, \citenamefont {Wu}, \citenamefont {Ma}, \citenamefont
  {Jiao}, \citenamefont {Zhang}, \citenamefont {Wu}, \citenamefont {Yuan},
  \citenamefont {Li},\ and\ \citenamefont {Zhu}}]{Zhang2020}%
  \BibitemOpen
  \bibfield  {author} {\bibinfo {author} {\bibfnamefont {J.}~\bibnamefont
  {Zhang}}, \bibinfo {author} {\bibfnamefont {W.~M.}\ \bibnamefont {Wang}},
  \bibinfo {author} {\bibfnamefont {X.~H.}\ \bibnamefont {Yang}}, \bibinfo
  {author} {\bibfnamefont {D.}~\bibnamefont {Wu}}, \bibinfo {author}
  {\bibfnamefont {Y.~Y.}\ \bibnamefont {Ma}}, \bibinfo {author} {\bibfnamefont
  {J.~L.}\ \bibnamefont {Jiao}}, \bibinfo {author} {\bibfnamefont
  {Z.}~\bibnamefont {Zhang}}, \bibinfo {author} {\bibfnamefont {F.~Y.}\
  \bibnamefont {Wu}}, \bibinfo {author} {\bibfnamefont {X.~H.}\ \bibnamefont
  {Yuan}}, \bibinfo {author} {\bibfnamefont {Y.~T.}\ \bibnamefont {Li}}, \ and\
  \bibinfo {author} {\bibfnamefont {J.~Q.}\ \bibnamefont {Zhu}},\ }\bibfield
  {title} {\enquote {\bibinfo {title} {{Double-cone ignition scheme for
  inertial confinement fusion}},}\ }\href {\doibase 10.1098/rsta.2020.0015}
  {\bibfield  {journal} {\bibinfo  {journal} {Philosophical Transactions of the
  Royal Society A: Mathematical, Physical and Engineering Sciences}\ }\textbf
  {\bibinfo {volume} {378}},\ \bibinfo {pages} {20200015} (\bibinfo {year}
  {2020}{\natexlab{a}})}\BibitemShut {NoStop}%
\bibitem [{\citenamefont {Lan}(2022)}]{Lan2022}%
  \BibitemOpen
  \bibfield  {author} {\bibinfo {author} {\bibfnamefont {K.}~\bibnamefont
  {Lan}},\ }\bibfield  {title} {\enquote {\bibinfo {title} {{Dream fusion in
  octahedral spherical hohlraum}},}\ }\href {\doibase 10.1063/5.0103362}
  {\bibfield  {journal} {\bibinfo  {journal} {Matter and Radiation at
  Extremes}\ }\textbf {\bibinfo {volume} {7}},\ \bibinfo {pages} {055701}
  (\bibinfo {year} {2022})}\BibitemShut {NoStop}%
\bibitem [{\citenamefont {Betti}\ \emph {et~al.}(2007)\citenamefont {Betti},
  \citenamefont {Zhou}, \citenamefont {Anderson}, \citenamefont {Perkins},
  \citenamefont {Theobald},\ and\ \citenamefont {Solodov}}]{Betti2007}%
  \BibitemOpen
  \bibfield  {author} {\bibinfo {author} {\bibfnamefont {R.}~\bibnamefont
  {Betti}}, \bibinfo {author} {\bibfnamefont {C.~D.}\ \bibnamefont {Zhou}},
  \bibinfo {author} {\bibfnamefont {K.~S.}\ \bibnamefont {Anderson}}, \bibinfo
  {author} {\bibfnamefont {L.~J.}\ \bibnamefont {Perkins}}, \bibinfo {author}
  {\bibfnamefont {W.}~\bibnamefont {Theobald}}, \ and\ \bibinfo {author}
  {\bibfnamefont {A.~A.}\ \bibnamefont {Solodov}},\ }\bibfield  {title}
  {\enquote {\bibinfo {title} {{Shock ignition of thermonuclear fuel with high
  areal density}},}\ }\href {\doibase 10.1103/PhysRevLett.98.155001} {\bibfield
   {journal} {\bibinfo  {journal} {Physical Review Letters}\ }\textbf {\bibinfo
  {volume} {98}},\ \bibinfo {pages} {155001} (\bibinfo {year}
  {2007})}\BibitemShut {NoStop}%
\bibitem [{\citenamefont {Baton}\ \emph {et~al.}(2020)\citenamefont {Baton},
  \citenamefont {Cola{\"{i}}tis}, \citenamefont {Rousseaux}, \citenamefont
  {Boutoux}, \citenamefont {Brygoo}, \citenamefont {Jacquet}, \citenamefont
  {Koenig}, \citenamefont {Batani}, \citenamefont {Casner}, \citenamefont
  {Bel}, \citenamefont {Raffestin}, \citenamefont {Tentori}, \citenamefont
  {Tikhonchuk}, \citenamefont {Trela}, \citenamefont {Reverdin}, \citenamefont
  {Le-Deroff}, \citenamefont {Theobald}, \citenamefont {Cristoforetti},
  \citenamefont {Gizzi}, \citenamefont {Koester}, \citenamefont {Labate},\ and\
  \citenamefont {Shigemori}}]{Baton2020}%
  \BibitemOpen
  \bibfield  {author} {\bibinfo {author} {\bibfnamefont {S.~D.}\ \bibnamefont
  {Baton}}, \bibinfo {author} {\bibfnamefont {A.}~\bibnamefont
  {Cola{\"{i}}tis}}, \bibinfo {author} {\bibfnamefont {C.}~\bibnamefont
  {Rousseaux}}, \bibinfo {author} {\bibfnamefont {G.}~\bibnamefont {Boutoux}},
  \bibinfo {author} {\bibfnamefont {S.}~\bibnamefont {Brygoo}}, \bibinfo
  {author} {\bibfnamefont {L.}~\bibnamefont {Jacquet}}, \bibinfo {author}
  {\bibfnamefont {M.}~\bibnamefont {Koenig}}, \bibinfo {author} {\bibfnamefont
  {D.}~\bibnamefont {Batani}}, \bibinfo {author} {\bibfnamefont
  {A.}~\bibnamefont {Casner}}, \bibinfo {author} {\bibfnamefont {E.~L.}\
  \bibnamefont {Bel}}, \bibinfo {author} {\bibfnamefont {D.}~\bibnamefont
  {Raffestin}}, \bibinfo {author} {\bibfnamefont {A.}~\bibnamefont {Tentori}},
  \bibinfo {author} {\bibfnamefont {V.}~\bibnamefont {Tikhonchuk}}, \bibinfo
  {author} {\bibfnamefont {J.}~\bibnamefont {Trela}}, \bibinfo {author}
  {\bibfnamefont {C.}~\bibnamefont {Reverdin}}, \bibinfo {author}
  {\bibfnamefont {L.}~\bibnamefont {Le-Deroff}}, \bibinfo {author}
  {\bibfnamefont {W.}~\bibnamefont {Theobald}}, \bibinfo {author}
  {\bibfnamefont {G.}~\bibnamefont {Cristoforetti}}, \bibinfo {author}
  {\bibfnamefont {L.~A.}\ \bibnamefont {Gizzi}}, \bibinfo {author}
  {\bibfnamefont {P.}~\bibnamefont {Koester}}, \bibinfo {author} {\bibfnamefont
  {L.}~\bibnamefont {Labate}}, \ and\ \bibinfo {author} {\bibfnamefont
  {K.}~\bibnamefont {Shigemori}},\ }\bibfield  {title} {\enquote {\bibinfo
  {title} {{Preliminary results from the LMJ-PETAL experiment on hot electrons
  characterization in the context of shock ignition}},}\ }\href {\doibase
  10.1016/j.hedp.2020.100796} {\bibfield  {journal} {\bibinfo  {journal} {High
  Energy Density Physics}\ }\textbf {\bibinfo {volume} {36}},\ \bibinfo {pages}
  {100796} (\bibinfo {year} {2020})}\BibitemShut {NoStop}%
\bibitem [{\citenamefont {Tentori}\ \emph {et~al.}(2021)\citenamefont
  {Tentori}, \citenamefont {Colaitis}, \citenamefont {Theobald}, \citenamefont
  {Casner}, \citenamefont {Raffestin}, \citenamefont {Ruocco}, \citenamefont
  {Trela}, \citenamefont {{Le Bel}}, \citenamefont {Anderson}, \citenamefont
  {Wei}, \citenamefont {Henderson}, \citenamefont {Peebles}, \citenamefont
  {Scott}, \citenamefont {Baton}, \citenamefont {Pikuz}, \citenamefont {Betti},
  \citenamefont {Khan}, \citenamefont {Woolsey}, \citenamefont {Zhang},\ and\
  \citenamefont {Batani}}]{Tentori2021}%
  \BibitemOpen
  \bibfield  {author} {\bibinfo {author} {\bibfnamefont {A.}~\bibnamefont
  {Tentori}}, \bibinfo {author} {\bibfnamefont {A.}~\bibnamefont {Colaitis}},
  \bibinfo {author} {\bibfnamefont {W.}~\bibnamefont {Theobald}}, \bibinfo
  {author} {\bibfnamefont {A.}~\bibnamefont {Casner}}, \bibinfo {author}
  {\bibfnamefont {D.}~\bibnamefont {Raffestin}}, \bibinfo {author}
  {\bibfnamefont {A.}~\bibnamefont {Ruocco}}, \bibinfo {author} {\bibfnamefont
  {J.}~\bibnamefont {Trela}}, \bibinfo {author} {\bibfnamefont
  {E.}~\bibnamefont {{Le Bel}}}, \bibinfo {author} {\bibfnamefont
  {K.}~\bibnamefont {Anderson}}, \bibinfo {author} {\bibfnamefont
  {M.}~\bibnamefont {Wei}}, \bibinfo {author} {\bibfnamefont {B.}~\bibnamefont
  {Henderson}}, \bibinfo {author} {\bibfnamefont {J.}~\bibnamefont {Peebles}},
  \bibinfo {author} {\bibfnamefont {R.}~\bibnamefont {Scott}}, \bibinfo
  {author} {\bibfnamefont {S.}~\bibnamefont {Baton}}, \bibinfo {author}
  {\bibfnamefont {S.~A.}\ \bibnamefont {Pikuz}}, \bibinfo {author}
  {\bibfnamefont {R.}~\bibnamefont {Betti}}, \bibinfo {author} {\bibfnamefont
  {M.}~\bibnamefont {Khan}}, \bibinfo {author} {\bibfnamefont {N.}~\bibnamefont
  {Woolsey}}, \bibinfo {author} {\bibfnamefont {S.}~\bibnamefont {Zhang}}, \
  and\ \bibinfo {author} {\bibfnamefont {D.}~\bibnamefont {Batani}},\
  }\bibfield  {title} {\enquote {\bibinfo {title} {{Experimental
  characterization of hot-electron emission and shock dynamics in the context
  of the shock ignition approach to inertial confinement fusion}},}\ }\href
  {\doibase 10.1063/5.0059651} {\bibfield  {journal} {\bibinfo  {journal}
  {Physics of Plasmas}\ }\textbf {\bibinfo {volume} {28}},\ \bibinfo {pages}
  {103302} (\bibinfo {year} {2021})}\BibitemShut {NoStop}%
\bibitem [{\citenamefont {Abu-shawareb}(2024)}]{Abu-shawareb2024}%
  \BibitemOpen
  \bibfield  {author} {\bibinfo {author} {\bibfnamefont {H.}~\bibnamefont
  {Abu-shawareb}},\ }\bibfield  {title} {\enquote {\bibinfo {title}
  {{Achievement of Target Gain Larger than Unity in an Inertial Fusion
  Experiment}},}\ }\href {\doibase 10.1103/PhysRevLett.132.065102} {\bibfield
  {journal} {\bibinfo  {journal} {Physical Review Letters}\ }\textbf {\bibinfo
  {volume} {132}},\ \bibinfo {pages} {65102} (\bibinfo {year}
  {2024})}\BibitemShut {NoStop}%
\bibitem [{\citenamefont {Batani}\ \emph {et~al.}(2023)\citenamefont {Batani},
  \citenamefont {Cola{\"{i}}tis}, \citenamefont {Consoli}, \citenamefont
  {Danson}, \citenamefont {Gizzi}, \citenamefont {Honrubia}, \citenamefont
  {K{\"{u}}hl}, \citenamefont {{Le Pape}}, \citenamefont {Miquel},
  \citenamefont {Perlado}, \citenamefont {Scott}, \citenamefont {Tatarakis},
  \citenamefont {Tikhonchuk},\ and\ \citenamefont {Volpe}}]{Batani2023}%
  \BibitemOpen
  \bibfield  {author} {\bibinfo {author} {\bibfnamefont {D.}~\bibnamefont
  {Batani}}, \bibinfo {author} {\bibfnamefont {A.}~\bibnamefont
  {Cola{\"{i}}tis}}, \bibinfo {author} {\bibfnamefont {F.}~\bibnamefont
  {Consoli}}, \bibinfo {author} {\bibfnamefont {C.~N.}\ \bibnamefont {Danson}},
  \bibinfo {author} {\bibfnamefont {L.~A.}\ \bibnamefont {Gizzi}}, \bibinfo
  {author} {\bibfnamefont {J.}~\bibnamefont {Honrubia}}, \bibinfo {author}
  {\bibfnamefont {T.}~\bibnamefont {K{\"{u}}hl}}, \bibinfo {author}
  {\bibfnamefont {S.}~\bibnamefont {{Le Pape}}}, \bibinfo {author}
  {\bibfnamefont {J.~L.}\ \bibnamefont {Miquel}}, \bibinfo {author}
  {\bibfnamefont {J.~M.}\ \bibnamefont {Perlado}}, \bibinfo {author}
  {\bibfnamefont {R.~H.}\ \bibnamefont {Scott}}, \bibinfo {author}
  {\bibfnamefont {M.}~\bibnamefont {Tatarakis}}, \bibinfo {author}
  {\bibfnamefont {V.}~\bibnamefont {Tikhonchuk}}, \ and\ \bibinfo {author}
  {\bibfnamefont {L.}~\bibnamefont {Volpe}},\ }\bibfield  {title} {\enquote
  {\bibinfo {title} {{Future for inertial-fusion energy in Europe: A
  roadmap}},}\ }\href {\doibase 10.1017/hpl.2023.80} {\bibfield  {journal}
  {\bibinfo  {journal} {High Power Laser Science and Engineering}\ }\textbf
  {\bibinfo {volume} {11}},\ \bibinfo {pages} {e83} (\bibinfo {year}
  {2023})}\BibitemShut {NoStop}%
\bibitem [{\citenamefont {Sui}\ and\ \citenamefont {Lan}(2024)}]{Sui2024}%
  \BibitemOpen
  \bibfield  {author} {\bibinfo {author} {\bibfnamefont {Z.}~\bibnamefont
  {Sui}}\ and\ \bibinfo {author} {\bibfnamefont {K.}~\bibnamefont {Lan}},\
  }\bibfield  {title} {\enquote {\bibinfo {title} {{Driver at 10 MJ and 1
  shot/30 min for inertial confinement fusion at high gain: Efficient, compact,
  low-cost, low laser-plasma instabilities, beam color selectable from
  2$\omega$/3$\omega$/4$\omega$, applicable to multiple laser fusion
  schemes}},}\ }\href {\doibase 10.1063/5.0216435} {\bibfield  {journal}
  {\bibinfo  {journal} {Matter and Radiation at Extremes}\ }\textbf {\bibinfo
  {volume} {9}},\ \bibinfo {pages} {1--5} (\bibinfo {year} {2024})}\BibitemShut
  {NoStop}%
\bibitem [{\citenamefont {Gopalaswamy}\ \emph {et~al.}(2024)\citenamefont
  {Gopalaswamy}, \citenamefont {Williams}, \citenamefont {Betti}, \citenamefont
  {Patel}, \citenamefont {Knauer}, \citenamefont {Lees}, \citenamefont {Cao},
  \citenamefont {Campbell}, \citenamefont {Farmakis}, \citenamefont {Ejaz},
  \citenamefont {Anderson}, \citenamefont {Epstein}, \citenamefont
  {Carroll-Nellenbeck}, \citenamefont {Igumenshchev}, \citenamefont {Marozas},
  \citenamefont {Radha}, \citenamefont {Solodov}, \citenamefont {Thomas},
  \citenamefont {Woo}, \citenamefont {Collins}, \citenamefont {Hu},
  \citenamefont {Scullin}, \citenamefont {Turnbull}, \citenamefont {Goncharov},
  \citenamefont {Churnetski}, \citenamefont {Forrest}, \citenamefont {Glebov},
  \citenamefont {Heuer}, \citenamefont {McClow}, \citenamefont {Shah},
  \citenamefont {Stoeckl}, \citenamefont {Theobald}, \citenamefont {Edgell},
  \citenamefont {Ivancic}, \citenamefont {Rosenberg}, \citenamefont {Regan},
  \citenamefont {Bredesen}, \citenamefont {Fella}, \citenamefont {Koch},
  \citenamefont {Janezic}, \citenamefont {Bonino}, \citenamefont {Harding},
  \citenamefont {Bauer}, \citenamefont {Sampat}, \citenamefont {Waxer},
  \citenamefont {Labuzeta}, \citenamefont {Morse}, \citenamefont
  {Gatu-Johnson}, \citenamefont {Petrasso}, \citenamefont {Frenje},
  \citenamefont {Murray}, \citenamefont {Serrato}, \citenamefont {Guzman},
  \citenamefont {Shuldberg}, \citenamefont {Farrell},\ and\ \citenamefont
  {Deeney}}]{Gopalaswamy2024}%
  \BibitemOpen
  \bibfield  {author} {\bibinfo {author} {\bibfnamefont {V.}~\bibnamefont
  {Gopalaswamy}}, \bibinfo {author} {\bibfnamefont {C.~A.}\ \bibnamefont
  {Williams}}, \bibinfo {author} {\bibfnamefont {R.}~\bibnamefont {Betti}},
  \bibinfo {author} {\bibfnamefont {D.}~\bibnamefont {Patel}}, \bibinfo
  {author} {\bibfnamefont {J.~P.}\ \bibnamefont {Knauer}}, \bibinfo {author}
  {\bibfnamefont {A.}~\bibnamefont {Lees}}, \bibinfo {author} {\bibfnamefont
  {D.}~\bibnamefont {Cao}}, \bibinfo {author} {\bibfnamefont {E.~M.}\
  \bibnamefont {Campbell}}, \bibinfo {author} {\bibfnamefont {P.}~\bibnamefont
  {Farmakis}}, \bibinfo {author} {\bibfnamefont {R.}~\bibnamefont {Ejaz}},
  \bibinfo {author} {\bibfnamefont {K.~S.}\ \bibnamefont {Anderson}}, \bibinfo
  {author} {\bibfnamefont {R.}~\bibnamefont {Epstein}}, \bibinfo {author}
  {\bibfnamefont {J.}~\bibnamefont {Carroll-Nellenbeck}}, \bibinfo {author}
  {\bibfnamefont {I.~V.}\ \bibnamefont {Igumenshchev}}, \bibinfo {author}
  {\bibfnamefont {J.~A.}\ \bibnamefont {Marozas}}, \bibinfo {author}
  {\bibfnamefont {P.~B.}\ \bibnamefont {Radha}}, \bibinfo {author}
  {\bibfnamefont {A.~A.}\ \bibnamefont {Solodov}}, \bibinfo {author}
  {\bibfnamefont {C.~A.}\ \bibnamefont {Thomas}}, \bibinfo {author}
  {\bibfnamefont {K.~M.}\ \bibnamefont {Woo}}, \bibinfo {author} {\bibfnamefont
  {T.~J.}\ \bibnamefont {Collins}}, \bibinfo {author} {\bibfnamefont {S.~X.}\
  \bibnamefont {Hu}}, \bibinfo {author} {\bibfnamefont {W.}~\bibnamefont
  {Scullin}}, \bibinfo {author} {\bibfnamefont {D.}~\bibnamefont {Turnbull}},
  \bibinfo {author} {\bibfnamefont {V.~N.}\ \bibnamefont {Goncharov}}, \bibinfo
  {author} {\bibfnamefont {K.}~\bibnamefont {Churnetski}}, \bibinfo {author}
  {\bibfnamefont {C.~J.}\ \bibnamefont {Forrest}}, \bibinfo {author}
  {\bibfnamefont {V.~Y.}\ \bibnamefont {Glebov}}, \bibinfo {author}
  {\bibfnamefont {P.~V.}\ \bibnamefont {Heuer}}, \bibinfo {author}
  {\bibfnamefont {H.}~\bibnamefont {McClow}}, \bibinfo {author} {\bibfnamefont
  {R.~C.}\ \bibnamefont {Shah}}, \bibinfo {author} {\bibfnamefont
  {C.}~\bibnamefont {Stoeckl}}, \bibinfo {author} {\bibfnamefont
  {W.}~\bibnamefont {Theobald}}, \bibinfo {author} {\bibfnamefont {D.~H.}\
  \bibnamefont {Edgell}}, \bibinfo {author} {\bibfnamefont {S.}~\bibnamefont
  {Ivancic}}, \bibinfo {author} {\bibfnamefont {M.~J.}\ \bibnamefont
  {Rosenberg}}, \bibinfo {author} {\bibfnamefont {S.~P.}\ \bibnamefont
  {Regan}}, \bibinfo {author} {\bibfnamefont {D.}~\bibnamefont {Bredesen}},
  \bibinfo {author} {\bibfnamefont {C.}~\bibnamefont {Fella}}, \bibinfo
  {author} {\bibfnamefont {M.}~\bibnamefont {Koch}}, \bibinfo {author}
  {\bibfnamefont {R.~T.}\ \bibnamefont {Janezic}}, \bibinfo {author}
  {\bibfnamefont {M.~J.}\ \bibnamefont {Bonino}}, \bibinfo {author}
  {\bibfnamefont {D.~R.}\ \bibnamefont {Harding}}, \bibinfo {author}
  {\bibfnamefont {K.~A.}\ \bibnamefont {Bauer}}, \bibinfo {author}
  {\bibfnamefont {S.}~\bibnamefont {Sampat}}, \bibinfo {author} {\bibfnamefont
  {L.~J.}\ \bibnamefont {Waxer}}, \bibinfo {author} {\bibfnamefont
  {M.}~\bibnamefont {Labuzeta}}, \bibinfo {author} {\bibfnamefont {S.~F.}\
  \bibnamefont {Morse}}, \bibinfo {author} {\bibfnamefont {M.}~\bibnamefont
  {Gatu-Johnson}}, \bibinfo {author} {\bibfnamefont {R.~D.}\ \bibnamefont
  {Petrasso}}, \bibinfo {author} {\bibfnamefont {J.~A.}\ \bibnamefont
  {Frenje}}, \bibinfo {author} {\bibfnamefont {J.}~\bibnamefont {Murray}},
  \bibinfo {author} {\bibfnamefont {B.}~\bibnamefont {Serrato}}, \bibinfo
  {author} {\bibfnamefont {D.}~\bibnamefont {Guzman}}, \bibinfo {author}
  {\bibfnamefont {C.}~\bibnamefont {Shuldberg}}, \bibinfo {author}
  {\bibfnamefont {M.}~\bibnamefont {Farrell}}, \ and\ \bibinfo {author}
  {\bibfnamefont {C.}~\bibnamefont {Deeney}},\ }\bibfield  {title} {\enquote
  {\bibinfo {title} {{Demonstration of a hydrodynamically equivalent burning
  plasma in direct-drive inertial confinement fusion}},}\ }\href {\doibase
  10.1038/s41567-023-02361-4} {\bibfield  {journal} {\bibinfo  {journal}
  {Nature Physics}\ }\textbf {\bibinfo {volume} {20}},\ \bibinfo {pages}
  {751--757} (\bibinfo {year} {2024})}\BibitemShut {NoStop}%
\bibitem [{\citenamefont {Williams}\ \emph {et~al.}(2024)\citenamefont
  {Williams}, \citenamefont {Betti}, \citenamefont {Gopalaswamy}, \citenamefont
  {Knauer}, \citenamefont {Forrest}, \citenamefont {Lees}, \citenamefont
  {Ejaz}, \citenamefont {Farmakis}, \citenamefont {Cao}, \citenamefont {Radha},
  \citenamefont {Anderson}, \citenamefont {Regan}, \citenamefont {Glebov},
  \citenamefont {Shah}, \citenamefont {Stoeckl}, \citenamefont {Ivancic},
  \citenamefont {Churnetski}, \citenamefont {Janezic}, \citenamefont {Fella},
  \citenamefont {Rosenberg}, \citenamefont {Bonino}, \citenamefont {Harding},
  \citenamefont {Shmayda}, \citenamefont {Carroll-Nellenback}, \citenamefont
  {Hu}, \citenamefont {Epstein}, \citenamefont {Collins}, \citenamefont
  {Thomas}, \citenamefont {Igumenshchev}, \citenamefont {Goncharov},
  \citenamefont {Theobald}, \citenamefont {Woo}, \citenamefont {Marozas},
  \citenamefont {Bauer}, \citenamefont {Sampat}, \citenamefont {Waxer},
  \citenamefont {Turnbull}, \citenamefont {Heuer}, \citenamefont {McClow},
  \citenamefont {Ceurvorst}, \citenamefont {Scullin}, \citenamefont {Edgell},
  \citenamefont {Koch}, \citenamefont {Bredesen}, \citenamefont {Johnson},
  \citenamefont {Frenje}, \citenamefont {Petrasso}, \citenamefont {Shuldberg},
  \citenamefont {Farrell}, \citenamefont {Murray}, \citenamefont {Guzman},
  \citenamefont {Serrato}, \citenamefont {Morse}, \citenamefont {Labuzeta},
  \citenamefont {Deeney},\ and\ \citenamefont {Campbell}}]{Williams2024}%
  \BibitemOpen
  \bibfield  {author} {\bibinfo {author} {\bibfnamefont {C.~A.}\ \bibnamefont
  {Williams}}, \bibinfo {author} {\bibfnamefont {R.}~\bibnamefont {Betti}},
  \bibinfo {author} {\bibfnamefont {V.}~\bibnamefont {Gopalaswamy}}, \bibinfo
  {author} {\bibfnamefont {J.~P.}\ \bibnamefont {Knauer}}, \bibinfo {author}
  {\bibfnamefont {C.~J.}\ \bibnamefont {Forrest}}, \bibinfo {author}
  {\bibfnamefont {A.}~\bibnamefont {Lees}}, \bibinfo {author} {\bibfnamefont
  {R.}~\bibnamefont {Ejaz}}, \bibinfo {author} {\bibfnamefont {P.~S.}\
  \bibnamefont {Farmakis}}, \bibinfo {author} {\bibfnamefont {D.}~\bibnamefont
  {Cao}}, \bibinfo {author} {\bibfnamefont {P.~B.}\ \bibnamefont {Radha}},
  \bibinfo {author} {\bibfnamefont {K.~S.}\ \bibnamefont {Anderson}}, \bibinfo
  {author} {\bibfnamefont {S.~P.}\ \bibnamefont {Regan}}, \bibinfo {author}
  {\bibfnamefont {V.~Y.}\ \bibnamefont {Glebov}}, \bibinfo {author}
  {\bibfnamefont {R.~C.}\ \bibnamefont {Shah}}, \bibinfo {author}
  {\bibfnamefont {C.}~\bibnamefont {Stoeckl}}, \bibinfo {author} {\bibfnamefont
  {S.}~\bibnamefont {Ivancic}}, \bibinfo {author} {\bibfnamefont
  {K.}~\bibnamefont {Churnetski}}, \bibinfo {author} {\bibfnamefont {R.~T.}\
  \bibnamefont {Janezic}}, \bibinfo {author} {\bibfnamefont {C.}~\bibnamefont
  {Fella}}, \bibinfo {author} {\bibfnamefont {M.~J.}\ \bibnamefont
  {Rosenberg}}, \bibinfo {author} {\bibfnamefont {M.~J.}\ \bibnamefont
  {Bonino}}, \bibinfo {author} {\bibfnamefont {D.~R.}\ \bibnamefont {Harding}},
  \bibinfo {author} {\bibfnamefont {W.~T.}\ \bibnamefont {Shmayda}}, \bibinfo
  {author} {\bibfnamefont {J.}~\bibnamefont {Carroll-Nellenback}}, \bibinfo
  {author} {\bibfnamefont {S.~X.}\ \bibnamefont {Hu}}, \bibinfo {author}
  {\bibfnamefont {R.}~\bibnamefont {Epstein}}, \bibinfo {author} {\bibfnamefont
  {T.~J.}\ \bibnamefont {Collins}}, \bibinfo {author} {\bibfnamefont {C.~A.}\
  \bibnamefont {Thomas}}, \bibinfo {author} {\bibfnamefont {I.~V.}\
  \bibnamefont {Igumenshchev}}, \bibinfo {author} {\bibfnamefont {V.~N.}\
  \bibnamefont {Goncharov}}, \bibinfo {author} {\bibfnamefont {W.}~\bibnamefont
  {Theobald}}, \bibinfo {author} {\bibfnamefont {K.~M.}\ \bibnamefont {Woo}},
  \bibinfo {author} {\bibfnamefont {J.~A.}\ \bibnamefont {Marozas}}, \bibinfo
  {author} {\bibfnamefont {K.~A.}\ \bibnamefont {Bauer}}, \bibinfo {author}
  {\bibfnamefont {S.}~\bibnamefont {Sampat}}, \bibinfo {author} {\bibfnamefont
  {L.~J.}\ \bibnamefont {Waxer}}, \bibinfo {author} {\bibfnamefont
  {D.}~\bibnamefont {Turnbull}}, \bibinfo {author} {\bibfnamefont {P.~V.}\
  \bibnamefont {Heuer}}, \bibinfo {author} {\bibfnamefont {H.}~\bibnamefont
  {McClow}}, \bibinfo {author} {\bibfnamefont {L.}~\bibnamefont {Ceurvorst}},
  \bibinfo {author} {\bibfnamefont {W.}~\bibnamefont {Scullin}}, \bibinfo
  {author} {\bibfnamefont {D.~H.}\ \bibnamefont {Edgell}}, \bibinfo {author}
  {\bibfnamefont {M.}~\bibnamefont {Koch}}, \bibinfo {author} {\bibfnamefont
  {D.}~\bibnamefont {Bredesen}}, \bibinfo {author} {\bibfnamefont {M.~G.}\
  \bibnamefont {Johnson}}, \bibinfo {author} {\bibfnamefont {J.~A.}\
  \bibnamefont {Frenje}}, \bibinfo {author} {\bibfnamefont {R.~D.}\
  \bibnamefont {Petrasso}}, \bibinfo {author} {\bibfnamefont {C.}~\bibnamefont
  {Shuldberg}}, \bibinfo {author} {\bibfnamefont {M.}~\bibnamefont {Farrell}},
  \bibinfo {author} {\bibfnamefont {J.}~\bibnamefont {Murray}}, \bibinfo
  {author} {\bibfnamefont {D.}~\bibnamefont {Guzman}}, \bibinfo {author}
  {\bibfnamefont {B.}~\bibnamefont {Serrato}}, \bibinfo {author} {\bibfnamefont
  {S.~F.}\ \bibnamefont {Morse}}, \bibinfo {author} {\bibfnamefont
  {M.}~\bibnamefont {Labuzeta}}, \bibinfo {author} {\bibfnamefont
  {C.}~\bibnamefont {Deeney}}, \ and\ \bibinfo {author} {\bibfnamefont {E.~M.}\
  \bibnamefont {Campbell}},\ }\bibfield  {title} {\enquote {\bibinfo {title}
  {{Demonstration of hot-spot fuel gain exceeding unity in direct-drive
  inertial confinement fusion implosions}},}\ }\href {\doibase
  10.1038/s41567-023-02363-2} {\bibfield  {journal} {\bibinfo  {journal}
  {Nature Physics}\ }\textbf {\bibinfo {volume} {20}},\ \bibinfo {pages}
  {758--764} (\bibinfo {year} {2024})}\BibitemShut {NoStop}%
\bibitem [{\citenamefont {Michel}\ \emph {et~al.}(2013)\citenamefont {Michel},
  \citenamefont {Maximov}, \citenamefont {Short}, \citenamefont {Delettrez},
  \citenamefont {Edgell}, \citenamefont {Hu}, \citenamefont {Igumenshchev},
  \citenamefont {Myatt}, \citenamefont {Solodov}, \citenamefont {Stoeckl},
  \citenamefont {Yaakobi},\ and\ \citenamefont {Froula}}]{Michel2013}%
  \BibitemOpen
  \bibfield  {author} {\bibinfo {author} {\bibfnamefont {D.~T.}\ \bibnamefont
  {Michel}}, \bibinfo {author} {\bibfnamefont {A.~V.}\ \bibnamefont {Maximov}},
  \bibinfo {author} {\bibfnamefont {R.~W.}\ \bibnamefont {Short}}, \bibinfo
  {author} {\bibfnamefont {J.~A.}\ \bibnamefont {Delettrez}}, \bibinfo {author}
  {\bibfnamefont {D.}~\bibnamefont {Edgell}}, \bibinfo {author} {\bibfnamefont
  {S.~X.}\ \bibnamefont {Hu}}, \bibinfo {author} {\bibfnamefont {I.~V.}\
  \bibnamefont {Igumenshchev}}, \bibinfo {author} {\bibfnamefont {J.~F.}\
  \bibnamefont {Myatt}}, \bibinfo {author} {\bibfnamefont {A.~A.}\ \bibnamefont
  {Solodov}}, \bibinfo {author} {\bibfnamefont {C.}~\bibnamefont {Stoeckl}},
  \bibinfo {author} {\bibfnamefont {B.}~\bibnamefont {Yaakobi}}, \ and\
  \bibinfo {author} {\bibfnamefont {D.~H.}\ \bibnamefont {Froula}},\ }\bibfield
   {title} {\enquote {\bibinfo {title} {{Measured hot-electron intensity
  thresholds quantified by a two-plasmon-decay resonant common-wave gain in
  various experimental configurations}},}\ }\href {\doibase 10.1063/1.4803090}
  {\bibfield  {journal} {\bibinfo  {journal} {Physics of Plasmas}\ }\textbf
  {\bibinfo {volume} {20}},\ \bibinfo {pages} {55703} (\bibinfo {year}
  {2013})}\BibitemShut {NoStop}%
\bibitem [{\citenamefont {Froula}\ \emph {et~al.}(2012)\citenamefont {Froula},
  \citenamefont {Yaakobi}, \citenamefont {Hu}, \citenamefont {Chang},
  \citenamefont {Craxton}, \citenamefont {Edgell}, \citenamefont {Follett},
  \citenamefont {Michel}, \citenamefont {Myatt}, \citenamefont {Seka},
  \citenamefont {Short}, \citenamefont {Solodov},\ and\ \citenamefont
  {Stoeckl}}]{Froula2012}%
  \BibitemOpen
  \bibfield  {author} {\bibinfo {author} {\bibfnamefont {D.~H.}\ \bibnamefont
  {Froula}}, \bibinfo {author} {\bibfnamefont {B.}~\bibnamefont {Yaakobi}},
  \bibinfo {author} {\bibfnamefont {S.~X.}\ \bibnamefont {Hu}}, \bibinfo
  {author} {\bibfnamefont {P.-Y.}\ \bibnamefont {Chang}}, \bibinfo {author}
  {\bibfnamefont {R.~S.}\ \bibnamefont {Craxton}}, \bibinfo {author}
  {\bibfnamefont {D.~H.}\ \bibnamefont {Edgell}}, \bibinfo {author}
  {\bibfnamefont {R.}~\bibnamefont {Follett}}, \bibinfo {author} {\bibfnamefont
  {D.~T.}\ \bibnamefont {Michel}}, \bibinfo {author} {\bibfnamefont {J.~F.}\
  \bibnamefont {Myatt}}, \bibinfo {author} {\bibfnamefont {W.}~\bibnamefont
  {Seka}}, \bibinfo {author} {\bibfnamefont {R.~W.}\ \bibnamefont {Short}},
  \bibinfo {author} {\bibfnamefont {A.}~\bibnamefont {Solodov}}, \ and\
  \bibinfo {author} {\bibfnamefont {C.}~\bibnamefont {Stoeckl}},\ }\bibfield
  {title} {\enquote {\bibinfo {title} {Saturation of the two-plasmon decay
  instability in long-scale-length plasmas relevant to direct-drive inertial
  confinement fusion},}\ }\href@noop {} {\bibfield  {journal} {\bibinfo
  {journal} {Physical Review Letters}\ }\textbf {\bibinfo {volume} {108}},\
  \bibinfo {pages} {165003} (\bibinfo {year} {2012})}\BibitemShut {NoStop}%
\bibitem [{\citenamefont {Seka}\ \emph {et~al.}(2009)\citenamefont {Seka},
  \citenamefont {Edgell}, \citenamefont {Myatt}, \citenamefont {Maximov},
  \citenamefont {Short}, \citenamefont {Goncharov},\ and\ \citenamefont
  {Baldis}}]{Seka2009a}%
  \BibitemOpen
  \bibfield  {author} {\bibinfo {author} {\bibfnamefont {W.}~\bibnamefont
  {Seka}}, \bibinfo {author} {\bibfnamefont {D.~H.}\ \bibnamefont {Edgell}},
  \bibinfo {author} {\bibfnamefont {J.~F.}\ \bibnamefont {Myatt}}, \bibinfo
  {author} {\bibfnamefont {A.~V.}\ \bibnamefont {Maximov}}, \bibinfo {author}
  {\bibfnamefont {R.~W.}\ \bibnamefont {Short}}, \bibinfo {author}
  {\bibfnamefont {V.~N.}\ \bibnamefont {Goncharov}}, \ and\ \bibinfo {author}
  {\bibfnamefont {H.~A.}\ \bibnamefont {Baldis}},\ }\bibfield  {title}
  {\enquote {\bibinfo {title} {{Two-plasmon-decay instability in direct-drive
  inertial confinement fusion experiments}},}\ }\href {\doibase
  10.1063/1.3125242} {\bibfield  {journal} {\bibinfo  {journal} {Physics of
  Plasmas}\ }\textbf {\bibinfo {volume} {16}},\ \bibinfo {pages} {052701}
  (\bibinfo {year} {2009})}\BibitemShut {NoStop}%
\bibitem [{\citenamefont {Turnbull}\ \emph
  {et~al.}(2020{\natexlab{a}})\citenamefont {Turnbull}, \citenamefont
  {Maximov}, \citenamefont {Cao}, \citenamefont {Christopherson}, \citenamefont
  {Edgell}, \citenamefont {Follett}, \citenamefont {Gopalaswamy}, \citenamefont
  {Knauer}, \citenamefont {Palastro}, \citenamefont {Shvydky}, \citenamefont
  {Stoeckl}, \citenamefont {Wen},\ and\ \citenamefont
  {Froula}}]{Turnbull2020b}%
  \BibitemOpen
  \bibfield  {author} {\bibinfo {author} {\bibfnamefont {D.}~\bibnamefont
  {Turnbull}}, \bibinfo {author} {\bibfnamefont {A.~V.}\ \bibnamefont
  {Maximov}}, \bibinfo {author} {\bibfnamefont {D.}~\bibnamefont {Cao}},
  \bibinfo {author} {\bibfnamefont {A.~R.}\ \bibnamefont {Christopherson}},
  \bibinfo {author} {\bibfnamefont {D.~H.}\ \bibnamefont {Edgell}}, \bibinfo
  {author} {\bibfnamefont {R.~K.}\ \bibnamefont {Follett}}, \bibinfo {author}
  {\bibfnamefont {V.}~\bibnamefont {Gopalaswamy}}, \bibinfo {author}
  {\bibfnamefont {J.~P.}\ \bibnamefont {Knauer}}, \bibinfo {author}
  {\bibfnamefont {J.~P.}\ \bibnamefont {Palastro}}, \bibinfo {author}
  {\bibfnamefont {A.}~\bibnamefont {Shvydky}}, \bibinfo {author} {\bibfnamefont
  {C.}~\bibnamefont {Stoeckl}}, \bibinfo {author} {\bibfnamefont
  {H.}~\bibnamefont {Wen}}, \ and\ \bibinfo {author} {\bibfnamefont {D.~H.}\
  \bibnamefont {Froula}},\ }\bibfield  {title} {\enquote {\bibinfo {title}
  {{Impact of spatiotemporal smoothing on the two-plasmon-decay
  instability}},}\ }\href {\doibase 10.1063/5.0019080} {\bibfield  {journal}
  {\bibinfo  {journal} {Physics of Plasmas}\ }\textbf {\bibinfo {volume}
  {27}},\ \bibinfo {pages} {102710} (\bibinfo {year}
  {2020}{\natexlab{a}})}\BibitemShut {NoStop}%
\bibitem [{\citenamefont {Regan}\ \emph {et~al.}(2010)\citenamefont {Regan},
  \citenamefont {Meezan}, \citenamefont {Suter}, \citenamefont {Strozzi},
  \citenamefont {Kruer}, \citenamefont {Meeker}, \citenamefont {Glenzer},
  \citenamefont {Seka}, \citenamefont {Stoeckl}, \citenamefont {Glebov},
  \citenamefont {Sangster}, \citenamefont {Meyerhofer}, \citenamefont
  {McCrory}, \citenamefont {Williams}, \citenamefont {Jones}, \citenamefont
  {Callahan}, \citenamefont {Rosen}, \citenamefont {Landen}, \citenamefont
  {Sorce},\ and\ \citenamefont {MacGowan}}]{Regan2010}%
  \BibitemOpen
  \bibfield  {author} {\bibinfo {author} {\bibfnamefont {S.~P.}\ \bibnamefont
  {Regan}}, \bibinfo {author} {\bibfnamefont {N.~B.}\ \bibnamefont {Meezan}},
  \bibinfo {author} {\bibfnamefont {L.~J.}\ \bibnamefont {Suter}}, \bibinfo
  {author} {\bibfnamefont {D.~J.}\ \bibnamefont {Strozzi}}, \bibinfo {author}
  {\bibfnamefont {W.~L.}\ \bibnamefont {Kruer}}, \bibinfo {author}
  {\bibfnamefont {D.}~\bibnamefont {Meeker}}, \bibinfo {author} {\bibfnamefont
  {S.~H.}\ \bibnamefont {Glenzer}}, \bibinfo {author} {\bibfnamefont
  {W.}~\bibnamefont {Seka}}, \bibinfo {author} {\bibfnamefont {C.}~\bibnamefont
  {Stoeckl}}, \bibinfo {author} {\bibfnamefont {V.~Y.}\ \bibnamefont {Glebov}},
  \bibinfo {author} {\bibfnamefont {T.~C.}\ \bibnamefont {Sangster}}, \bibinfo
  {author} {\bibfnamefont {D.~D.}\ \bibnamefont {Meyerhofer}}, \bibinfo
  {author} {\bibfnamefont {R.~L.}\ \bibnamefont {McCrory}}, \bibinfo {author}
  {\bibfnamefont {E.~A.}\ \bibnamefont {Williams}}, \bibinfo {author}
  {\bibfnamefont {O.~S.}\ \bibnamefont {Jones}}, \bibinfo {author}
  {\bibfnamefont {D.~A.}\ \bibnamefont {Callahan}}, \bibinfo {author}
  {\bibfnamefont {M.~D.}\ \bibnamefont {Rosen}}, \bibinfo {author}
  {\bibfnamefont {O.~L.}\ \bibnamefont {Landen}}, \bibinfo {author}
  {\bibfnamefont {C.}~\bibnamefont {Sorce}}, \ and\ \bibinfo {author}
  {\bibfnamefont {B.~J.}\ \bibnamefont {MacGowan}},\ }\bibfield  {title}
  {\enquote {\bibinfo {title} {{Suprathermal electrons generated by the
  two-plasmon-decay instability in gas-filled Hohlraums}},}\ }\href {\doibase
  10.1063/1.3309481} {\bibfield  {journal} {\bibinfo  {journal} {Physics of
  Plasmas}\ }\textbf {\bibinfo {volume} {17}},\ \bibinfo {pages} {020703}
  (\bibinfo {year} {2010})}\BibitemShut {NoStop}%
\bibitem [{\citenamefont {Cao}\ \emph {et~al.}(2022)\citenamefont {Cao},
  \citenamefont {Patel}, \citenamefont {Lees}, \citenamefont {Stoeckl},
  \citenamefont {Rosenberg}, \citenamefont {Gopalaswamy}, \citenamefont {Wen},
  \citenamefont {Huang}, \citenamefont {Shvydky}, \citenamefont {Betti},\ and\
  \citenamefont {Ren}}]{Cao2022}%
  \BibitemOpen
  \bibfield  {author} {\bibinfo {author} {\bibfnamefont {S.~H.}\ \bibnamefont
  {Cao}}, \bibinfo {author} {\bibfnamefont {D.}~\bibnamefont {Patel}}, \bibinfo
  {author} {\bibfnamefont {A.}~\bibnamefont {Lees}}, \bibinfo {author}
  {\bibfnamefont {C.}~\bibnamefont {Stoeckl}}, \bibinfo {author} {\bibfnamefont
  {M.~J.}\ \bibnamefont {Rosenberg}}, \bibinfo {author} {\bibfnamefont
  {V.}~\bibnamefont {Gopalaswamy}}, \bibinfo {author} {\bibfnamefont
  {H.}~\bibnamefont {Wen}}, \bibinfo {author} {\bibfnamefont {H.}~\bibnamefont
  {Huang}}, \bibinfo {author} {\bibfnamefont {A.}~\bibnamefont {Shvydky}},
  \bibinfo {author} {\bibfnamefont {R.}~\bibnamefont {Betti}}, \ and\ \bibinfo
  {author} {\bibfnamefont {C.}~\bibnamefont {Ren}},\ }\bibfield  {title}
  {\enquote {\bibinfo {title} {{Predicting hot electron generation in inertial
  confinement fusion with particle-in-cell simulations}},}\ }\href {\doibase
  10.1103/PhysRevE.106.055214} {\bibfield  {journal} {\bibinfo  {journal}
  {Physical Review E}\ }\textbf {\bibinfo {volume} {106}},\ \bibinfo {pages}
  {1--6} (\bibinfo {year} {2022})}\BibitemShut {NoStop}%
\bibitem [{\citenamefont {Cao}\ and\ \citenamefont {Ren}(2023)}]{Cao2023}%
  \BibitemOpen
  \bibfield  {author} {\bibinfo {author} {\bibfnamefont {S.~H.}\ \bibnamefont
  {Cao}}\ and\ \bibinfo {author} {\bibfnamefont {C.}~\bibnamefont {Ren}},\
  }\bibfield  {title} {\enquote {\bibinfo {title} {{Evolution and hot electron
  generation of laser-plasma instabilities in direct-drive inertial confinement
  fusion}},}\ }\href {\doibase 10.1063/5.0161865} {\bibfield  {journal}
  {\bibinfo  {journal} {Physics of Plasmas}\ }\textbf {\bibinfo {volume}
  {30}},\ \bibinfo {pages} {092701} (\bibinfo {year} {2023})}\BibitemShut
  {NoStop}%
\bibitem [{\citenamefont {Rosenberg}\ \emph {et~al.}(2018)\citenamefont
  {Rosenberg}, \citenamefont {Solodov}, \citenamefont {Myatt}, \citenamefont
  {Seka}, \citenamefont {Michel}, \citenamefont {Hohenberger}, \citenamefont
  {Short}, \citenamefont {Epstein}, \citenamefont {Regan}, \citenamefont
  {Campbell}, \citenamefont {Chapman}, \citenamefont {Goyon}, \citenamefont
  {Ralph}, \citenamefont {Barrios}, \citenamefont {Moody},\ and\ \citenamefont
  {Bates}}]{Rosenberg2018}%
  \BibitemOpen
  \bibfield  {author} {\bibinfo {author} {\bibfnamefont {M.~J.}\ \bibnamefont
  {Rosenberg}}, \bibinfo {author} {\bibfnamefont {A.~A.}\ \bibnamefont
  {Solodov}}, \bibinfo {author} {\bibfnamefont {J.~F.}\ \bibnamefont {Myatt}},
  \bibinfo {author} {\bibfnamefont {W.}~\bibnamefont {Seka}}, \bibinfo {author}
  {\bibfnamefont {P.}~\bibnamefont {Michel}}, \bibinfo {author} {\bibfnamefont
  {M.}~\bibnamefont {Hohenberger}}, \bibinfo {author} {\bibfnamefont {R.~W.}\
  \bibnamefont {Short}}, \bibinfo {author} {\bibfnamefont {R.}~\bibnamefont
  {Epstein}}, \bibinfo {author} {\bibfnamefont {S.~P.}\ \bibnamefont {Regan}},
  \bibinfo {author} {\bibfnamefont {E.~M.}\ \bibnamefont {Campbell}}, \bibinfo
  {author} {\bibfnamefont {T.}~\bibnamefont {Chapman}}, \bibinfo {author}
  {\bibfnamefont {C.}~\bibnamefont {Goyon}}, \bibinfo {author} {\bibfnamefont
  {J.~E.}\ \bibnamefont {Ralph}}, \bibinfo {author} {\bibfnamefont {M.~A.}\
  \bibnamefont {Barrios}}, \bibinfo {author} {\bibfnamefont {J.~D.}\
  \bibnamefont {Moody}}, \ and\ \bibinfo {author} {\bibfnamefont {J.~W.}\
  \bibnamefont {Bates}},\ }\bibfield  {title} {\enquote {\bibinfo {title}
  {{Origins and Scaling of Hot-Electron Preheat in Ignition-Scale Direct-Drive
  Inertial Confinement Fusion Experiments}},}\ }\href {\doibase
  10.1103/PhysRevLett.120.055001} {\bibfield  {journal} {\bibinfo  {journal}
  {Physical Review Letters}\ }\textbf {\bibinfo {volume} {120}},\ \bibinfo
  {pages} {055001} (\bibinfo {year} {2018})}\BibitemShut {NoStop}%
\bibitem [{\citenamefont {Dewald}\ \emph {et~al.}(2016)\citenamefont {Dewald},
  \citenamefont {Hartemann}, \citenamefont {Michel}, \citenamefont {Milovich},
  \citenamefont {Hohenberger}, \citenamefont {Pak}, \citenamefont {Landen},
  \citenamefont {Divol}, \citenamefont {Robey}, \citenamefont {Hurricane},
  \citenamefont {D{\"{o}}ppner}, \citenamefont {Albert}, \citenamefont
  {Bachmann}, \citenamefont {Meezan}, \citenamefont {Mackinnon}, \citenamefont
  {Callahan},\ and\ \citenamefont {Edwards}}]{Dewald2016}%
  \BibitemOpen
  \bibfield  {author} {\bibinfo {author} {\bibfnamefont {E.~L.}\ \bibnamefont
  {Dewald}}, \bibinfo {author} {\bibfnamefont {F.}~\bibnamefont {Hartemann}},
  \bibinfo {author} {\bibfnamefont {P.}~\bibnamefont {Michel}}, \bibinfo
  {author} {\bibfnamefont {J.}~\bibnamefont {Milovich}}, \bibinfo {author}
  {\bibfnamefont {M.}~\bibnamefont {Hohenberger}}, \bibinfo {author}
  {\bibfnamefont {A.}~\bibnamefont {Pak}}, \bibinfo {author} {\bibfnamefont
  {O.~L.}\ \bibnamefont {Landen}}, \bibinfo {author} {\bibfnamefont
  {L.}~\bibnamefont {Divol}}, \bibinfo {author} {\bibfnamefont {H.~F.}\
  \bibnamefont {Robey}}, \bibinfo {author} {\bibfnamefont {O.~A.}\ \bibnamefont
  {Hurricane}}, \bibinfo {author} {\bibfnamefont {T.}~\bibnamefont
  {D{\"{o}}ppner}}, \bibinfo {author} {\bibfnamefont {F.}~\bibnamefont
  {Albert}}, \bibinfo {author} {\bibfnamefont {B.}~\bibnamefont {Bachmann}},
  \bibinfo {author} {\bibfnamefont {N.~B.}\ \bibnamefont {Meezan}}, \bibinfo
  {author} {\bibfnamefont {A.~J.}\ \bibnamefont {Mackinnon}}, \bibinfo {author}
  {\bibfnamefont {D.}~\bibnamefont {Callahan}}, \ and\ \bibinfo {author}
  {\bibfnamefont {M.~J.}\ \bibnamefont {Edwards}},\ }\bibfield  {title}
  {\enquote {\bibinfo {title} {{Generation and Beaming of Early Hot Electrons
  onto the Capsule in Laser-Driven Ignition Hohlraums}},}\ }\href {\doibase
  10.1103/PhysRevLett.116.075003} {\bibfield  {journal} {\bibinfo  {journal}
  {Physical Review Letters}\ }\textbf {\bibinfo {volume} {116}},\ \bibinfo
  {pages} {075003} (\bibinfo {year} {2016})}\BibitemShut {NoStop}%
\bibitem [{\citenamefont {Turnbull}\ \emph
  {et~al.}(2020{\natexlab{b}})\citenamefont {Turnbull}, \citenamefont
  {Maximov}, \citenamefont {Edgell}, \citenamefont {Seka}, \citenamefont
  {Follett}, \citenamefont {Palastro}, \citenamefont {Cao}, \citenamefont
  {Goncharov}, \citenamefont {Stoeckl},\ and\ \citenamefont
  {Froula}}]{turnbull2020}%
  \BibitemOpen
  \bibfield  {author} {\bibinfo {author} {\bibfnamefont {D.}~\bibnamefont
  {Turnbull}}, \bibinfo {author} {\bibfnamefont {A.~V.}\ \bibnamefont
  {Maximov}}, \bibinfo {author} {\bibfnamefont {D.~H.}\ \bibnamefont {Edgell}},
  \bibinfo {author} {\bibfnamefont {W.}~\bibnamefont {Seka}}, \bibinfo {author}
  {\bibfnamefont {R.~K.}\ \bibnamefont {Follett}}, \bibinfo {author}
  {\bibfnamefont {J.~P.}\ \bibnamefont {Palastro}}, \bibinfo {author}
  {\bibfnamefont {D.}~\bibnamefont {Cao}}, \bibinfo {author} {\bibfnamefont
  {V.~N.}\ \bibnamefont {Goncharov}}, \bibinfo {author} {\bibfnamefont
  {C.}~\bibnamefont {Stoeckl}}, \ and\ \bibinfo {author} {\bibfnamefont
  {D.~H.}\ \bibnamefont {Froula}},\ }\bibfield  {title} {\enquote {\bibinfo
  {title} {Anomalous {{Absorption}} by the {{Two-Plasmon Decay
  Instability}}},}\ }\href@noop {} {\bibfield  {journal} {\bibinfo  {journal}
  {Physical Review Letters}\ }\textbf {\bibinfo {volume} {124}},\ \bibinfo
  {pages} {185001} (\bibinfo {year} {2020}{\natexlab{b}})}\BibitemShut
  {NoStop}%
\bibitem [{\citenamefont {Wen}\ \emph {et~al.}(2015)\citenamefont {Wen},
  \citenamefont {Yan}, \citenamefont {Maximov},\ and\ \citenamefont
  {Ren}}]{Wen2015}%
  \BibitemOpen
  \bibfield  {author} {\bibinfo {author} {\bibfnamefont {H.}~\bibnamefont
  {Wen}}, \bibinfo {author} {\bibfnamefont {R.}~\bibnamefont {Yan}}, \bibinfo
  {author} {\bibfnamefont {A.~V.}\ \bibnamefont {Maximov}}, \ and\ \bibinfo
  {author} {\bibfnamefont {C.}~\bibnamefont {Ren}},\ }\bibfield  {title}
  {\enquote {\bibinfo {title} {{Linear regime of two-plasmon decay and
  stimulated Raman scattering instability near the quarter-critical density in
  plasmas}},}\ }\href {\doibase 10.1063/1.4919959} {\bibfield  {journal}
  {\bibinfo  {journal} {Physics of Plasmas}\ }\textbf {\bibinfo {volume}
  {22}},\ \bibinfo {pages} {052704} (\bibinfo {year} {2015})}\BibitemShut
  {NoStop}%
\bibitem [{\citenamefont {Follett}\ \emph {et~al.}(2016)\citenamefont
  {Follett}, \citenamefont {Delettrez}, \citenamefont {Edgell}, \citenamefont
  {Goncharov}, \citenamefont {Henchen}, \citenamefont {Katz}, \citenamefont
  {Michel}, \citenamefont {Myatt}, \citenamefont {Shaw}, \citenamefont
  {Solodov}, \citenamefont {Stoeckl}, \citenamefont {Yaakobi},\ and\
  \citenamefont {Froula}}]{Follett2016}%
  \BibitemOpen
  \bibfield  {author} {\bibinfo {author} {\bibfnamefont {R.~K.}\ \bibnamefont
  {Follett}}, \bibinfo {author} {\bibfnamefont {J.~A.}\ \bibnamefont
  {Delettrez}}, \bibinfo {author} {\bibfnamefont {D.~H.}\ \bibnamefont
  {Edgell}}, \bibinfo {author} {\bibfnamefont {V.~N.}\ \bibnamefont
  {Goncharov}}, \bibinfo {author} {\bibfnamefont {R.~J.}\ \bibnamefont
  {Henchen}}, \bibinfo {author} {\bibfnamefont {J.}~\bibnamefont {Katz}},
  \bibinfo {author} {\bibfnamefont {D.~T.}\ \bibnamefont {Michel}}, \bibinfo
  {author} {\bibfnamefont {J.~F.}\ \bibnamefont {Myatt}}, \bibinfo {author}
  {\bibfnamefont {J.}~\bibnamefont {Shaw}}, \bibinfo {author} {\bibfnamefont
  {A.~A.}\ \bibnamefont {Solodov}}, \bibinfo {author} {\bibfnamefont
  {C.}~\bibnamefont {Stoeckl}}, \bibinfo {author} {\bibfnamefont
  {B.}~\bibnamefont {Yaakobi}}, \ and\ \bibinfo {author} {\bibfnamefont
  {D.~H.}\ \bibnamefont {Froula}},\ }\bibfield  {title} {\enquote {\bibinfo
  {title} {{Two-plasmon decay mitigation in direct-drive
  inertial-confinement-fusion experiments using multilayer targets}},}\ }\href
  {\doibase 10.1103/PhysRevLett.116.155002} {\bibfield  {journal} {\bibinfo
  {journal} {Physical Review Letters}\ }\textbf {\bibinfo {volume} {116}},\
  \bibinfo {pages} {155002} (\bibinfo {year} {2016})}\BibitemShut {NoStop}%
\bibitem [{\citenamefont {Yao}\ \emph {et~al.}(2024)\citenamefont {Yao},
  \citenamefont {Li}, \citenamefont {Hao}, \citenamefont {Yan}, \citenamefont
  {Wang}, \citenamefont {Lei}, \citenamefont {Ding},\ and\ \citenamefont
  {Zheng}}]{Yao2024}%
  \BibitemOpen
  \bibfield  {author} {\bibinfo {author} {\bibfnamefont {C.}~\bibnamefont
  {Yao}}, \bibinfo {author} {\bibfnamefont {J.}~\bibnamefont {Li}}, \bibinfo
  {author} {\bibfnamefont {L.}~\bibnamefont {Hao}}, \bibinfo {author}
  {\bibfnamefont {R.}~\bibnamefont {Yan}}, \bibinfo {author} {\bibfnamefont
  {C.}~\bibnamefont {Wang}}, \bibinfo {author} {\bibfnamefont {A.}~\bibnamefont
  {Lei}}, \bibinfo {author} {\bibfnamefont {Y.-K.}\ \bibnamefont {Ding}}, \
  and\ \bibinfo {author} {\bibfnamefont {J.}~\bibnamefont {Zheng}},\ }\bibfield
   {title} {\enquote {\bibinfo {title} {{Anomalous hot electron generation from
  two-plasmon decay instability driven by broadband laser pulses with intensity
  modulations}},}\ }\href {\doibase 10.1088/1741-4326/ad6c62} {\bibfield
  {journal} {\bibinfo  {journal} {Nuclear Fusion}\ }\textbf {\bibinfo {volume}
  {64}},\ \bibinfo {pages} {106013} (\bibinfo {year} {2024})}\BibitemShut
  {NoStop}%
\bibitem [{\citenamefont {Lei}\ \emph {et~al.}(2024)\citenamefont {Lei},
  \citenamefont {Kang}, \citenamefont {Zhao}, \citenamefont {Liu},
  \citenamefont {An}, \citenamefont {Xiong}, \citenamefont {Wang},
  \citenamefont {Xie}, \citenamefont {Tu}, \citenamefont {Xu}, \citenamefont
  {Zhou}, \citenamefont {Fang}, \citenamefont {Wang}, \citenamefont {Xia},
  \citenamefont {Feng}, \citenamefont {Zhao}, \citenamefont {Ji}, \citenamefont
  {Cui}, \citenamefont {Zhou}, \citenamefont {Liu}, \citenamefont {Zheng},
  \citenamefont {Wang}, \citenamefont {Gao}, \citenamefont {Huang},\ and\
  \citenamefont {Fu}}]{Lei2024}%
  \BibitemOpen
  \bibfield  {author} {\bibinfo {author} {\bibfnamefont {A.}~\bibnamefont
  {Lei}}, \bibinfo {author} {\bibfnamefont {N.}~\bibnamefont {Kang}}, \bibinfo
  {author} {\bibfnamefont {Y.}~\bibnamefont {Zhao}}, \bibinfo {author}
  {\bibfnamefont {H.}~\bibnamefont {Liu}}, \bibinfo {author} {\bibfnamefont
  {H.}~\bibnamefont {An}}, \bibinfo {author} {\bibfnamefont {J.}~\bibnamefont
  {Xiong}}, \bibinfo {author} {\bibfnamefont {R.}~\bibnamefont {Wang}},
  \bibinfo {author} {\bibfnamefont {Z.}~\bibnamefont {Xie}}, \bibinfo {author}
  {\bibfnamefont {Y.}~\bibnamefont {Tu}}, \bibinfo {author} {\bibfnamefont
  {G.}~\bibnamefont {Xu}}, \bibinfo {author} {\bibfnamefont {X.}~\bibnamefont
  {Zhou}}, \bibinfo {author} {\bibfnamefont {Z.}~\bibnamefont {Fang}}, \bibinfo
  {author} {\bibfnamefont {W.}~\bibnamefont {Wang}}, \bibinfo {author}
  {\bibfnamefont {L.}~\bibnamefont {Xia}}, \bibinfo {author} {\bibfnamefont
  {W.}~\bibnamefont {Feng}}, \bibinfo {author} {\bibfnamefont {X.}~\bibnamefont
  {Zhao}}, \bibinfo {author} {\bibfnamefont {L.}~\bibnamefont {Ji}}, \bibinfo
  {author} {\bibfnamefont {Y.}~\bibnamefont {Cui}}, \bibinfo {author}
  {\bibfnamefont {S.}~\bibnamefont {Zhou}}, \bibinfo {author} {\bibfnamefont
  {Z.}~\bibnamefont {Liu}}, \bibinfo {author} {\bibfnamefont {C.}~\bibnamefont
  {Zheng}}, \bibinfo {author} {\bibfnamefont {L.}~\bibnamefont {Wang}},
  \bibinfo {author} {\bibfnamefont {Y.}~\bibnamefont {Gao}}, \bibinfo {author}
  {\bibfnamefont {X.}~\bibnamefont {Huang}}, \ and\ \bibinfo {author}
  {\bibfnamefont {S.}~\bibnamefont {Fu}},\ }\bibfield  {title} {\enquote
  {\bibinfo {title} {{Reduction of Backward Scatterings at the Low-Coherence
  Kunwu Laser Facility}},}\ }\href {\doibase 10.1103/PhysRevLett.132.035102}
  {\bibfield  {journal} {\bibinfo  {journal} {Physical Review Letters}\
  }\textbf {\bibinfo {volume} {132}},\ \bibinfo {pages} {35102} (\bibinfo
  {year} {2024})}\BibitemShut {NoStop}%
\bibitem [{\citenamefont {Wang}\ \emph {et~al.}(2024)\citenamefont {Wang},
  \citenamefont {An}, \citenamefont {Fang}, \citenamefont {Xiong},
  \citenamefont {Xie}, \citenamefont {Wang}, \citenamefont {He}, \citenamefont
  {Jia}, \citenamefont {Wang}, \citenamefont {Zheng}, \citenamefont {Xia},
  \citenamefont {Feng}, \citenamefont {Shi}, \citenamefont {Wang},
  \citenamefont {Sun}, \citenamefont {Gao},\ and\ \citenamefont
  {Fu}}]{Wang2024}%
  \BibitemOpen
  \bibfield  {author} {\bibinfo {author} {\bibfnamefont {P.}~\bibnamefont
  {Wang}}, \bibinfo {author} {\bibfnamefont {H.}~\bibnamefont {An}}, \bibinfo
  {author} {\bibfnamefont {Z.}~\bibnamefont {Fang}}, \bibinfo {author}
  {\bibfnamefont {J.}~\bibnamefont {Xiong}}, \bibinfo {author} {\bibfnamefont
  {Z.}~\bibnamefont {Xie}}, \bibinfo {author} {\bibfnamefont {C.}~\bibnamefont
  {Wang}}, \bibinfo {author} {\bibfnamefont {Z.}~\bibnamefont {He}}, \bibinfo
  {author} {\bibfnamefont {G.}~\bibnamefont {Jia}}, \bibinfo {author}
  {\bibfnamefont {R.}~\bibnamefont {Wang}}, \bibinfo {author} {\bibfnamefont
  {S.}~\bibnamefont {Zheng}}, \bibinfo {author} {\bibfnamefont
  {L.}~\bibnamefont {Xia}}, \bibinfo {author} {\bibfnamefont {W.}~\bibnamefont
  {Feng}}, \bibinfo {author} {\bibfnamefont {H.}~\bibnamefont {Shi}}, \bibinfo
  {author} {\bibfnamefont {W.}~\bibnamefont {Wang}}, \bibinfo {author}
  {\bibfnamefont {J.}~\bibnamefont {Sun}}, \bibinfo {author} {\bibfnamefont
  {Y.}~\bibnamefont {Gao}}, \ and\ \bibinfo {author} {\bibfnamefont
  {S.}~\bibnamefont {Fu}},\ }\bibfield  {title} {\enquote {\bibinfo {title}
  {{Backward scattering of laser plasma interactions from hundreds-of-joules
  broadband laser on thick target}},}\ }\href {\doibase 10.1063/5.0122406}
  {\bibfield  {journal} {\bibinfo  {journal} {Matter and Radiation at
  Extremes}\ }\textbf {\bibinfo {volume} {9}},\ \bibinfo {pages} {015602}
  (\bibinfo {year} {2024})}\BibitemShut {NoStop}%
\bibitem [{\citenamefont {Liu}\ \emph {et~al.}(2024)\citenamefont {Liu},
  \citenamefont {Deng}, \citenamefont {Wang}, \citenamefont {Zhang},
  \citenamefont {Meng}, \citenamefont {Wang}, \citenamefont {Gao},
  \citenamefont {Cai},\ and\ \citenamefont {Zhu}}]{Liu2024}%
  \BibitemOpen
  \bibfield  {author} {\bibinfo {author} {\bibfnamefont {Q.~K.}\ \bibnamefont
  {Liu}}, \bibinfo {author} {\bibfnamefont {L.}~\bibnamefont {Deng}}, \bibinfo
  {author} {\bibfnamefont {Q.}~\bibnamefont {Wang}}, \bibinfo {author}
  {\bibfnamefont {X.}~\bibnamefont {Zhang}}, \bibinfo {author} {\bibfnamefont
  {F.~Q.}\ \bibnamefont {Meng}}, \bibinfo {author} {\bibfnamefont {Y.~P.}\
  \bibnamefont {Wang}}, \bibinfo {author} {\bibfnamefont {Y.~Q.}\ \bibnamefont
  {Gao}}, \bibinfo {author} {\bibfnamefont {H.~B.}\ \bibnamefont {Cai}}, \ and\
  \bibinfo {author} {\bibfnamefont {S.~P.}\ \bibnamefont {Zhu}},\ }\bibfield
  {title} {\enquote {\bibinfo {title} {{Electron kinetic effects in
  back-stimulated Raman scattering bursts driven by broadband laser pulses}},}\
  }\href {\doibase 10.1063/5.0189529} {\bibfield  {journal} {\bibinfo
  {journal} {Matter and Radiation at Extremes}\ }\textbf {\bibinfo {volume}
  {9}},\ \bibinfo {pages} {047402} (\bibinfo {year} {2024})}\BibitemShut
  {NoStop}%
\bibitem [{\citenamefont {Ma}\ \emph {et~al.}(2021)\citenamefont {Ma},
  \citenamefont {Li}, \citenamefont {Weng}, \citenamefont {Yew}, \citenamefont
  {Kawata}, \citenamefont {Gibbon}, \citenamefont {Sheng},\ and\ \citenamefont
  {Zhang}}]{Ma2021}%
  \BibitemOpen
  \bibfield  {author} {\bibinfo {author} {\bibfnamefont {H.~H.}\ \bibnamefont
  {Ma}}, \bibinfo {author} {\bibfnamefont {X.~F.}\ \bibnamefont {Li}}, \bibinfo
  {author} {\bibfnamefont {S.~M.}\ \bibnamefont {Weng}}, \bibinfo {author}
  {\bibfnamefont {S.~H.}\ \bibnamefont {Yew}}, \bibinfo {author} {\bibfnamefont
  {S.}~\bibnamefont {Kawata}}, \bibinfo {author} {\bibfnamefont
  {P.}~\bibnamefont {Gibbon}}, \bibinfo {author} {\bibfnamefont {Z.~M.}\
  \bibnamefont {Sheng}}, \ and\ \bibinfo {author} {\bibfnamefont
  {J.}~\bibnamefont {Zhang}},\ }\bibfield  {title} {\enquote {\bibinfo {title}
  {{Mitigating parametric instabilities in plasmas by sunlight-like lasers}},}\
  }\href {\doibase 10.1063/5.0054653} {\bibfield  {journal} {\bibinfo
  {journal} {Matter and Radiation at Extremes}\ }\textbf {\bibinfo {volume}
  {6}},\ \bibinfo {pages} {55902} (\bibinfo {year} {2021})}\BibitemShut
  {NoStop}%
\bibitem [{\citenamefont {Follett}\ \emph {et~al.}(2021)\citenamefont
  {Follett}, \citenamefont {Shaw}, \citenamefont {Myatt}, \citenamefont {Wen},
  \citenamefont {Froula},\ and\ \citenamefont {Palastro}}]{Follett2021}%
  \BibitemOpen
  \bibfield  {author} {\bibinfo {author} {\bibfnamefont {R.~K.}\ \bibnamefont
  {Follett}}, \bibinfo {author} {\bibfnamefont {J.~G.}\ \bibnamefont {Shaw}},
  \bibinfo {author} {\bibfnamefont {J.~F.}\ \bibnamefont {Myatt}}, \bibinfo
  {author} {\bibfnamefont {H.}~\bibnamefont {Wen}}, \bibinfo {author}
  {\bibfnamefont {D.~H.}\ \bibnamefont {Froula}}, \ and\ \bibinfo {author}
  {\bibfnamefont {J.~P.}\ \bibnamefont {Palastro}},\ }\bibfield  {title}
  {\enquote {\bibinfo {title} {Thresholds of absolute two-plasmon-decay and
  stimulated {{Raman}} scattering instabilities driven by multiple broadband
  lasers},}\ }\href@noop {} {\bibfield  {journal} {\bibinfo  {journal} {Physics
  of Plasmas}\ }\textbf {\bibinfo {volume} {28}},\ \bibinfo {pages} {032103}
  (\bibinfo {year} {2021})}\BibitemShut {NoStop}%
\bibitem [{\citenamefont {Gao}\ \emph {et~al.}(2020)\citenamefont {Gao},
  \citenamefont {Cui}, \citenamefont {Ji}, \citenamefont {Rao}, \citenamefont
  {Zhao}, \citenamefont {Li}, \citenamefont {Liu}, \citenamefont {Feng},
  \citenamefont {Xia}, \citenamefont {Liu}, \citenamefont {Shi}, \citenamefont
  {Du}, \citenamefont {Liu}, \citenamefont {Li}, \citenamefont {Wang},
  \citenamefont {Zhang}, \citenamefont {Shan}, \citenamefont {Hua},
  \citenamefont {Ma}, \citenamefont {Sun}, \citenamefont {Chen}, \citenamefont
  {Huang}, \citenamefont {Zhu}, \citenamefont {Pei}, \citenamefont {Sui},\ and\
  \citenamefont {Fu}}]{Gao2020}%
  \BibitemOpen
  \bibfield  {author} {\bibinfo {author} {\bibfnamefont {Y.}~\bibnamefont
  {Gao}}, \bibinfo {author} {\bibfnamefont {Y.}~\bibnamefont {Cui}}, \bibinfo
  {author} {\bibfnamefont {L.}~\bibnamefont {Ji}}, \bibinfo {author}
  {\bibfnamefont {D.}~\bibnamefont {Rao}}, \bibinfo {author} {\bibfnamefont
  {X.}~\bibnamefont {Zhao}}, \bibinfo {author} {\bibfnamefont {F.}~\bibnamefont
  {Li}}, \bibinfo {author} {\bibfnamefont {D.}~\bibnamefont {Liu}}, \bibinfo
  {author} {\bibfnamefont {W.}~\bibnamefont {Feng}}, \bibinfo {author}
  {\bibfnamefont {L.}~\bibnamefont {Xia}}, \bibinfo {author} {\bibfnamefont
  {J.}~\bibnamefont {Liu}}, \bibinfo {author} {\bibfnamefont {H.}~\bibnamefont
  {Shi}}, \bibinfo {author} {\bibfnamefont {P.}~\bibnamefont {Du}}, \bibinfo
  {author} {\bibfnamefont {J.}~\bibnamefont {Liu}}, \bibinfo {author}
  {\bibfnamefont {X.}~\bibnamefont {Li}}, \bibinfo {author} {\bibfnamefont
  {T.}~\bibnamefont {Wang}}, \bibinfo {author} {\bibfnamefont {T.}~\bibnamefont
  {Zhang}}, \bibinfo {author} {\bibfnamefont {C.}~\bibnamefont {Shan}},
  \bibinfo {author} {\bibfnamefont {Y.}~\bibnamefont {Hua}}, \bibinfo {author}
  {\bibfnamefont {W.}~\bibnamefont {Ma}}, \bibinfo {author} {\bibfnamefont
  {X.}~\bibnamefont {Sun}}, \bibinfo {author} {\bibfnamefont {X.}~\bibnamefont
  {Chen}}, \bibinfo {author} {\bibfnamefont {X.}~\bibnamefont {Huang}},
  \bibinfo {author} {\bibfnamefont {J.}~\bibnamefont {Zhu}}, \bibinfo {author}
  {\bibfnamefont {W.}~\bibnamefont {Pei}}, \bibinfo {author} {\bibfnamefont
  {Z.}~\bibnamefont {Sui}}, \ and\ \bibinfo {author} {\bibfnamefont
  {S.}~\bibnamefont {Fu}},\ }\bibfield  {title} {\enquote {\bibinfo {title}
  {{Development of low-coherence high-power laser drivers for inertial
  confinement fusion}},}\ }\href {\doibase 10.1063/5.0009319} {\bibfield
  {journal} {\bibinfo  {journal} {Matter and Radiation at Extremes}\ }\textbf
  {\bibinfo {volume} {5}},\ \bibinfo {pages} {065201} (\bibinfo {year}
  {2020})}\BibitemShut {NoStop}%
\bibitem [{\citenamefont {Zhao}\ \emph {et~al.}(2017)\citenamefont {Zhao},
  \citenamefont {Weng}, \citenamefont {Chen}, \citenamefont {Zheng},
  \citenamefont {Zhuo}, \citenamefont {Ren}, \citenamefont {Sheng},\ and\
  \citenamefont {Zhang}}]{Zhao2017}%
  \BibitemOpen
  \bibfield  {author} {\bibinfo {author} {\bibfnamefont {Y.}~\bibnamefont
  {Zhao}}, \bibinfo {author} {\bibfnamefont {S.}~\bibnamefont {Weng}}, \bibinfo
  {author} {\bibfnamefont {M.}~\bibnamefont {Chen}}, \bibinfo {author}
  {\bibfnamefont {J.}~\bibnamefont {Zheng}}, \bibinfo {author} {\bibfnamefont
  {H.}~\bibnamefont {Zhuo}}, \bibinfo {author} {\bibfnamefont {C.}~\bibnamefont
  {Ren}}, \bibinfo {author} {\bibfnamefont {Z.}~\bibnamefont {Sheng}}, \ and\
  \bibinfo {author} {\bibfnamefont {J.}~\bibnamefont {Zhang}},\ }\bibfield
  {title} {\enquote {\bibinfo {title} {Effective suppression of parametric
  instabilities with decoupled broadband lasers in plasma},}\ }\href@noop {}
  {\bibfield  {journal} {\bibinfo  {journal} {Physics of Plasmas}\ }\textbf
  {\bibinfo {volume} {24}},\ \bibinfo {pages} {112102} (\bibinfo {year}
  {2017})}\BibitemShut {NoStop}%
\bibitem [{\citenamefont {MacPhee}\ \emph {et~al.}(2010)\citenamefont
  {MacPhee}, \citenamefont {Divol}, \citenamefont {Kemp}, \citenamefont {Akli},
  \citenamefont {Beg}, \citenamefont {Chen}, \citenamefont {Chen},
  \citenamefont {Hey}, \citenamefont {Fedosejevs}, \citenamefont {Freeman},
  \citenamefont {Henesian}, \citenamefont {Key}, \citenamefont {{Le Pape}},
  \citenamefont {Link}, \citenamefont {Ma}, \citenamefont {MacKinnon},
  \citenamefont {Ovchinnikov}, \citenamefont {Patel}, \citenamefont {Phillips},
  \citenamefont {Stephens}, \citenamefont {Tabak}, \citenamefont {Town},
  \citenamefont {Tsui}, \citenamefont {{Van Woerkom}}, \citenamefont {Wei},\
  and\ \citenamefont {Wilks}}]{Macphee2010}%
  \BibitemOpen
  \bibfield  {author} {\bibinfo {author} {\bibfnamefont {A.~G.}\ \bibnamefont
  {MacPhee}}, \bibinfo {author} {\bibfnamefont {L.}~\bibnamefont {Divol}},
  \bibinfo {author} {\bibfnamefont {A.~J.}\ \bibnamefont {Kemp}}, \bibinfo
  {author} {\bibfnamefont {K.~U.}\ \bibnamefont {Akli}}, \bibinfo {author}
  {\bibfnamefont {F.~N.}\ \bibnamefont {Beg}}, \bibinfo {author} {\bibfnamefont
  {C.~D.}\ \bibnamefont {Chen}}, \bibinfo {author} {\bibfnamefont
  {H.}~\bibnamefont {Chen}}, \bibinfo {author} {\bibfnamefont {D.~S.}\
  \bibnamefont {Hey}}, \bibinfo {author} {\bibfnamefont {R.~J.}\ \bibnamefont
  {Fedosejevs}}, \bibinfo {author} {\bibfnamefont {R.~R.}\ \bibnamefont
  {Freeman}}, \bibinfo {author} {\bibfnamefont {M.}~\bibnamefont {Henesian}},
  \bibinfo {author} {\bibfnamefont {M.~H.}\ \bibnamefont {Key}}, \bibinfo
  {author} {\bibfnamefont {S.}~\bibnamefont {{Le Pape}}}, \bibinfo {author}
  {\bibfnamefont {A.}~\bibnamefont {Link}}, \bibinfo {author} {\bibfnamefont
  {T.}~\bibnamefont {Ma}}, \bibinfo {author} {\bibfnamefont {A.~J.}\
  \bibnamefont {MacKinnon}}, \bibinfo {author} {\bibfnamefont {V.~M.}\
  \bibnamefont {Ovchinnikov}}, \bibinfo {author} {\bibfnamefont {P.~K.}\
  \bibnamefont {Patel}}, \bibinfo {author} {\bibfnamefont {T.~W.}\ \bibnamefont
  {Phillips}}, \bibinfo {author} {\bibfnamefont {R.~B.}\ \bibnamefont
  {Stephens}}, \bibinfo {author} {\bibfnamefont {M.}~\bibnamefont {Tabak}},
  \bibinfo {author} {\bibfnamefont {R.}~\bibnamefont {Town}}, \bibinfo {author}
  {\bibfnamefont {Y.~Y.}\ \bibnamefont {Tsui}}, \bibinfo {author}
  {\bibfnamefont {L.~D.}\ \bibnamefont {{Van Woerkom}}}, \bibinfo {author}
  {\bibfnamefont {M.~S.}\ \bibnamefont {Wei}}, \ and\ \bibinfo {author}
  {\bibfnamefont {S.~C.}\ \bibnamefont {Wilks}},\ }\bibfield  {title} {\enquote
  {\bibinfo {title} {{Limitation on prepulse level for cone-guided
  fast-ignition inertial confinement Fusion}},}\ }\href {\doibase
  10.1103/PhysRevLett.104.055002} {\bibfield  {journal} {\bibinfo  {journal}
  {Physical Review Letters}\ }\textbf {\bibinfo {volume} {104}},\ \bibinfo
  {pages} {055002} (\bibinfo {year} {2010})}\BibitemShut {NoStop}%
\bibitem [{\citenamefont {Li}\ \emph {et~al.}(2013)\citenamefont {Li},
  \citenamefont {Davies}, \citenamefont {Ma}, \citenamefont {Mori},
  \citenamefont {Ren}, \citenamefont {Solodov}, \citenamefont {Theobald},\ and\
  \citenamefont {Tonge}}]{Li2013}%
  \BibitemOpen
  \bibfield  {author} {\bibinfo {author} {\bibfnamefont {J.}~\bibnamefont
  {Li}}, \bibinfo {author} {\bibfnamefont {J.~R.}\ \bibnamefont {Davies}},
  \bibinfo {author} {\bibfnamefont {T.}~\bibnamefont {Ma}}, \bibinfo {author}
  {\bibfnamefont {W.~B.}\ \bibnamefont {Mori}}, \bibinfo {author}
  {\bibfnamefont {C.}~\bibnamefont {Ren}}, \bibinfo {author} {\bibfnamefont
  {A.~A.}\ \bibnamefont {Solodov}}, \bibinfo {author} {\bibfnamefont
  {W.}~\bibnamefont {Theobald}}, \ and\ \bibinfo {author} {\bibfnamefont
  {J.}~\bibnamefont {Tonge}},\ }\bibfield  {title} {\enquote {\bibinfo {title}
  {{Hot-electron generation from laser-pre-plasma interactions in cone-guided
  fast ignition}},}\ }\href {\doibase 10.1063/1.4807040} {\bibfield  {journal}
  {\bibinfo  {journal} {Physics of Plasmas}\ }\textbf {\bibinfo {volume}
  {20}},\ \bibinfo {pages} {052706} (\bibinfo {year} {2013})}\BibitemShut
  {NoStop}%
\bibitem [{\citenamefont {Ebrahim}\ \emph {et~al.}(1980)\citenamefont
  {Ebrahim}, \citenamefont {Baldis}, \citenamefont {Joshi},\ and\ \citenamefont
  {Benesch}}]{Ebrahim1980}%
  \BibitemOpen
  \bibfield  {author} {\bibinfo {author} {\bibfnamefont {N.~A.}\ \bibnamefont
  {Ebrahim}}, \bibinfo {author} {\bibfnamefont {H.~A.}\ \bibnamefont {Baldis}},
  \bibinfo {author} {\bibfnamefont {C.}~\bibnamefont {Joshi}}, \ and\ \bibinfo
  {author} {\bibfnamefont {R.}~\bibnamefont {Benesch}},\ }\bibfield  {title}
  {\enquote {\bibinfo {title} {{Hot electron generation by the two-plasmon
  decay instability in the laser-plasma interaction at 10.6 m}},}\ }\href
  {\doibase 10.1103/PhysRevLett.45.1179} {\bibfield  {journal} {\bibinfo
  {journal} {Physical Review Letters}\ }\textbf {\bibinfo {volume} {45}},\
  \bibinfo {pages} {1179--1182} (\bibinfo {year} {1980})}\BibitemShut {NoStop}%
\bibitem [{\citenamefont {Yaakobi}\ \emph {et~al.}(2013)\citenamefont
  {Yaakobi}, \citenamefont {Solodov}, \citenamefont {Myatt}, \citenamefont
  {Delettrez}, \citenamefont {Stoeckl},\ and\ \citenamefont
  {Froula}}]{Yaakobi2013}%
  \BibitemOpen
  \bibfield  {author} {\bibinfo {author} {\bibfnamefont {B.}~\bibnamefont
  {Yaakobi}}, \bibinfo {author} {\bibfnamefont {A.~A.}\ \bibnamefont
  {Solodov}}, \bibinfo {author} {\bibfnamefont {J.~F.}\ \bibnamefont {Myatt}},
  \bibinfo {author} {\bibfnamefont {J.~A.}\ \bibnamefont {Delettrez}}, \bibinfo
  {author} {\bibfnamefont {C.}~\bibnamefont {Stoeckl}}, \ and\ \bibinfo
  {author} {\bibfnamefont {D.~H.}\ \bibnamefont {Froula}},\ }\bibfield  {title}
  {\enquote {\bibinfo {title} {{Measurements of the divergence of fast
  electrons in laser-irradiated spherical targets}},}\ }\href {\doibase
  10.1063/1.4824008} {\bibfield  {journal} {\bibinfo  {journal} {Physics of
  Plasmas}\ }\textbf {\bibinfo {volume} {20}},\ \bibinfo {pages} {092706}
  (\bibinfo {year} {2013})}\BibitemShut {NoStop}%
\bibitem [{\citenamefont {Yan}\ \emph {et~al.}(2012)\citenamefont {Yan},
  \citenamefont {Ren}, \citenamefont {Li}, \citenamefont {Maximov},
  \citenamefont {Mori}, \citenamefont {Sheng},\ and\ \citenamefont
  {Tsung}}]{Yan2012}%
  \BibitemOpen
  \bibfield  {author} {\bibinfo {author} {\bibfnamefont {R.}~\bibnamefont
  {Yan}}, \bibinfo {author} {\bibfnamefont {C.}~\bibnamefont {Ren}}, \bibinfo
  {author} {\bibfnamefont {J.}~\bibnamefont {Li}}, \bibinfo {author}
  {\bibfnamefont {A.~V.}\ \bibnamefont {Maximov}}, \bibinfo {author}
  {\bibfnamefont {W.~B.}\ \bibnamefont {Mori}}, \bibinfo {author}
  {\bibfnamefont {Z.-M.}\ \bibnamefont {Sheng}}, \ and\ \bibinfo {author}
  {\bibfnamefont {F.~S.}\ \bibnamefont {Tsung}},\ }\bibfield  {title} {\enquote
  {\bibinfo {title} {Generating energetic electrons through staged acceleration
  in the two-plasmon-decay instability in inertial confinement fusion},}\
  }\href@noop {} {\bibfield  {journal} {\bibinfo  {journal} {Physical Review
  Letters}\ }\textbf {\bibinfo {volume} {108}},\ \bibinfo {pages} {175002}
  (\bibinfo {year} {2012})}\BibitemShut {NoStop}%
\bibitem [{\citenamefont {Yan}\ \emph {et~al.}(2009)\citenamefont {Yan},
  \citenamefont {Maximov}, \citenamefont {Ren},\ and\ \citenamefont
  {Tsung}}]{Yan2009}%
  \BibitemOpen
  \bibfield  {author} {\bibinfo {author} {\bibfnamefont {R.}~\bibnamefont
  {Yan}}, \bibinfo {author} {\bibfnamefont {A.~V.}\ \bibnamefont {Maximov}},
  \bibinfo {author} {\bibfnamefont {C.}~\bibnamefont {Ren}}, \ and\ \bibinfo
  {author} {\bibfnamefont {F.~S.}\ \bibnamefont {Tsung}},\ }\bibfield  {title}
  {\enquote {\bibinfo {title} {{Growth and Saturation of Convective Modes of
  the Two-Plasmon Decay Instability in Inertial Confinement Fusion}},}\ }\href
  {\doibase 10.1103/PhysRevLett.103.175002} {\bibfield  {journal} {\bibinfo
  {journal} {Physical Review Letters}\ }\textbf {\bibinfo {volume} {103}},\
  \bibinfo {pages} {175002} (\bibinfo {year} {2009})}\BibitemShut {NoStop}%
\bibitem [{\citenamefont {Zhang}\ \emph
  {et~al.}(2020{\natexlab{b}})\citenamefont {Zhang}, \citenamefont {Krauland},
  \citenamefont {Peebles}, \citenamefont {Li}, \citenamefont {Beg},
  \citenamefont {Alexander}, \citenamefont {Theobald}, \citenamefont {Betti},
  \citenamefont {Haberberger}, \citenamefont {Campbell}, \citenamefont {Yan},
  \citenamefont {Borwick}, \citenamefont {Ren},\ and\ \citenamefont
  {Wei}}]{Zhang2020a}%
  \BibitemOpen
  \bibfield  {author} {\bibinfo {author} {\bibfnamefont {S.}~\bibnamefont
  {Zhang}}, \bibinfo {author} {\bibfnamefont {C.~M.}\ \bibnamefont {Krauland}},
  \bibinfo {author} {\bibfnamefont {J.}~\bibnamefont {Peebles}}, \bibinfo
  {author} {\bibfnamefont {J.}~\bibnamefont {Li}}, \bibinfo {author}
  {\bibfnamefont {F.~N.}\ \bibnamefont {Beg}}, \bibinfo {author} {\bibfnamefont
  {N.}~\bibnamefont {Alexander}}, \bibinfo {author} {\bibfnamefont
  {W.}~\bibnamefont {Theobald}}, \bibinfo {author} {\bibfnamefont
  {R.}~\bibnamefont {Betti}}, \bibinfo {author} {\bibfnamefont
  {D.}~\bibnamefont {Haberberger}}, \bibinfo {author} {\bibfnamefont {E.~M.}\
  \bibnamefont {Campbell}}, \bibinfo {author} {\bibfnamefont {R.}~\bibnamefont
  {Yan}}, \bibinfo {author} {\bibfnamefont {E.}~\bibnamefont {Borwick}},
  \bibinfo {author} {\bibfnamefont {C.}~\bibnamefont {Ren}}, \ and\ \bibinfo
  {author} {\bibfnamefont {M.~S.}\ \bibnamefont {Wei}},\ }\bibfield  {title}
  {\enquote {\bibinfo {title} {{Experimental study of hot electron generation
  in shock ignition relevant high-intensity regime with large scale hot
  plasmas}},}\ }\href {\doibase 10.1063/1.5119250} {\bibfield  {journal}
  {\bibinfo  {journal} {Physics of Plasmas}\ }\textbf {\bibinfo {volume}
  {27}},\ \bibinfo {pages} {23111} (\bibinfo {year}
  {2020}{\natexlab{b}})}\BibitemShut {NoStop}%
\bibitem [{\citenamefont {Li}\ \emph {et~al.}(2020)\citenamefont {Li},
  \citenamefont {Zhang}, \citenamefont {Krauland}, \citenamefont {Wen},
  \citenamefont {Beg}, \citenamefont {Ren},\ and\ \citenamefont
  {Wei}}]{Li2020}%
  \BibitemOpen
  \bibfield  {author} {\bibinfo {author} {\bibfnamefont {J.}~\bibnamefont
  {Li}}, \bibinfo {author} {\bibfnamefont {S.}~\bibnamefont {Zhang}}, \bibinfo
  {author} {\bibfnamefont {C.~M.}\ \bibnamefont {Krauland}}, \bibinfo {author}
  {\bibfnamefont {H.}~\bibnamefont {Wen}}, \bibinfo {author} {\bibfnamefont
  {F.~N.}\ \bibnamefont {Beg}}, \bibinfo {author} {\bibfnamefont
  {C.}~\bibnamefont {Ren}}, \ and\ \bibinfo {author} {\bibfnamefont {M.~S.}\
  \bibnamefont {Wei}},\ }\bibfield  {title} {\enquote {\bibinfo {title} {{Pump
  depletion and hot electron generation in long density scale length plasma
  with shock ignition high intensity laser}},}\ }\href {\doibase
  10.1103/PhysRevE.101.033206} {\bibfield  {journal} {\bibinfo  {journal}
  {Physical Review E}\ }\textbf {\bibinfo {volume} {101}},\ \bibinfo {pages}
  {033206} (\bibinfo {year} {2020})}\BibitemShut {NoStop}%
\bibitem [{\citenamefont {Liu}\ \emph {et~al.}(2017)\citenamefont {Liu},
  \citenamefont {Kang}, \citenamefont {Zhou}, \citenamefont {An}, \citenamefont
  {Fang}, \citenamefont {Xiong}, \citenamefont {Li}, \citenamefont {Lei},\ and\
  \citenamefont {Lin}}]{Liu2017}%
  \BibitemOpen
  \bibfield  {author} {\bibinfo {author} {\bibfnamefont {H.}~\bibnamefont
  {Liu}}, \bibinfo {author} {\bibfnamefont {N.}~\bibnamefont {Kang}}, \bibinfo
  {author} {\bibfnamefont {S.}~\bibnamefont {Zhou}}, \bibinfo {author}
  {\bibfnamefont {H.}~\bibnamefont {An}}, \bibinfo {author} {\bibfnamefont
  {Z.}~\bibnamefont {Fang}}, \bibinfo {author} {\bibfnamefont {J.}~\bibnamefont
  {Xiong}}, \bibinfo {author} {\bibfnamefont {K.}~\bibnamefont {Li}}, \bibinfo
  {author} {\bibfnamefont {A.}~\bibnamefont {Lei}}, \ and\ \bibinfo {author}
  {\bibfnamefont {Z.}~\bibnamefont {Lin}},\ }\bibfield  {title} {\enquote
  {\bibinfo {title} {{Emission properties of suprathermal electrons produced by
  laser-plasma interactions}},}\ }\href {\doibase 10.1017/S0263034617000702}
  {\bibfield  {journal} {\bibinfo  {journal} {Laser and Particle Beams}\
  }\textbf {\bibinfo {volume} {35}},\ \bibinfo {pages} {663--669} (\bibinfo
  {year} {2017})}\BibitemShut {NoStop}%
\bibitem [{\citenamefont {Myatt}\ \emph {et~al.}(2014)\citenamefont {Myatt},
  \citenamefont {Zhang}, \citenamefont {Short}, \citenamefont {Maximov},
  \citenamefont {Seka}, \citenamefont {Froula}, \citenamefont {Edgell},
  \citenamefont {Michel}, \citenamefont {Igumenshchev}, \citenamefont {Hinkel},
  \citenamefont {Michel},\ and\ \citenamefont {Moody}}]{Myatt2014}%
  \BibitemOpen
  \bibfield  {author} {\bibinfo {author} {\bibfnamefont {J.~F.}\ \bibnamefont
  {Myatt}}, \bibinfo {author} {\bibfnamefont {J.}~\bibnamefont {Zhang}},
  \bibinfo {author} {\bibfnamefont {R.~W.}\ \bibnamefont {Short}}, \bibinfo
  {author} {\bibfnamefont {A.~V.}\ \bibnamefont {Maximov}}, \bibinfo {author}
  {\bibfnamefont {W.}~\bibnamefont {Seka}}, \bibinfo {author} {\bibfnamefont
  {D.~H.}\ \bibnamefont {Froula}}, \bibinfo {author} {\bibfnamefont {D.~H.}\
  \bibnamefont {Edgell}}, \bibinfo {author} {\bibfnamefont {D.~T.}\
  \bibnamefont {Michel}}, \bibinfo {author} {\bibfnamefont {I.~V.}\
  \bibnamefont {Igumenshchev}}, \bibinfo {author} {\bibfnamefont {D.~E.}\
  \bibnamefont {Hinkel}}, \bibinfo {author} {\bibfnamefont {P.}~\bibnamefont
  {Michel}}, \ and\ \bibinfo {author} {\bibfnamefont {J.~D.}\ \bibnamefont
  {Moody}},\ }\bibfield  {title} {\enquote {\bibinfo {title} {{Multiple-beam
  laser-plasma interactions in inertial confinement fusion}},}\ }\href
  {\doibase 10.1063/1.4878623} {\bibfield  {journal} {\bibinfo  {journal}
  {Physics of Plasmas}\ }\textbf {\bibinfo {volume} {21}},\ \bibinfo {pages}
  {055501} (\bibinfo {year} {2014})}\BibitemShut {NoStop}%
\bibitem [{\citenamefont {Yan}, \citenamefont {Li},\ and\ \citenamefont
  {Ren}(2014)}]{Yan2014}%
  \BibitemOpen
  \bibfield  {author} {\bibinfo {author} {\bibfnamefont {R.}~\bibnamefont
  {Yan}}, \bibinfo {author} {\bibfnamefont {J.}~\bibnamefont {Li}}, \ and\
  \bibinfo {author} {\bibfnamefont {C.}~\bibnamefont {Ren}},\ }\bibfield
  {title} {\enquote {\bibinfo {title} {{Intermittent laser-plasma interactions
  and hot electron generation in shock ignition}},}\ }\href {\doibase
  10.1063/1.4882682} {\bibfield  {journal} {\bibinfo  {journal} {Physics of
  Plasmas}\ }\textbf {\bibinfo {volume} {21}},\ \bibinfo {pages} {062705}
  (\bibinfo {year} {2014})}\BibitemShut {NoStop}%
\bibitem [{\citenamefont {Michel}\ \emph {et~al.}(2015)\citenamefont {Michel},
  \citenamefont {Divol}, \citenamefont {Dewald}, \citenamefont {Milovich},
  \citenamefont {Hohenberger}, \citenamefont {Jones}, \citenamefont {Hopkins},
  \citenamefont {Berger}, \citenamefont {Kruer},\ and\ \citenamefont
  {Moody}}]{Michel2015}%
  \BibitemOpen
  \bibfield  {author} {\bibinfo {author} {\bibfnamefont {P.}~\bibnamefont
  {Michel}}, \bibinfo {author} {\bibfnamefont {L.}~\bibnamefont {Divol}},
  \bibinfo {author} {\bibfnamefont {E.~L.}\ \bibnamefont {Dewald}}, \bibinfo
  {author} {\bibfnamefont {J.~L.}\ \bibnamefont {Milovich}}, \bibinfo {author}
  {\bibfnamefont {M.}~\bibnamefont {Hohenberger}}, \bibinfo {author}
  {\bibfnamefont {O.~S.}\ \bibnamefont {Jones}}, \bibinfo {author}
  {\bibfnamefont {L.~B.}\ \bibnamefont {Hopkins}}, \bibinfo {author}
  {\bibfnamefont {R.~L.}\ \bibnamefont {Berger}}, \bibinfo {author}
  {\bibfnamefont {W.~L.}\ \bibnamefont {Kruer}}, \ and\ \bibinfo {author}
  {\bibfnamefont {J.~D.}\ \bibnamefont {Moody}},\ }\bibfield  {title} {\enquote
  {\bibinfo {title} {{Multibeam Stimulated Raman Scattering in Inertial
  Confinement Fusion Conditions}},}\ }\href {\doibase
  10.1103/PhysRevLett.115.055003} {\bibfield  {journal} {\bibinfo  {journal}
  {Physical Review Letters}\ }\textbf {\bibinfo {volume} {115}},\ \bibinfo
  {pages} {055003} (\bibinfo {year} {2015})}\BibitemShut {NoStop}%
\bibitem [{\citenamefont {Michel}\ \emph {et~al.}(2012)\citenamefont {Michel},
  \citenamefont {Maximov}, \citenamefont {Short}, \citenamefont {Hu},
  \citenamefont {Myatt}, \citenamefont {Seka}, \citenamefont {Solodov},
  \citenamefont {Yaakobi},\ and\ \citenamefont {Froula}}]{Michel2012a}%
  \BibitemOpen
  \bibfield  {author} {\bibinfo {author} {\bibfnamefont {D.~T.}\ \bibnamefont
  {Michel}}, \bibinfo {author} {\bibfnamefont {A.~V.}\ \bibnamefont {Maximov}},
  \bibinfo {author} {\bibfnamefont {R.~W.}\ \bibnamefont {Short}}, \bibinfo
  {author} {\bibfnamefont {S.~X.}\ \bibnamefont {Hu}}, \bibinfo {author}
  {\bibfnamefont {J.~F.}\ \bibnamefont {Myatt}}, \bibinfo {author}
  {\bibfnamefont {W.}~\bibnamefont {Seka}}, \bibinfo {author} {\bibfnamefont
  {A.~A.}\ \bibnamefont {Solodov}}, \bibinfo {author} {\bibfnamefont
  {B.}~\bibnamefont {Yaakobi}}, \ and\ \bibinfo {author} {\bibfnamefont
  {D.~H.}\ \bibnamefont {Froula}},\ }\bibfield  {title} {\enquote {\bibinfo
  {title} {{Experimental validation of the two-plasmon-decay common-wave
  process}},}\ }\href {\doibase 10.1103/PhysRevLett.109.155007} {\bibfield
  {journal} {\bibinfo  {journal} {Physical Review Letters}\ }\textbf {\bibinfo
  {volume} {109}},\ \bibinfo {pages} {155007} (\bibinfo {year}
  {2012})}\BibitemShut {NoStop}%
\bibitem [{\citenamefont {Lian}\ \emph {et~al.}(2022)\citenamefont {Lian},
  \citenamefont {Ji}, \citenamefont {Yan}, \citenamefont {Cao}, \citenamefont
  {Ren}, \citenamefont {Wan}, \citenamefont {Yang}, \citenamefont {Ding},\ and\
  \citenamefont {Zheng}}]{Lian2022}%
  \BibitemOpen
  \bibfield  {author} {\bibinfo {author} {\bibfnamefont {C.~W.}\ \bibnamefont
  {Lian}}, \bibinfo {author} {\bibfnamefont {Y.}~\bibnamefont {Ji}}, \bibinfo
  {author} {\bibfnamefont {R.}~\bibnamefont {Yan}}, \bibinfo {author}
  {\bibfnamefont {S.~H.}\ \bibnamefont {Cao}}, \bibinfo {author} {\bibfnamefont
  {C.}~\bibnamefont {Ren}}, \bibinfo {author} {\bibfnamefont {Z.~H.}\
  \bibnamefont {Wan}}, \bibinfo {author} {\bibfnamefont {D.}~\bibnamefont
  {Yang}}, \bibinfo {author} {\bibfnamefont {Y.~K.}\ \bibnamefont {Ding}}, \
  and\ \bibinfo {author} {\bibfnamefont {J.}~\bibnamefont {Zheng}},\ }\bibfield
   {title} {\enquote {\bibinfo {title} {{Two plasmon decay instability
  stimulated by large-incidence-angle laser in inertial confinement fusion}},}\
  }\href {\doibase 10.1088/1361-6587/ac7b47} {\bibfield  {journal} {\bibinfo
  {journal} {Plasma Physics and Controlled Fusion}\ }\textbf {\bibinfo {volume}
  {64}},\ \bibinfo {pages} {085009} (\bibinfo {year} {2022})}\BibitemShut
  {NoStop}%
\bibitem [{\citenamefont {Zhou}\ \emph {et~al.}(2023)\citenamefont {Zhou},
  \citenamefont {Cao}, \citenamefont {Lian}, \citenamefont {Ji}, \citenamefont
  {Yan}, \citenamefont {Li}, \citenamefont {Yang}, \citenamefont {Hao},
  \citenamefont {Ren},\ and\ \citenamefont {Zheng}}]{Zhou2023}%
  \BibitemOpen
  \bibfield  {author} {\bibinfo {author} {\bibfnamefont {F.~X.}\ \bibnamefont
  {Zhou}}, \bibinfo {author} {\bibfnamefont {S.~H.}\ \bibnamefont {Cao}},
  \bibinfo {author} {\bibfnamefont {C.~W.}\ \bibnamefont {Lian}}, \bibinfo
  {author} {\bibfnamefont {Y.}~\bibnamefont {Ji}}, \bibinfo {author}
  {\bibfnamefont {R.}~\bibnamefont {Yan}}, \bibinfo {author} {\bibfnamefont
  {J.}~\bibnamefont {Li}}, \bibinfo {author} {\bibfnamefont {D.}~\bibnamefont
  {Yang}}, \bibinfo {author} {\bibfnamefont {L.}~\bibnamefont {Hao}}, \bibinfo
  {author} {\bibfnamefont {C.}~\bibnamefont {Ren}}, \ and\ \bibinfo {author}
  {\bibfnamefont {J.}~\bibnamefont {Zheng}},\ }\bibfield  {title} {\enquote
  {\bibinfo {title} {{Large-incidence-angle multiple-beam two-plasmon decay
  instability in inertial confinement fusion}},}\ }\href {\doibase
  10.1063/5.0162495} {\bibfield  {journal} {\bibinfo  {journal} {Physics of
  Plasmas}\ }\textbf {\bibinfo {volume} {30}},\ \bibinfo {pages} {092702}
  (\bibinfo {year} {2023})}\BibitemShut {NoStop}%
\bibitem [{\citenamefont {Fonseca}\ \emph {et~al.}(2002)\citenamefont
  {Fonseca}, \citenamefont {Silva}, \citenamefont {Tsung}, \citenamefont
  {Decyk}, \citenamefont {Lu}, \citenamefont {Ren}, \citenamefont {Mori},
  \citenamefont {Deng}, \citenamefont {Lee}, \citenamefont {Katsouleas},\ and\
  \citenamefont {Adam}}]{Fonseca2002}%
  \BibitemOpen
  \bibfield  {author} {\bibinfo {author} {\bibfnamefont {R.~A.}\ \bibnamefont
  {Fonseca}}, \bibinfo {author} {\bibfnamefont {L.~O.}\ \bibnamefont {Silva}},
  \bibinfo {author} {\bibfnamefont {F.~S.}\ \bibnamefont {Tsung}}, \bibinfo
  {author} {\bibfnamefont {V.~K.}\ \bibnamefont {Decyk}}, \bibinfo {author}
  {\bibfnamefont {W.}~\bibnamefont {Lu}}, \bibinfo {author} {\bibfnamefont
  {C.}~\bibnamefont {Ren}}, \bibinfo {author} {\bibfnamefont {W.~B.}\
  \bibnamefont {Mori}}, \bibinfo {author} {\bibfnamefont {S.}~\bibnamefont
  {Deng}}, \bibinfo {author} {\bibfnamefont {S.}~\bibnamefont {Lee}}, \bibinfo
  {author} {\bibfnamefont {T.}~\bibnamefont {Katsouleas}}, \ and\ \bibinfo
  {author} {\bibfnamefont {J.~C.}\ \bibnamefont {Adam}},\ }\enquote {\bibinfo
  {title} {Computational science --- iccs 2002: International conference
  amsterdam, the netherlands, april 21--24, 2002 proceedings, part iii},}\ \
  (\bibinfo  {publisher} {Springer Berlin Heidelberg},\ \bibinfo {address}
  {Berlin, Heidelberg},\ \bibinfo {year} {2002})\ Chap.\ \bibinfo {chapter}
  {OSIRIS: A Three-Dimensional, Fully Relativistic Particle in Cell Code for
  Modeling Plasma Based Accelerators}, pp.\ \bibinfo {pages}
  {342--351}\BibitemShut {NoStop}%
\bibitem [{\citenamefont {Simon}\ \emph {et~al.}(1983)\citenamefont {Simon},
  \citenamefont {Short}, \citenamefont {Williams},\ and\ \citenamefont
  {Dewandre}}]{Simon1983}%
  \BibitemOpen
  \bibfield  {author} {\bibinfo {author} {\bibfnamefont {A.}~\bibnamefont
  {Simon}}, \bibinfo {author} {\bibfnamefont {R.~W.}\ \bibnamefont {Short}},
  \bibinfo {author} {\bibfnamefont {E.~A.}\ \bibnamefont {Williams}}, \ and\
  \bibinfo {author} {\bibfnamefont {T.}~\bibnamefont {Dewandre}},\ }\bibfield
  {title} {\enquote {\bibinfo {title} {{On the inhomogeneous two-plasmon
  instability}},}\ }\href {\doibase 10.1063/1.864037} {\bibfield  {journal}
  {\bibinfo  {journal} {Physics of Fluids}\ }\textbf {\bibinfo {volume} {26}},\
  \bibinfo {pages} {3107--3118} (\bibinfo {year} {1983})}\BibitemShut {NoStop}%
\bibitem [{\citenamefont {Wen}\ \emph {et~al.}(2019)\citenamefont {Wen},
  \citenamefont {Tsung}, \citenamefont {Mori}, \citenamefont {Fonseca},\ and\
  \citenamefont {Silva}}]{Wen2019}%
  \BibitemOpen
  \bibfield  {author} {\bibinfo {author} {\bibfnamefont {H.}~\bibnamefont
  {Wen}}, \bibinfo {author} {\bibfnamefont {F.~S.}\ \bibnamefont {Tsung}},
  \bibinfo {author} {\bibfnamefont {W.~B.}\ \bibnamefont {Mori}}, \bibinfo
  {author} {\bibfnamefont {R.~A.}\ \bibnamefont {Fonseca}}, \ and\ \bibinfo
  {author} {\bibfnamefont {L.~O.}\ \bibnamefont {Silva}},\ }\bibfield  {title}
  {\enquote {\bibinfo {title} {{Petascale particle-in-cell simulations of
  kinetic effects in inertial fusion energy plasmas}},}\ }\href {\doibase
  10.1088/1361-6587/ab019a} {\bibfield  {journal} {\bibinfo  {journal} {Plasma
  Physics and Controlled Fusion}\ }\textbf {\bibinfo {volume} {61}},\ \bibinfo
  {pages} {044007} (\bibinfo {year} {2019})}\BibitemShut {NoStop}%
\bibitem [{\citenamefont {Zhang}\ \emph {et~al.}(2014)\citenamefont {Zhang},
  \citenamefont {Myatt}, \citenamefont {Short}, \citenamefont {Maximov},
  \citenamefont {Vu}, \citenamefont {Dubois},\ and\ \citenamefont
  {Russell}}]{Zhang2014}%
  \BibitemOpen
  \bibfield  {author} {\bibinfo {author} {\bibfnamefont {J.}~\bibnamefont
  {Zhang}}, \bibinfo {author} {\bibfnamefont {J.~F.}\ \bibnamefont {Myatt}},
  \bibinfo {author} {\bibfnamefont {R.~W.}\ \bibnamefont {Short}}, \bibinfo
  {author} {\bibfnamefont {A.~V.}\ \bibnamefont {Maximov}}, \bibinfo {author}
  {\bibfnamefont {H.~X.}\ \bibnamefont {Vu}}, \bibinfo {author} {\bibfnamefont
  {D.~F.}\ \bibnamefont {Dubois}}, \ and\ \bibinfo {author} {\bibfnamefont
  {D.~A.}\ \bibnamefont {Russell}},\ }\bibfield  {title} {\enquote {\bibinfo
  {title} {{Multiple beam two-plasmon decay: Linear threshold to nonlinear
  saturation in three dimensions}},}\ }\href {\doibase
  10.1103/PhysRevLett.113.105001} {\bibfield  {journal} {\bibinfo  {journal}
  {Physical Review Letters}\ }\textbf {\bibinfo {volume} {113}},\ \bibinfo
  {pages} {105001} (\bibinfo {year} {2014})}\BibitemShut {NoStop}%
\bibitem [{\citenamefont {Myatt}\ \emph {et~al.}(2012)\citenamefont {Myatt},
  \citenamefont {Zhang}, \citenamefont {Delettrez}, \citenamefont {Maximov},
  \citenamefont {Short}, \citenamefont {Seka}, \citenamefont {Edgell},
  \citenamefont {Dubois}, \citenamefont {Russell},\ and\ \citenamefont
  {Vu}}]{Myatt2012}%
  \BibitemOpen
  \bibfield  {author} {\bibinfo {author} {\bibfnamefont {J.~F.}\ \bibnamefont
  {Myatt}}, \bibinfo {author} {\bibfnamefont {J.}~\bibnamefont {Zhang}},
  \bibinfo {author} {\bibfnamefont {J.~a.}\ \bibnamefont {Delettrez}}, \bibinfo
  {author} {\bibfnamefont {a.~V.}\ \bibnamefont {Maximov}}, \bibinfo {author}
  {\bibfnamefont {R.~W.}\ \bibnamefont {Short}}, \bibinfo {author}
  {\bibfnamefont {W.}~\bibnamefont {Seka}}, \bibinfo {author} {\bibfnamefont
  {D.~H.}\ \bibnamefont {Edgell}}, \bibinfo {author} {\bibfnamefont {D.~F.}\
  \bibnamefont {Dubois}}, \bibinfo {author} {\bibfnamefont {D.~a.}\
  \bibnamefont {Russell}}, \ and\ \bibinfo {author} {\bibfnamefont {H.~X.}\
  \bibnamefont {Vu}},\ }\bibfield  {title} {\enquote {\bibinfo {title} {{The
  dynamics of hot-electron heating in direct-drive-implosion experiments caused
  by two-plasmon-decay instability}},}\ }\href {\doibase 10.1063/1.3683004}
  {\bibfield  {journal} {\bibinfo  {journal} {Physics of Plasmas}\ }\textbf
  {\bibinfo {volume} {19}},\ \bibinfo {pages} {022707} (\bibinfo {year}
  {2012})}\BibitemShut {NoStop}%
\bibitem [{\citenamefont {Rosenbluth}, \citenamefont {White},\ and\
  \citenamefont {Liu}(1973)}]{Rosenbluth1973}%
  \BibitemOpen
  \bibfield  {author} {\bibinfo {author} {\bibfnamefont {M.~N.}\ \bibnamefont
  {Rosenbluth}}, \bibinfo {author} {\bibfnamefont {R.~B.}\ \bibnamefont
  {White}}, \ and\ \bibinfo {author} {\bibfnamefont {C.~S.}\ \bibnamefont
  {Liu}},\ }\bibfield  {title} {\enquote {\bibinfo {title} {{Temporal evolution
  of a three-wave parametric instability}},}\ }\href {\doibase
  10.1103/PhysRevLett.31.1190} {\bibfield  {journal} {\bibinfo  {journal}
  {Physical Review Letters}\ }\textbf {\bibinfo {volume} {31}},\ \bibinfo
  {pages} {1190--1193} (\bibinfo {year} {1973})}\BibitemShut {NoStop}%
\bibitem [{\citenamefont {Afeyan}\ and\ \citenamefont
  {Williams}(1985)}]{afeyan1985}%
  \BibitemOpen
  \bibfield  {author} {\bibinfo {author} {\bibfnamefont {B.~B.}\ \bibnamefont
  {Afeyan}}\ and\ \bibinfo {author} {\bibfnamefont {E.~A.}\ \bibnamefont
  {Williams}},\ }\bibfield  {title} {\enquote {\bibinfo {title} {{Stimulated
  Raman sidescattering with the effects of oblique incidence}},}\ }\href
  {\doibase 10.1063/1.865340} {\bibfield  {journal} {\bibinfo  {journal}
  {Physics of Fluids}\ }\textbf {\bibinfo {volume} {28}},\ \bibinfo {pages}
  {3397--3408} (\bibinfo {year} {1985})}\BibitemShut {NoStop}%
\bibitem [{\citenamefont {Seka}\ \emph {et~al.}(2014)\citenamefont {Seka},
  \citenamefont {Myatt}, \citenamefont {Short}, \citenamefont {Froula},
  \citenamefont {Katz}, \citenamefont {Goncharov},\ and\ \citenamefont
  {Igumenshchev}}]{Seka2014}%
  \BibitemOpen
  \bibfield  {author} {\bibinfo {author} {\bibfnamefont {W.}~\bibnamefont
  {Seka}}, \bibinfo {author} {\bibfnamefont {J.~F.}\ \bibnamefont {Myatt}},
  \bibinfo {author} {\bibfnamefont {R.~W.}\ \bibnamefont {Short}}, \bibinfo
  {author} {\bibfnamefont {D.~H.}\ \bibnamefont {Froula}}, \bibinfo {author}
  {\bibfnamefont {J.}~\bibnamefont {Katz}}, \bibinfo {author} {\bibfnamefont
  {V.~N.}\ \bibnamefont {Goncharov}}, \ and\ \bibinfo {author} {\bibfnamefont
  {I.~V.}\ \bibnamefont {Igumenshchev}},\ }\bibfield  {title} {\enquote
  {\bibinfo {title} {{Nonuniformly driven two-plasmon-decay instability in
  direct-drive implosions}},}\ }\href {\doibase 10.1103/PhysRevLett.112.145001}
  {\bibfield  {journal} {\bibinfo  {journal} {Physical Review Letters}\
  }\textbf {\bibinfo {volume} {112}},\ \bibinfo {pages} {145001} (\bibinfo
  {year} {2014})}\BibitemShut {NoStop}%
\end{thebibliography}%

\end{document}